\documentclass[structabstract]{aa}  

\usepackage{graphicx}
\usepackage{natbib}
\usepackage{morefloats}

\bibpunct{(}{)}{;}{a}{}{,} 
\usepackage{txfonts}

\begin{document}

 \title{A search for steep spectrum radio relics and halos with the GMRT}

 \titlerunning{Observations of diffuse steep-spectrum sources}

   \author{R.~J. van Weeren\inst{1}
         \and H.~J.~A. R\"ottgering\inst{1}
          \and M. Br\"uggen \inst{2}
          \and A. Cohen \inst{3}
          }

   \institute{Leiden Observatory, Leiden University,
              P.O. Box 9513, NL-2300 RA Leiden\\
              \email{rvweeren@strw.leidenuniv.nl}
                 \and             Jacobs University Bremen, PO Box 750 561, 28725 Bremen
                   \and   Naval Research Laboratory, Code 7213, Washington, DC 20375, USA             }


\abstract
   {Diffuse radio emission, in the form of radio halos and relics, traces regions in clusters with shocks or turbulence, probably produced by cluster mergers. The shocks and turbulence are important for the total energetics and detailed temperature distribution within the intracluster medium (ICM).  Only a small fraction of clusters exhibit diffuse radio emission, whereas a large majority of well-studied clusters shows clear substructure in the ICM. Some models of diffuse radio emission in clusters indicate that virtually all clusters should contain diffuse radio sources with a steep spectrum.  External accretion shocks associated with filamentary structures of galaxies could also accelerate electrons to relativistic energies and hence produce diffuse synchrotron emitting regions. The detection of radio emission from such filaments is important for our understanding of the origin of the Warm-Hot Intergalactic Medium (WHIM), and relativistic electrons and magnetic fields in the cosmic web. Here we report on Giant Metrewave Radio Telescope (GMRT) observations of a sample of steep spectrum sources from the 74~MHz~VLSS survey.    These sources are diffuse on scales~$\gtrsim~15\arcsec$, and not clearly associated with nearby ($z~\lesssim~0.1$) galaxies.  
}
   { 
   The main aim of the observations is to search for diffuse radio emission associated with galaxy clusters or the cosmic web. 
   }
   {We have carried out GMRT~610~MHz continuum observations of unidentified diffuse steep spectrum sources. 
   }
   { We have constructed a sample of diffuse steep spectrum sources, selected from the 74~MHz VLSS survey. We identified eight diffuse radio sources probably all located in clusters. We found five radio relics, one cluster with a giant radio halo and a radio relic, and one radio mini-halo. The giant radio halo has the highest radio power ($P_{1.4}$) known to date. By complementing our observations with measurements from the literature we find correlations between the physical size of relics and the spectral index, in the sense that smaller relics have steeper spectra. Furthermore, larger relics are mostly located in the outskirts of clusters while smaller relics are located closer to the cluster center.   
   }
{
}
   \maketitle

\section{Introduction}

Studies of large-scale structure (LSS) formation have made significant advances during the last 
decade. It has been found that nearly all massive clusters have undergone at least several mergers in their 
history and that presently clusters are still in the process of accreting matter. A significant fraction of the 
accreting mass is in the form of (smaller) clusters and galaxy groups. Cluster mergers are the most energetic 
events in the present day Universe, with kinetic energies of the order of $10^{63}-10^{64}$ erg, which are dissipated in giant 
shock waves and turbulence. An important aspect is the total energy budget and the detailed temperature distribution within the ICM, both of them are affected by the merger history of a cluster \citep[e.g.,][]{2008SSRv..134..311D}. 

Diffuse steep spectrum radio emission is observed in about 50 massive merging and post-merging galaxy clusters \citep[see the review by][and references therein]{2008SSRv..134...93F}. This diffuse emission is difficult to detect due to its low surface brightness and steep spectral index\footnote{$F_{\nu} \propto \nu^{\alpha}$, with $\alpha$ the spectral index}, so the fraction of clusters hosting diffuse radio emission is probably larger than we currently know. The diffuse emission in clusters is commonly divided into three main classes \citep{1996IAUS..175..333F}. \emph{Radio Halos} are extended ($\gtrsim 1$~Mpc) diffuse unpolarized ($\lesssim 5\%$) sources, located in the center of clusters. They have a regular smooth appearance,  and follow the thermal X-ray emission. \emph{Radio Relics} are elongated structures with an irregular morphology, mostly located in the periphery of clusters. Relics can be highly polarized ($10-50\%$). Several different subclasses have been identified \citep{2004rcfg.proc..335K}. Most known radio relics and halos are found in clusters which show signs of a current or recent merger. This supports the scenario in which the relativistic electrons are accelerated by merger-induced shocks or turbulence. However, \emph{Radio Mini-halos} are not associated with merging clusters. They are found in the centers of cool core clusters \citep[e.g.,][]{1991A&ARv...2..191F,  2006PhR...427....1P} and are associated with the central cluster galaxy and typically have sizes $\lesssim 500$~kpc, with the diffuse emission surrounding the central cluster galaxy \citep[e.g.,][]{2009A&A...499..371G}.

Two different mechanisms for in-situ acceleration of particles have been proposed to explain relics in clusters: (i) adiabatic compression of fossil radio plasma by a passing shock wave producing a so called radio ``phoenix'' \citep{2001A&A...366...26E, 2002MNRAS.331.1011E}, or (ii) diffusive shock acceleration (DSA) by the Fermi-I process \cite[e.g.,][]{1983RPPh...46..973D, 1987PhR...154....1B, 1991SSRv...58..259J, 1998A&A...332..395E, 2001RPPh...64..429M}. In the first scenario, radio relics should have toroidal and complex filamentary morphologies. These relics are capable of producing very steep, curved radio spectra due to  inverse Compton (IC) and synchrotron losses. In the DSA scenario the electrons are accelerated by multiple crossings of the shock front (in a first order Fermi process). These relics have large sizes (Mpc) and are direct tracers of shock fronts in clusters. The spectral index is determined by the balance between the continuous acceleration at the shock front and energy losses in the post-shock
region.

The  diffuse emission within clusters reveals the presence of relativistic electrons and magnetic fields on scales $\sim 1$~Mpc. Spectral aging, due to synchrotron and IC losses of the emitting electrons may explain the steep spectra. If the electrons are injected via Fermi acceleration (DSA), their energy follows a power-law distribution. The power-law index of the injected electrons is related to the Mach number of the shocks \citep[e.g.,][]{2007MNRAS.375...77H}: shocks with a low Mach number have steeper radio spectra. 
Clearly, low-frequency surveys are needed to locate and study these sources \citep{2006MNRAS.369.1577C, 2007MNRAS.378.1565C, 2008A&A...480..687C}. Interestingly, \cite{2008Natur.455..944B} discovered a radio halo in the cluster \object{Abell~521}, which was previously known to host a radio relic, with a spectral index of $\sim-2.1$, suggesting the existence of a population of diffuse source in clusters with spectral indices~$< -1.5$.

Numerical simulations show the development of various types of shocks during structure formation~\citep{2000ApJ...542..608M}. These shocks differ in their location with respect to the cluster center and Mach numbers \citep{2000ApJ...542..608M, 2002MNRAS.337..199M, 2003ApJ...593..599R, 2006MNRAS.367..113P, 2009MNRAS.395.1333V}. External accretion shocks have $\mathcal{M} \gg 1$ and process the low-density, unshocked intergalactic medium (IGM). This results in relatively flat spectral indices of about $-0.5$ at the location of the shock front. Further away from the shock front the spectral index steepens due to synchrotron and IC losses. 
Internal shocks, (i.e., merger and flow shocks) occur within the cluster. The Mach numbers of these shocks are lower resulting in steeper spectral indices. 
Binary merger shocks are the result of a cluster merging with a another cluster or a large sub-structure. In this case a double radio relic is expected \citep[e.g.,][]{1999ApJ...518..603R, 2008MNRAS.391.1511H, 2009arXiv0908.0728V}.


As pointed out by \cite{1999ApJ...514....1C}, hydrodynamic models indicate that up to half of the baryons at present time should have temperatures in the range of $10^{5}-10^{7}$~K. Unfortunately, studying the abundance and distribution of this WHIM is very challenging, since its main tracers are highly excited Oxygen lines which are difficult to observe \citep[e.g.,][]{2005Natur.433..495N}. A fraction of the accretion shocks will be supersonic and can accelerate energetic electrons up to energies of $10^{18}-10^{19}$ eV \citep[e.g.,][]{1995ApJ...454...60N, 1996ApJ...456..422K, 2008ICRC....4..555I}. In the presence of magnetic fields, such electrons will emit faint diffuse synchrotron radiation.  The detection of these radio filaments is very important as this would provide a probe of the WHIM. 
Recent magnetohydrodynamical modeling indicates that detecting radio emission from the filamentary cosmic web should be possible \citep[e.g.,][]{2004ApJ...617..281K, 2007MNRAS.375...77H}. \cite{2008MNRAS.391.1511H, 2008MNRAS.385.1242P, 2008MNRAS.385.1211P} however find that in the outskirts of clusters (at a few times the virial radius) or filaments, external accretion shocks cause little radio emission, owing to the low density of both magnetic field energy and cosmic ray (CR) particles there \citep{2001ApJ...562..233M}. They are therefore difficult to detect even with the sensitivity of upcoming radio telescopes such as LOFAR.  Relic emission from internal accretion shocks occur in a higher density environment so that they should be detected with current radio facilities.

When searching for radio halos, relics and filaments in low-frequency radio surveys, various other steep spectrum sources are also present. These include ultra-steep (angular size $\lesssim 15\arcsec$) spectrum sources \citep[USS, see ][for a review]{2008A&ARv..15...67M} associated with high-z radio galaxies (HzRG), ``fossil'' or ``dying'' FR-I \citep{1974MNRAS.167P..31F} radio sources, and ``head-tail'' sources, the last two having a steep spectrum due to spectral aging of the radio emission. In high-resolution ($\lesssim 5\arcsec$) observations, for example at $1.4$~GHz with the Very Large Array (VLA), diffuse objects will be resolved out due to missing short baselines. This provides a method for selecting diffuse radio sources associated with galaxy clusters or the cosmic web. FR-I sources can be partly excluded by removing sources that are clearly associated with individual galaxies.

The $74$~MHz VLA low-frequency Sky Survey (VLSS), \cite{2007AJ....134.1245C}, covers about $3\pi$~steradians of sky north of $\delta=-30\degr$. The resolution of the survey is 80\arcsec~(FWHM) and the rms noise level is about 0.1~Jy~beam$^{-1}$. The source catalog contains roughly 70,\ 000 sources with a point source detection limit of $0.7$~Jy beam$^{-1}$. A new calibration algorithm \citep{2004SPIE.5489..180C} was used to remove the ionospheric distortions, which can be severe at this low-frequency. The $1.4$ GHz~NRAO VLA Sky Survey (NVSS), \cite{1998AJ....115.1693C}, covers the entire sky above $\delta = -40\degr$. The NVSS images have a rms noise of about 0.45~mJy~beam$^{-1}$, and a resolution of 45\arcsec (FWHM). The catalog contains about $2 \times 10^{6}$ sources above a flux of $\sim 2.5$~mJy. 

In this paper we present radio continuum observations of 26~diffuse (angular size $\gtrsim 15\arcsec$) steep spectrum sources  selected from the VLSS survey with the GMRT at 610~MHz. The main aim of this project is to determine the morphology of the sources and search for diffuse structures which could be associated with shock fronts or turbulence in clusters, and accretion shocks onto filaments of galaxies. 

The layout of this paper is as follows. In Sect.~\ref{sec:selection} we discuss the sample selection, this is followed by an overview of the observations and data reduction in Sect.~\ref{sec:obs-reduction}. In Sect.~\ref{sec:results} we present the radio maps of the most interesting sources and discuss these sources individually. By combining our radio observations with data from the literature (X-ray and optical observations) we have tried to classify the sources. In Sect. \ref{sec:spectra}, spectral indices are modeled using our flux measurements combined with literature values. We end with a discussion and conclusions in Sects.~\ref{sec:discussion} and~\ref{sec:conclusion}.

Throughout this paper we assume a $\Lambda$CDM cosmology with $H_{0} = 71$ km s$^{-1}$ Mpc$^{-1}$, $\Omega_{m} = 0.3$, and $\Omega_{\Lambda} = 0.7$.

\section{Sample Selection}
\label{sec:selection}
Spectral indices (between 1400 and 74~MHz) were calculated for all sources in the VLSS survey. We selected sources with $\alpha \leq -1.35$ which resulted in a total of $176$ sources. This cutoff is somewhat arbitrary, but a significant lower cutoff resulted in a very small number of sources selected while a higher cutoff would select too many sources for follow-up observations. VLA B-array 1.4~GHz $\sim5$ min snapshot observations were carried out on March 25--29 and May 10, 2005 of a subsample of 68 from the 176 sources. Two intermediate frequencies (IFs) with a bandwidth of 50~MHz each were used, centered at $1385$ and $1465$~MHz. From this $68$ sources, $36$ were found to be resolved out. This showed that these sources had extended emission on scales $\gtrsim 15\arcsec$. From the 36 sources 13 were identified with known nearby galaxies. The remaining 23 sources were included in the sample. We have also searched for additional sources with $\alpha \leq -1.15$, by making use of the $1.4$~GHz~FIRST survey \citep[5\arcsec~FWHM, ][]{1995ApJ...450..559B}. Sources with a FIRST flux at least 8 times lower than the 1.4~GHz NVSS flux were initially selected and visually inspected to remove obvious double lobe sources. The spectral index cutoff of $-1.15$ was chosen because higher values resulted in too many double lobes to be selected for visual inspection. Furthermore, most known radio halos and relics have spectral indices steeper than this value. The amount of flux resolved out in the selection criterium was a tradeoff as lower values also resulted in too many sources to be selected for visual inspection. 
After visual inspection we found three additional sources which showed the presence of diffuse emission. None of these sources were clearly associated with nearby individual galaxies in the POSS-II or SDSS surveys \citep{2009ApJS..182..543A}. The final list of sources and their coordinates are given in Table~\ref{tab:results}.

\section{Observations \& Data Reduction}
\label{sec:obs-reduction}
High-sensitivity radio observations at $610$~MHz were carried out with the GMRT in February and November 2008 of a sample of $26$ diffuse steep spectrum radio sources. 
We divided a total $102$~hours of observation time evenly between the 26 sources. A total of $32$~MHz bandwidth was recorded, using both  the upper (USB) and lower sidebands (LSB) which included both RR and LL polarizations. The data were collected in spectral line mode with $128$~channels per sideband (IF), resulting in a spectral resolution of $125$~kHz per channel. To increase UV-coverage we cycled between various sources, typically spending $40$~min on a sources before moving to the next source. However, due to scheduling constraints it was only possible to do this for about half of our sources. The observations resulted in a net on-source time of $\sim 3$~hours, after flagging certain time-ranges which were affected by radio frequency interference (RFI) or had other problems.

The data were reduced and analyzed with the NRAO Astronomical Image Processing System (AIPS) package. 
 Bandpass calibration was carried out using the standard flux calibrators: 3C48, 3C147, and 3C286.  Fluxes of 29.43 Jy (3C48), 38.26 Jy (3C147), and 21.07 Jy (3C286) at $610$~MHz were assigned to these sources using the \cite{perleyandtaylor} extension to the \cite{1977A&A....61...99B} scale. A set of $6$ channels free of RFI was taken to normalize the bandpass (channel $15-20$) for each antenna. Strong RFI was removed automatically (with the AIPS task `FLGIT'). The data was then visually inspected for remaining low-level RFI using the AIPS tasks `SPFLG' and `TVFLG'. After that an initial phase and amplitude calibration was carried out using the bandpass and secondary calibrators, where we also transferred the flux densities from the primary calibrators to the secondary calibrators. The found solutions were then transferred to the target sources. We have not chosen to average any channels in order to minimize the effects of bandwidth smearing and to aid further removal of RFI.  The first and last few channels of the data were discarded as they were noisy. 

For making images we used the polyhedron method \citep{1989ASPC....6..259P, 1992A&A...261..353C} to minimize the effects of non-coplanar baselines. Both USB and LSB were simultaneously gridded, imaged and cleaned. We used a total of 199 facets to cover $\sim 2$ times the full primary beam. This made the removal of sidelobes from strong sources far away from the field center possible. After a first round of imaging, in some cases ``ripples'' were seen in the maps which were subsequently removed after identifying the corresponding baseline(s). Several rounds of phase self-calibration were carried out before doing a final amplitude and phase selfcalibration.  Images were made using robust weighting \citep[robust = 0.5,][]{briggs_phd} and corrected for the primary beam attenuation. Images were cleaned to $3$ times the rms noise level to minimize clean bias effects.

The thermal noise in each map is expected\footnote{http://www.gmrt.ncra.tifr.res.in/gmrt\_hpage/

Users/doc/manual/UsersManual/node13.html} to be
\begin{equation}
\sigma_{\rm thermal} = \frac{\sqrt{2}T_{s}}{G\sqrt{n(n-1) N_{IF} \Delta\nu t_{\rm int} }} \mbox{ ,}
\end{equation}
with $T_{s} = 92$ K the system temperature, $G = 0.32$ K Jy$^{-1}$ the antenna gain, $n \approx 28$ the number of working antennas, $N_{IF}=2$ the number of sidebands used (both recording RR and LL polarizations), $\Delta\nu = 13.5$~MHz the bandwidth per sideband, and $t_{\rm int}$ the net integration time. The expected thermal noise for $3$ hrs integration time is about 32 $\mu$Jy~beam$^{-1}$, where we have taken into account that typically $20\%$ of the data is flagged due to RFI. The noise levels in our maps range from 40 to 202 $\mu$Jy~beam$^{-1}$. The noise levels dependent on the UV-coverage and the presence of strong confusing sources in the field of view limiting the dynamic range. The lowest noise level of 40 $\mu$Jy~beam$^{-1}$ is quite close to the thermal noise level. The uncertainty in the calibration of the absolute flux-scale is between $5-10\%$, see \cite{2004ApJ...612..974C}.

\section{Results}All sources, except \object{VLSS~J2229.1$-$0136}, were detected in the GMRT images. The NVSS counterpart of \object{VLSS J2229.1$-$0136} has a position offset of about 40\arcsec, the flux density in the NVSS and VLSS surveys of these sources is also close to the detection threshold. We therefore conclude that \object{VLSS~J2229.1$-$0136} is most likely a noise peak as the GMRT observations should have easily detected the source. From here on we have left out \object{VLSS~J2229.1$-$0136} in the further discussions.
A summary of the beam parameters and noise levels for the maps is given in Table~\ref{tab:results}. To identify optical counterparts overlays were made using SDSS and POSS-II images. Spectroscopic redshifts were included from the literature. For sources without a spectroscopic redshift, but having SDSS DR7 coverage we took the SDSS photometric redshift. For other sources we used that $K$ and $R$ magnitudes of massive elliptical galaxies correlate with redshift \citep[$K$-$z$ and $R$-$z$ relations, e.g.,][]{2003MNRAS.339..173W, 2007A&A...464..879D}.  

\label{sec:results}
We have also checked for any X-ray counterparts to the radio sources.  In the next subsection we describe the most interesting sources that appear to be related to shocks or turbulence in clusters or filaments. The other sources are discussed in Appendix~\ref{sec:fr-i}.

\begin{table*}
\begin{center}
\caption{Source properties and results}
\begin{tabular}{lllllllll}
\hline
\hline
Source Name &RA& DEC & Beam size & $\sigma_{\mathrm{rms}}$ & S$_{610}^{j, k}$   &$\alpha_{74 \mathrm{MHz}}^{1400 \mathrm{MHz}}$ & LAS$^{h}$ & Redshift \\
                         & (J2000)& (J2000)& arcsec        &  $\mu$Jy  &  mJy              &  & arcsec & z \\
\hline
VLSS J0004.9$-$3457 &00 04 53.63 &$-$34 56 34.1  & 11.4 $\times$ 6.6 & 73&$158 \pm 16$&$-1.40 \pm 0.04$ &116 & $0.29 \pm 0.08^{a}$\\ 

VLSS J0227.4$-$1642  &  02 27 26.74  &  $-$16 42 47.2     & 5.3 $\times$ 4.0 &  94    &  $37.0 \pm 6.2$ &$-1.50 \pm 0.06$ &89 &$\gtrsim 0.7$ \\ 
VLSS J0250.5$-$1247 & 02 50 30.83   &  $-$12 47 30.3 &   9.4 $\times$ 3.7 &  118   & $33.2 \pm 3.3 $    &$-1.94 \pm 0.07$&21 &$\gtrsim 0.7$ \\
VLSS J0511.6+0254     &05 11 37.79 &+02 54 19.4 & 8.7 $\times$ 6.1& 139& $98.0 \pm 10.3$ & $-1.42 \pm 0.05$& 45 & $ 0.20 \pm 0.05^{a}$\\ 

VLSS J0516.2+0103     &05 16 17.89 &+01 03 40.1 & 8.1 $\times$ 6.5 &79&$17.4 \pm 4.4 $& $-1.73 \pm 0.06$& 35 & $\gtrsim 0.7$\\ 

VLSS J0646.8$-$0722 &06 46 52.10   &  $-$07 22 37.9    & 5.7  $\times$ 4.8&  63  &   $120\pm 13$     &$-1.70 \pm 0.05$ & 59& $0.23 \pm 0.06^{a}$\\ 
VLSS J0717.5+3745$^{i}$    &07 17 30.92 &+37 45 29.7 & 8.2 $\times$ 6.0 &78& $501 \pm 50 $&$-1.15 \pm 0.04$ & 171 & 0.5548$^{f}$ \\ 

VLSS J0915.7+2511$^{i}$    & 09 15 41.51& +25 11 48.2  & 8.6 $\times$5.9 &134&  $194  \pm 21$  &$-1.52 \pm 0.04$ & 63 &$0.324^{c}$ \\
VLSS J1117.1+7003  & 11 17 06.46   &     +70 03 57.1   & 7.8 $\times$ 4.3 &  52  &  $9.1 \pm 0.9$  &$-1.87 \pm 0.07$ &18 &$\gtrsim 0.7$ \\  
VLSS J1133.7+2324  & 11 33 44.73 &+23 24 50.6&6.6 $\times$ 3.9& 51 &    $80.6 \pm 10.1$&$-1.69 \pm 0.06$ &84 & $0.61 \pm 0.16^{d}$\\ 

VLSS J1431.8+1331  &  14 31 49.90 &+13 31 54.1&5.3  $\times$ 4.8 & 51  & $120 \pm 13$    &$-2.03 \pm 0.05$ &101 & $0.159936^{c}$ \\ 
VLSS J1515.1+0424$^{i}$    &  15 15 09.18 &+04 24 41.1 &7.6 $\times$ 5.4&79 & $192 \pm 33$& $-1.50 \pm 0.05$& 169 & 0.0972$^{g}$ \\ 

VLSS J1636.5+3326  &16 36 34.93 &+33 26 33.9& 5.2 $\times$ 4.7 & 40 &   $16.1 \pm 1.9$  &$-1.84 \pm 0.14$& 28&$0.65 \pm 0.19^{d}$\\  
VLSS J1710.5+6844  &17 10 35.03 &+68 44 58.4  & 8.9 $\times$ 4.7 & 63  & $ 23.3 \pm 2.8 $ &$-1.75 \pm 0.06$ &45&$0.28 \pm 0.08^{a}$ \\  
VLSS J1930.4+1048  &19 30 27.19 &+10 48 02.5  & 5.3 $\times$ 4.8 & 56 &  $67.4 \pm 7.0 $  & $-1.90 \pm 0.06$    &101  &   \ldots     \\  
VLSS J2043.9$-$1118  &20 43 58.46 &$-$11 18 45.6 & 5.8 $\times$ 4.2  &  77 &  $53.9 \pm 6.3$  &$-1.74 \pm0.05$ &41 & $0.43 \pm 0.15^{b}$ \\  
VLSS J2044.7+0447  & 20 44 43.64 &+04 47 24.5  &10.5 $\times$ 7.5 & 202 &  $57.7 \pm 5.8  $ & $-1.55 \pm 0.06$&71 & $0.46 \pm 0.15^{b}$  \\  
VLSS J2122.9+0012  & 21 22 54.14 &+00 12 03.4  & 5.8 $\times$ 3.8  & 81  & $24.0 \pm 2.4$  & $-2.00 \pm 0.08$&12 &$\gtrsim 1.4$ \\ 
VLSS J2209.5+1546   &22 09 32.91& +15 46 29.9 &6.9 $\times$ 6.2 & 64& $36.4 \pm 3.8$ & $-1.56 \pm 0.07$ & 61 & $\gtrsim 0.7$\\ 

VLSS J2213.2+3411  & 22 13 12.45 &+34 11 51.6   & 5.5 $\times$  4.6 & 94 &  $127 \pm 13$   &$-1.55 \pm 0.04$ &62&$1.6 \pm 0.5^{a, e}$\\  

VLSS J2217.5+5943  & 22 17 30.39 &+59 43 05.3  & 6.3 $\times$ 4.3 & 47  &  $79.9 \pm 8.0$  & $-2.20 \pm 0.06$   &   99&   \ldots     \\  
VLSS J2229.1$-$0136  &22 29 11.95 &$-$01 36 58.8 & 8.6 $\times$  4.2 & 101 &   \ldots   & \ldots  &   \ldots   & \ldots   \\ 

VLSS J2241.3$-$1626  & 22 41 22.57& $-$16 25 35.7  &6.1 $\times$ 5.8&129 &$69.7 \pm 7.3$ &$-1.44 \pm 0.06$ & 47 &  $\gtrsim 0.7$\\ 

VLSS J2341.1+1231  &23 41 06.91 &+12 31 36.9  & 5.5 $\times$ 4.4  & 50  & $117  \pm 14$  & $-1.70 \pm 0.04$&115&$0.62 \pm 0.15^{b}$  \\  
VLSS J2345.2+2157  & 23 45 15.47& +21 57 55.1  & 4.8 $\times$ 4.2 & 63 &   $79.3 \pm 8.4$   &$-2.14 \pm 0.05$ &70&$0.23 \pm 0.06^{a}$\\ 
VLSS J2357.0+0441     &23 57 05.54 &+04 41 14.7&7.8 $\times$ 4.9&82& \ldots  & \ldots  & \ldots & \ldots\\

\hline
\hline
\end{tabular}
\label{tab:results}
\end{center}
$^{a}$ redshift estimated using the fitted Hubble-K relation from \cite{2003MNRAS.339..173W}\\
$^{b}$ redshift estimated using the fitted Hubble-R relation from \cite{2007A&A...464..879D} \\
$^{c}$ spectroscopic redshift from SDSS DR7\\
$^{d}$ median photometric redshift from SDSS DR7\\
$^{e}$ using the K$_{s}$ band magnitude from \cite{2002AJ....123..637D}\\
$^{f}$  \cite{2003MNRAS.339..913E}\\
$^{g}$  \cite{1999ApJS..125...35S}\\
$^{h}$ LAS = largest angular size\\
$^{i}$ resolved out in the 1.4~GHz FIRST survey\\
$^{j}$ fluxes are extracted from the regions indicated in the figures by dotted lines. \\
$^{k}$ for the compact sources (LAS $< 45\arcsec$) the fluxes were measured by fitting single Gaussians to the sources after convolving the maps to the 45\arcsec NVSS resolution\\
\end{table*}

\subsection{Individual sources}

\subsubsection{VLSS J1133.7+2324}  
\begin{figure}

\begin{center}
\includegraphics[angle =90, trim =0cm 0cm 0cm 0cm,width=0.5\textwidth]{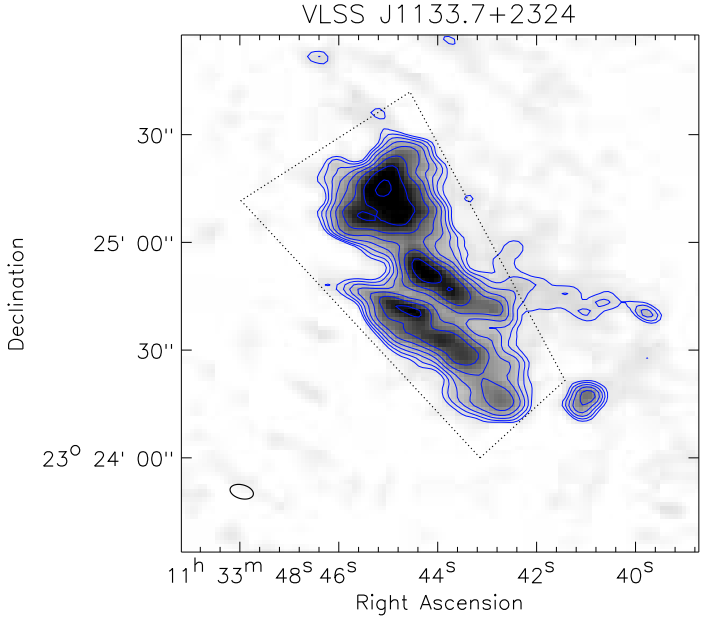}
\end{center}
\caption{GMRT 610 MHz radio map.  The Gaussian beam size is shown at the bottom left corner. Contour levels are drawn at $\sqrt{[1,2,4,6,8,16,32,...]}  \times 4\sigma_{\mathrm{rms}}$, see Table~\ref{tab:results}. The dotted lines indicate the region where the 610~MHz flux (S$_{610}$), reported in Table~\ref{tab:results}, has been extracted.}
\label{fig:radiomap5}
\end{figure}

\begin{figure}
    \begin{center}
      \includegraphics[angle = 90, trim =0cm 0cm 0cm 0cm,width=0.5\textwidth]{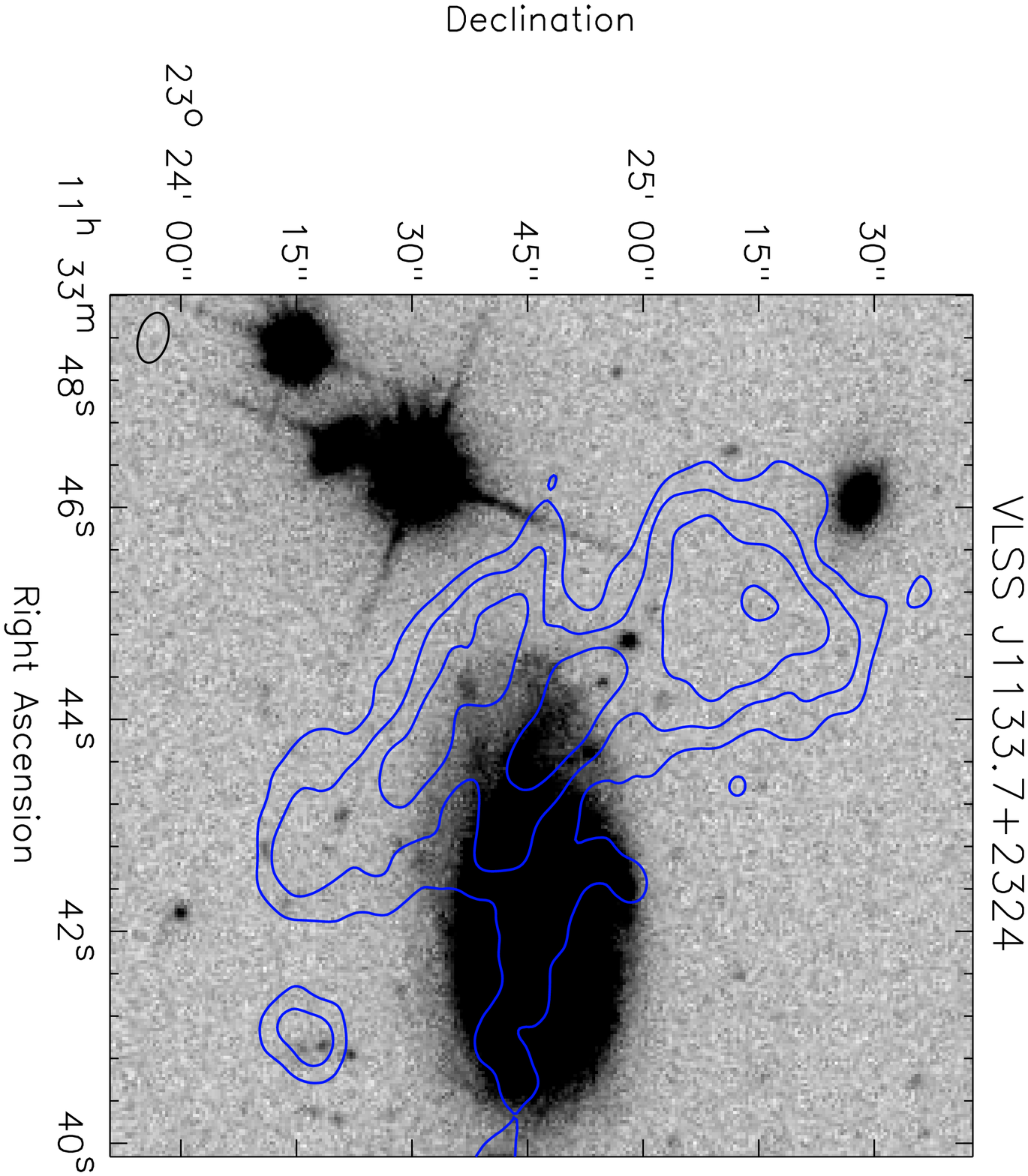}
       \end{center}
      \caption{SDSS r band image overlaid with GMRT 610~MHz contours. Contour levels are drawn at $\sqrt{[1,8, 32, 128, 512,...]} \times 4\sigma_{\mathrm{rms}}$.}
            \label{fig:radiomap5_o}
 \end{figure}

The radio emission shows two parallel filamentary structures and a northern patch of diffuse emission connected with the western filamentary component.  Some faint radio emission is extending towards the west. This extension coincides with the galaxy \object{UGC~6544} (\object{MCG+04-27-065}; \object{PGC 35694}) located a z=0.02385 \citep{1997AJ....113.1197H}. This galaxy was classified as a Sbc spiral by \cite{1973UGC...C...0000N} and \cite{2003A&A...412...45P}. The source is also listed in the IRAF Faint Source Catalog (\object{F11311+2341}) and detected at $60$ and $100$ $\mu$m \citep{1990IRASF.C......0M}. The galaxy has been included in 21 cm hydrogen emission line surveys. \cite{2005ApJS..160..149S} give a self-absorption corrected line flux of $2.97 \pm 0.32$ Jy km s$^{-1}$ and a line-width of $372 \pm 3$ km s$^{-1}$.The galaxy also has a small UV-excess \citep{2002PNAOJ...6..107M}.

Using the IRAS fluxes we can estimate the star forming rate (SFR) in this galaxy \citep{1997ApJ...478..144S} by calculating the Far Infrared Luminosity ($L_{\mathrm{FIR}}$)
\begin{equation}
L_{\mathrm{FIR}} = 3.94 \times 10^{5}(2.58 S_{60\mu \mathrm{m}} + S_{100\mu \mathrm{m}})r(S_{60\mu \mathrm{m}}/S_{100\mu \mathrm{m}})D_{L}^{2} \mbox{ ,}
\end{equation}
with $L_{\mathrm{FIR}}$ in $L_{\sun}$, the flux $S$ in Jy, $D_{L}$ the luminosity distance in Mpc, and $r(S_{60}/S_{100})$ a color correction factor \citep{1985cgqo.book.....L} which is of the order of one. This gives a SFR of $\sim 3 M_{\sun}$yr$^{-1}$, using SFR ($M_{\sun} \mbox{yr}^{-1}$) = $L_{\mathrm{FIR}}/(5.8 \times 10^{9} L_{\sun})$, \cite{1998ApJ...498..541K}. 
The radio flux of the galaxy is difficult to measure as part of the radio emission from the galaxy overlaps with the emission from the filamentary source. For the part which does not overlap with the steep spectrum source we measure a flux of about $6$~mJy. We adopt a total flux for the galaxy of $10$~mJy, assuming we missed about 40\% of the previously reported flux. The luminosity at $1.49$~GHz (L$_{1.49\,\mathrm{GHz}}$) is $L = 4\pi D_{L}^{2} S_{\nu}(1+z)^{-\alpha-1}$, with $S_{\nu}$ the observed flux at the rest frequency, and $\alpha \sim -0.5$. This results in $L_{1.49\,\mathrm{GHz}} = 8.2 \times 10^{21}$ W~Hz$^{-1}$. Using the FIR radio correlation \citep{1991ApJ...376...95C} we find a luminosity of $\sim 10^{22}$~W~Hz$^{-1}$, in good agreement.

The galaxy is unlikely to be associated with the filamentary radio structure to the east, given the above calculation, the steep spectral index, morphology, and spatial offset. In the background (partly behind the spiral galaxy) there is an overdensity of faint red galaxies. These have a median photometric redshift of about 0.6 (SDSS DR7) and follow roughly the filamentary radio source. This is probably a cluster or a galaxy filament, with the foreground galaxy hiding part of the cluster/filament. Based on the morphology, the radio source is then classified as a relic. If the steep spectrum radio source is located at a distance of $z~=~0.6$, the LAS of 84\arcsec corresponds to a physical size of of 660~kpc.

\subsubsection{VLSS J1431.8+1331} 
\begin{figure}
    \begin{center}
      \includegraphics[angle = 90, trim =0cm 0cm 0cm 0cm,width=0.5\textwidth]{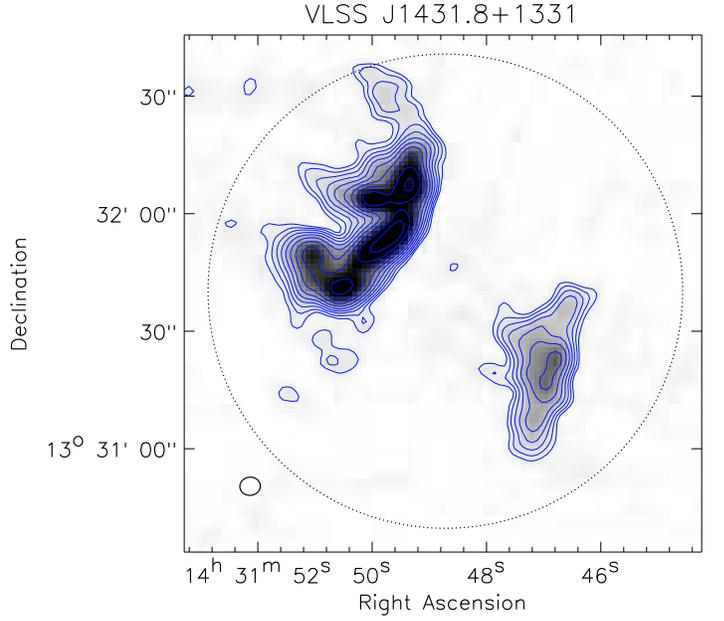}
       \end{center}
      \caption{GMRT 610~MHz radio map. Contour levels are drawn as in Fig.~\ref{fig:radiomap5}. }
            \label{fig:radiomap6}
 \end{figure}
 \begin{figure}
    \begin{center}
      \includegraphics[angle = 90, trim =0cm 0cm 0cm 0cm,width=0.5\textwidth]{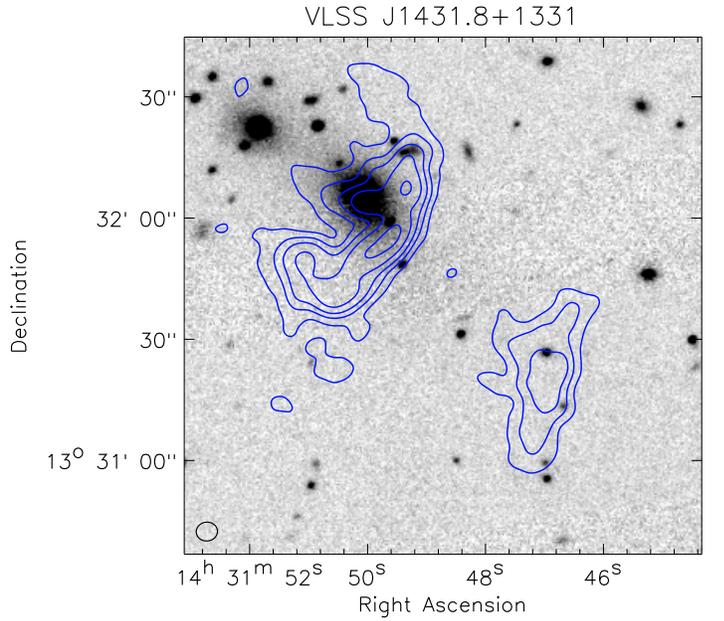}
       \end{center}
      \caption{SDSS r band image overlaid with GMRT 610~MHz contours. Contour levels are drawn as in Fig.~\ref{fig:radiomap5_o}.}
             \label{fig:radiomap6_o}
 \end{figure}
The radio emission shows two distinct components, the brightest one located to the east. This source has an optical counterpart, the cD galaxy of the cluster \object{MaxBCG~J217.95869+13.53470} \citep{2007ApJ...660..239K}. The cluster is  also detected in X-rays by ROSAT \citep{1999A&A...349..389V, 2000IAUC.7432R...1V} as source \object{1RXS~J143150.6+133256}. A SDSS DR7 spectrum puts the cD galaxy at a distance of $z=0.160$. Three other sources in the cluster have a measured redshifts of  $0.15962$, $0.159153$, and $0.164794$. At this redshift the radio emission corresponds to a physical size of $170$ kpc for the eastern and $125$ kpc for the western component. The eastern component is an irregular curved structure with several bright knots. One of these corresponds with the nucleus of the cD galaxy. The morphology suggests that we are either seeing the interaction of radio plasma from the central AGN with the surrounding ICM or a small central radio source with a bright relic tracing a shock front (seen in projection).

The western fainter component does not seem to be associated with any galaxy. The source could either be remnant radio emission from a previous AGN episode or the signature of a shock wave. In the former case, the spectral index should be steeper then eastern component because of spectral aging.

\subsubsection{VLSS J2217.5+5943, 24P73} 
\begin{figure}
    \begin{center}
      \includegraphics[angle = 90, trim =0cm 0cm 0cm 0cm,width=0.5\textwidth]{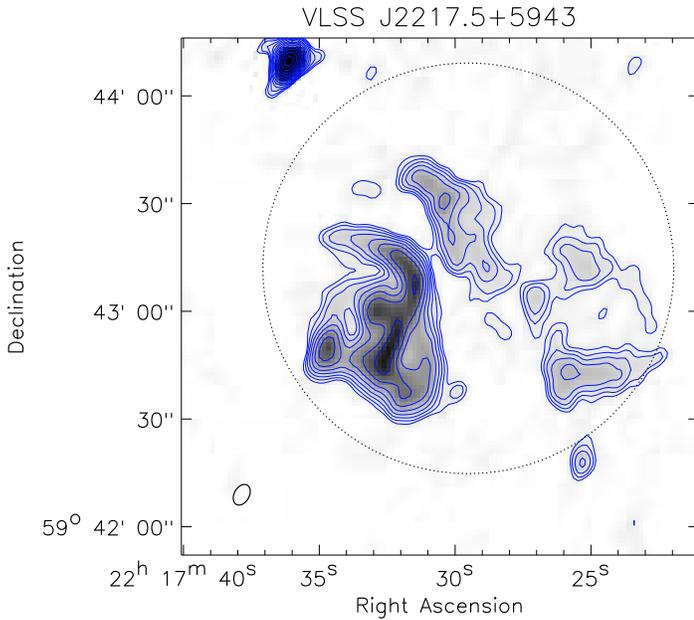}
       \end{center}
      \caption{GMRT 610~MHz radio map. Contour levels are drawn as in Fig.~\ref{fig:radiomap6}. }
             \label{fig:radiomap14}
 \end{figure}
This source was found to have an ultra-steep spectrum ($\alpha=-2.58 \pm 0.14$) by the Synthesis Telescope of the Dominion Radio Observatory (DRAO) Galactic plane survey at 408~MHz and 1.42~GHz \citep{1989JRASC..83..105H,1990A&AS...82..113J}. Since this source was located in the galactic plane, \cite{1990A&AS...82..113J} suggested the sources might be a pulsar. However, they noted that the source seemed to be slightly extended at $408$~MHz. At $1.42$~GHz the source was resolved into two distinct components. Both of these components were also listed as a single source \object{25P23}. The source was also detected in the $38$~MHz~8C survey \citep{1990MNRAS.244..233R,1995MNRAS.274..447H}. Subsequent L-band ($1.4$ GHz) and X-band ($8.4$ GHz) observations with the VLA by \cite{1994A&AS..104..481G}, separated the source clearly into two components. A compact northern component and a southern diffuse component. The X-band observations only detected the northern component. By comparing the fluxes of the VLA and DRAO observations they concluded that the southern diffuse component had an ultra-steep spectrum and provided the bulk of the emission at low frequencies. Since the source was resolved this ruled out a pulsar identification. In fact it was suggested that this source might be a radio halo or relic located in a galaxy cluster behind the galactic plane. 

GMRT observations of \object{24P73} show a complex filamentary source. The source has some similarities with the relic sources in Abell~85 and Abell~133 \citep{2001AJ....122.1172S}. POSS-II and 2MASS images covering the area do not show the presence of any cluster. However this is not unexpected given the extinction of $A_{B}=6.7$ \citep{1998ApJ...500..525S}. Given the steep spectral index and morphology we conclude that the sources is a radio relic. Deep NIR imaging will be necessary to identify the galaxy cluster associated with the relic. 

\subsubsection{VLSS J0004.9$-$3457} 
\begin{figure}
    \begin{center}
      \includegraphics[angle = 90, trim =0cm 0cm 0cm 0cm,width=0.5\textwidth]{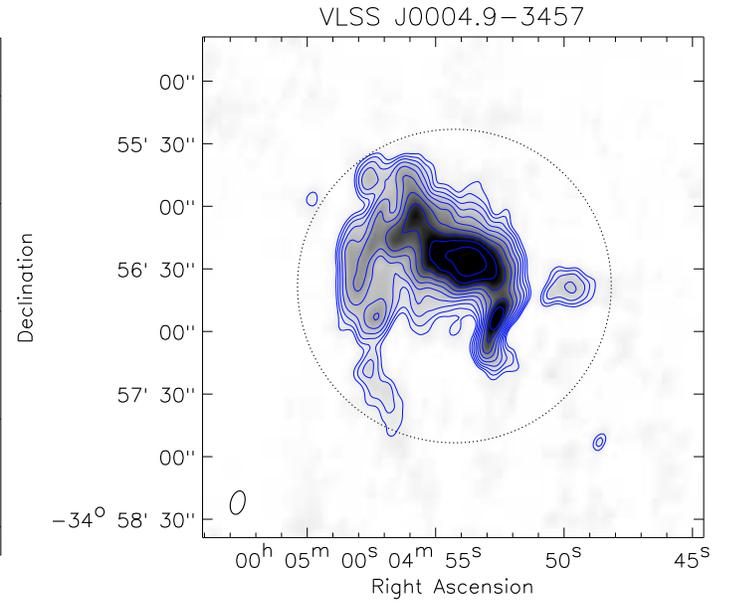}
       \end{center}
      \caption{GMRT 610~MHz radio map. Contour levels are drawn as in Fig.~\ref{fig:radiomap5}.}
             \label{fig:radiomap18}
 \end{figure}
 \begin{figure}
    \begin{center}
      \includegraphics[angle = 90, trim =0cm 0cm 0cm 0cm,width=0.5\textwidth]{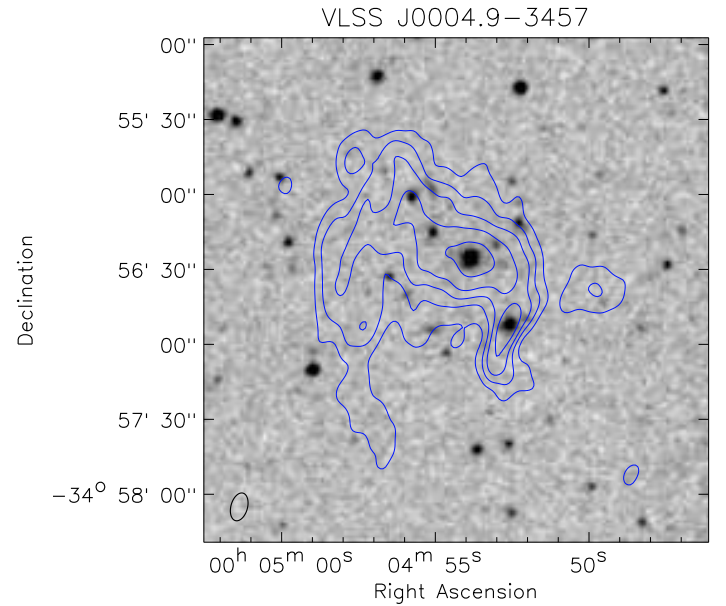}
       \end{center}
      \caption{POSS-II red image overlaid with GMRT 610~MHz contours. Contour levels are drawn as in Fig.~\ref{fig:radiomap5_o}.}
            \label{fig:radiomap18_o}
 \end{figure}
The central component of the radio source is associated with a K=14.86~mag elliptical galaxy.  An overdensity of galaxies indicates the presence of a cluster, candidate \object{B02291} \citep{2001A&A...379...21Z} from radio-optically selected clusters of galaxies. We estimate a redshift for the cluster of $0.29 \pm 0.08$ (using the K-band magnitude and the K-z relation). The cluster may be linked to a larger cluster about 3\arcmin~towards the southwest at a redshift of $0.33 \pm 0.09$. The X-ray source \object{2E 0002.2-3515} \citep{1996yCat.9013....0H} is probably associated with this cluster of galaxies visible in POSS-II images. The diffuse radio emission surrounding the central elliptical galaxy resembles a mini-halo. No X-ray emission from the ICM of the cluster is detected in the ROSAT All Sky Survey. This implies that the cluster is not very massive. The object is similar to the mini-halo in the cluster \object{MRC 0116+111} located at $z = 0.131$ \citep{2002IAUS..199..159G, 2009MNRAS.tmp.1144B}, which was also not detected in the ROSAT All Sky Survey. The size of the mini-halo is about $200$~kpc, similar to other mini-halos \citep[e.g.,][]{2009A&A...499..371G}. The radio image shows a bright knot at RA 00$^\mathrm{h}$ 04$^\mathrm{m}$ 52.5$^\mathrm{s}$ DEC~$-$34\degr~56\arcmin~55\arcsec, to the south of the main component. This knot has an optical counterpart. 
To the east an arc-like structure extends from the central component and bends towards the south. The arc is not associated with an optical counterpart. The origin of the arc is unclear, it could be part of a disturbed FR-I source, or a relic-like structure of fossil radio plasma.

\subsubsection{VLSS J0717.5+3745, MACS~J0717.5+3745}   
The radio source is associated with the massive X-ray luminous cluster \object{MACS J0717.5+3745} at z=0.5548, with an overall ICM temperature of 11.6~keV \citep{2001ApJ...553..668E, 2003MNRAS.339..913E, 2007ApJ...661L..33E}. The cluster shows a pronounced substructure in optical images. \cite{2004ApJ...609L..49E} reported the discovery of a large-scale filamentary structure of galaxies connected to the cluster. NVSS, WENSS and VLSS images reveal the presence of a steep-spectrum radio source ($\alpha =-1.15$). The radio source was classified as a radio relic by \cite{2003MNRAS.339..913E}. \cite{2009ApJ...693L..56M} presented X-ray (Chandra) and optical observations (Hubble Space Telescope, ACS; Keck-II, DEIMOS) of the cluster. They found the cluster to be an active triple merger. Temperatures in the cluster exceeding $20$~keV were found in some regions. Regions with a lower temperature of $\sim5$~keV were found at the position of two subclusters, probably remnants of cool cores. 
\begin{figure}
    \begin{center}
      \includegraphics[angle = 90, trim =0cm 0cm 0cm 0cm,width=0.5\textwidth]{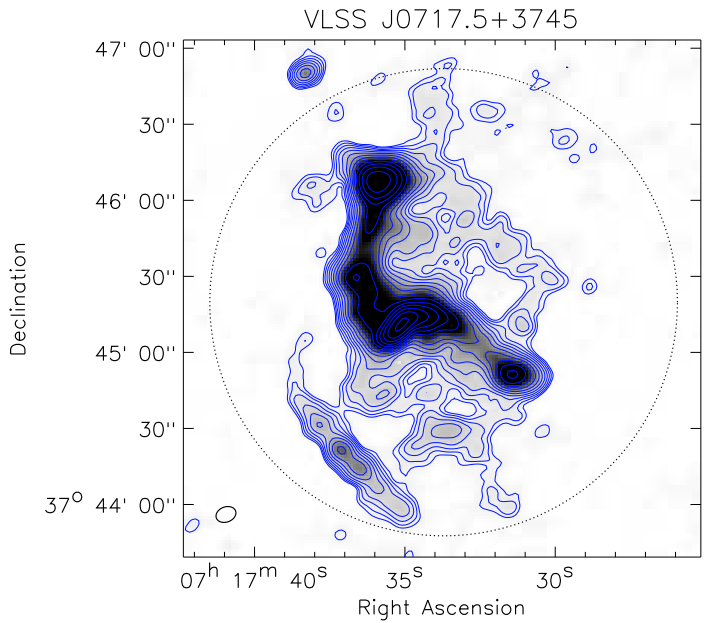}
       \end{center}
      \caption{GMRT 610~MHz radio map. Contour levels are drawn as in Fig.~\ref{fig:radiomap5}.}
            \label{fig:radiomap21}
 \end{figure}
\begin{figure}
    \begin{center}
      \includegraphics[angle = 90, trim =0cm 0cm 0cm 0cm,width=0.5\textwidth]{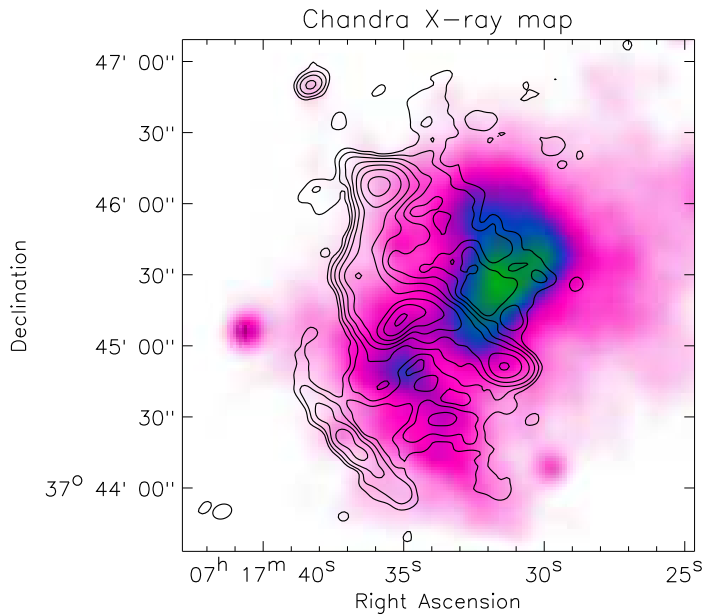}
       \end{center}
      \caption{Chandra X-ray map overlaid with radio contours at 610~MHz from the GMRT. The color scale represents the X-ray emission from $0.5-7.0$~keV. The image has been adaptively smoothed using the TARA$^{a}$ package using a minimal significance of $5$. Contour levels are drawn at $[1, 2, 4, 8, 16, 32,  ...]~\times$ $0.312$~mJy~beam$^{-1}$.}
$^{a}$ http://www.astro.psu.edu/xray/docs/TARA
            \label{fig: xray_s21}
 \end{figure}
\begin{figure}
    \begin{center}
      \includegraphics[angle = 90, trim =0cm 0cm 0cm 0cm,width=0.5\textwidth]{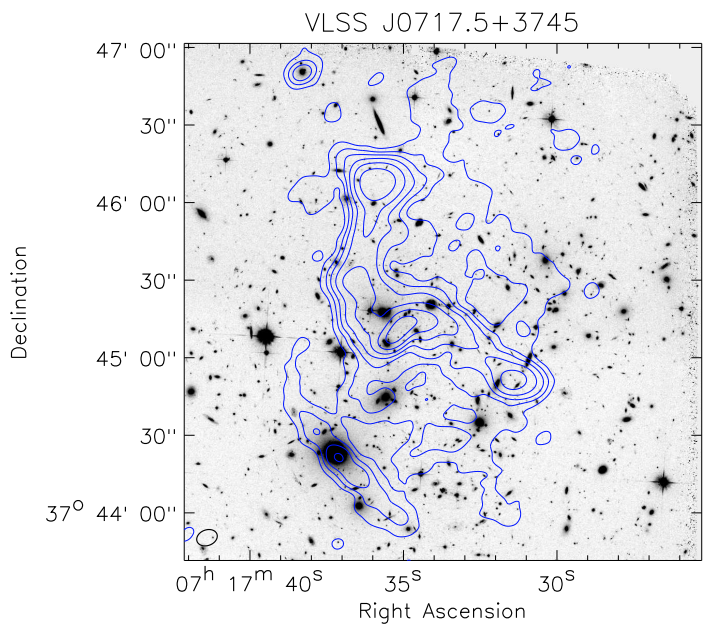}
       \end{center}
      \caption{HST F814W ACS image overlaid with GMRT 610~MHz contours. Contour levels are drawn as in Fig.~\ref{fig:radiomap5_o}.} 
            \label{fig:radiomap21_o}
 \end{figure}

Here we shortly describe our results for this cluster, a more detailed analysis in combination with additional archival VLA observations has been presented in a separate paper \citep{2009arXiv0905.3650V}, see also \cite{2009A&A...503..707B}. Our radio maps reveal a complex source, consisting of different components. The main component is a twisted structure, with enhanced regions of radio emission in the north and southwest of the cluster. These regions are connected by two bridges of emission to a bright central elongated source. The structure has a linear size of 700~kpc. No obvious counterparts are visible for both the north and south-west components. The presence of the two radio bridges suggest a relation with the central component. Although, no obvious counterpart was found for the central component by \cite{2003MNRAS.339..913E}, we identify an elliptical galaxy at RA~07$^\mathrm{h}$~17$^\mathrm{m}$~35$^\mathrm{s}$.5, DEC~+37\degr~45\arcmin~05\farcs5 as a possible counterpart. If this is indeed the case then the source could be a wide-angle tail (WAT) source with the two bridges being the tails of the central source and the north and south-west components the hotspots. The eastern boundary of the structure is sharp, while on the other side some faint emission is seen extending further westwards. Diffuse radio emission is also seen to the south of the main structure. The diffuse emission within the cluster has a total size of about 1.2~Mpc and given that it roughly follows the X-ray emission we classify it as a radio halo. Clearly, the emission is not associated with individual sources. Using a spectral index of $-1.2$, typical for radio halos, we estimate the radio power ($P_{1.4}$) to be $5 \times 10^{25}$~W~Hz$^{-1}$. This makes it the most powerful radio known to date, in agreement  with the X-ray luminosity-radio power ($L_{\mathrm{X}}-P_{1.4}$) and temperature-radio power ($T-P_{1.4}$) correlations \citep{2000ApJ...544..686L,2002A&A...396...83E,2006MNRAS.369.1577C}.

Towards the south a fainter linear structure is seen. A compact core, located halfway the linear structure is associated with a bright elliptical foreground galaxy (RA~07$^\mathrm{h}$~17$^\mathrm{m}$~37$^\mathrm{s}$.2, DEC~+37\degr~44\arcmin~23\arcsec).  The source is probably a FR-I source associated with the compact core.

Interestingly, the main twisted radio structure is located in between the brightest X-ray emission of the cluster. The main cool core visible in the X-ray image has no radio emission associated with it. The central radio structure also coincides with regions having a significantly higher X-ray temperature $\gtrsim15$~keV. The ICM temperature and X-ray morphology of the cluster favor of a relic-like interpretation. We therefore conclude that the twisted radio structure is a giant relic tracing a shock front linked to the merger activity of the system. 


\subsubsection{VLSS J0915.7+2511}   
\begin{figure}
    \begin{center}
      \includegraphics[angle = 90, trim =0cm 0cm 0cm 0cm,width=0.5\textwidth]{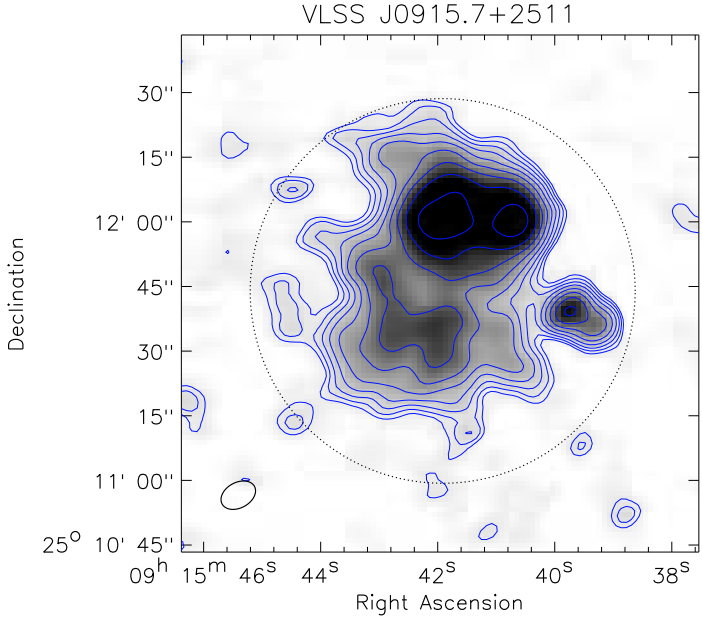}
       \end{center}
      \caption{GMRT 610~MHz radio map. Contour levels are drawn as in Fig.~\ref{fig:radiomap5}.}
                 \label{fig:radiomap22}
 \end{figure}
\begin{figure}
    \begin{center}
      \includegraphics[angle = 90, trim =0cm 0cm 0cm 0cm,width=0.5\textwidth]{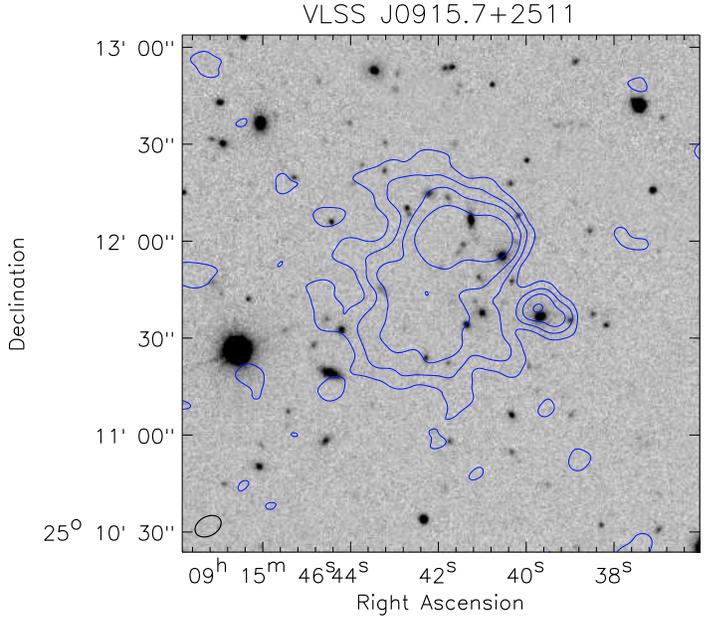}
       \end{center}
      \caption{SDSS r band image overlaid with GMRT 610~MHz contours. Contour levels are drawn as in Fig.~\ref{fig:radiomap5_o}.}
           \label{fig:radiomap22_o}
 \end{figure}

The radio map shows a diffuse region of emission associated with a cluster of galaxies (\object{MaxBCG~J138.91895+25.19876}) at a redshift of $0.324$. The radio source consists of a northern component and a fainter southern one. To the west a source is associated with an elliptical galaxy (RA~09$^\mathrm{h}$~15$^\mathrm{m}$~39$^\mathrm{s}$.7, DEC~+25\degr~11\arcmin~37\arcsec). A few possible counterparts are visible in SDSS DR7 images for the northern diffuse component. The southern diffuse component has no obvious optical counterpart. This could be a radio relic with a projected size of about 190~kpc. High-resolution observations will be needed to confirm this interpretation.

\subsubsection{VLSS~J1515.1+0424, Abell 2048}    
\begin{figure}
    \begin{center}
      \includegraphics[angle = 90, trim =0cm 0cm 0cm 0cm,width=0.5\textwidth]{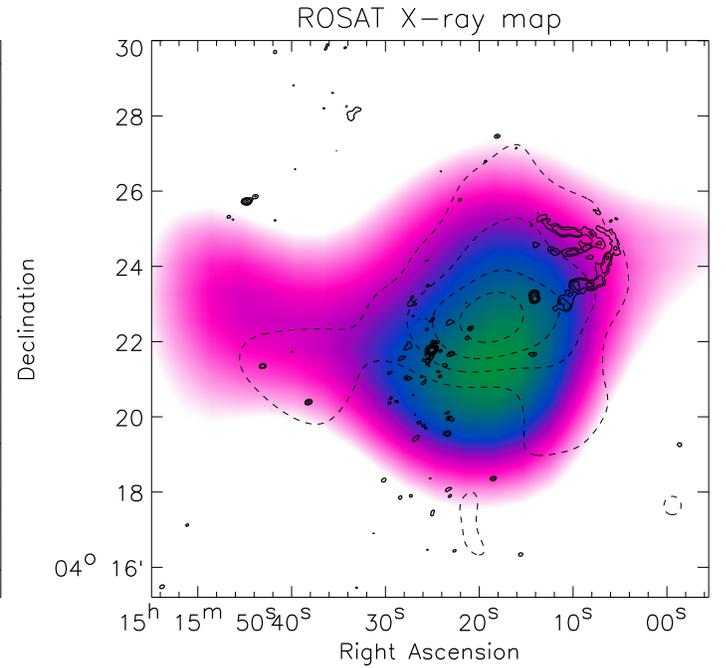}
       \end{center}
      \caption{X-ray emission from ROSAT in the $0.1- 2.0$~keV energy band. The image has been convolved with a circular Gaussian of 225\arcsec. The solid contours represent the radio emission at 610 MHz from the GMRT. The radio contours are drawn at $[1, 2, 4, 8, 16, 32,  ...]  \times 7\sigma_{\mathrm{rms}}$. Dashed contours show the galaxy distribution from SDSS DR7. Only galaxies with a photometric redshift between $0.06+z_{\mathrm{err}}<z<0.15-z_{\mathrm{err}}$ were selected from the catalogs, with $z_{\mathrm{err}}$ the error in the photometric redshift and $z=0.0972$ The galaxy isodensity contours are drawn at $[6, 9, 12, 15,...]$ galaxies arcmin$^{-2}$.}
       \label{fig:xraymap25}
 \end{figure}

The radio image shows a filamentary radio source in the periphery of the cluster \object{Abell~2048} at a redshift of 0.0972. The source consists of three elongated structures orientated roughly east-west. The three structures connect to the east. 
The source could be a complex double WAT source. However, the radio structure itself does not seem to be connected to any particular galaxy. The source has a projected size of 310~kpc, if located in the cluster. In the southeast a compact double-lobe source is associated with a large elliptical galaxy  (RA~15$^\mathrm{h}$~15$^\mathrm{m}$~14$^\mathrm{s}$.1, DEC~+04\degr~23\arcmin~10\arcsec) located in the cluster. On the other side of the cluster a 0.19~Jy source (\object{PMN~J1515+0421}) limits the dynamic range. The cluster is also detected in X-rays as \object{RX~J1515.2+0421} \citep{1998A&AS..127..145B}. A substructure on the east-side of the main cluster is visible, hinting at a possible cluster merger, see Fig.~\ref{fig:xraymap25}. Given the location of the radio source at the edge of the cluster, the lack of a connection with a single optical counterpart, the steep radio spectrum, and the morphology, we classify the source as a peripheral radio relic.

\begin{figure}
    \begin{center}
      \includegraphics[angle = 90, trim =0cm 0cm 0cm 0cm,width=0.5\textwidth]{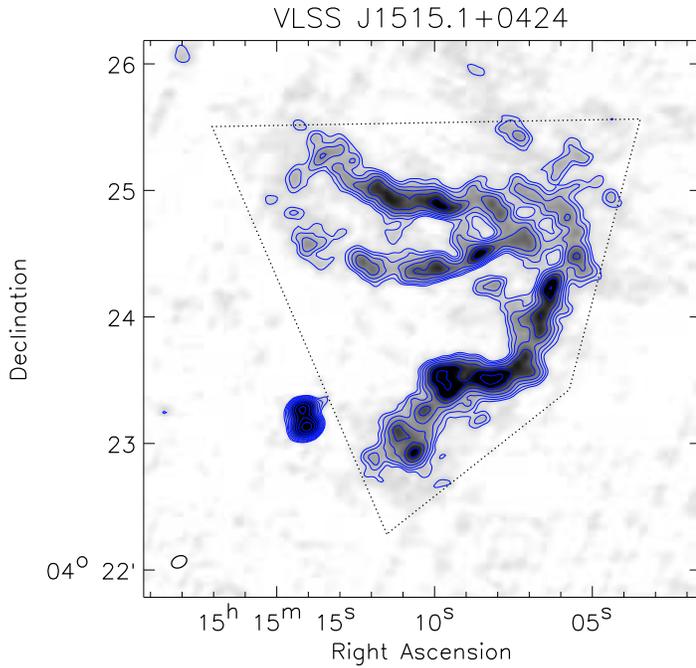}
       \end{center}
      \caption{GMRT 610~MHz radio map. Contour levels are drawn as in Fig.~\ref{fig:radiomap5}.}
            \label{fig:radiomap25}
 \end{figure}
\begin{figure}
    \begin{center}
      \includegraphics[angle = 90, trim =0cm 0cm 0cm 0cm,width=0.5\textwidth]{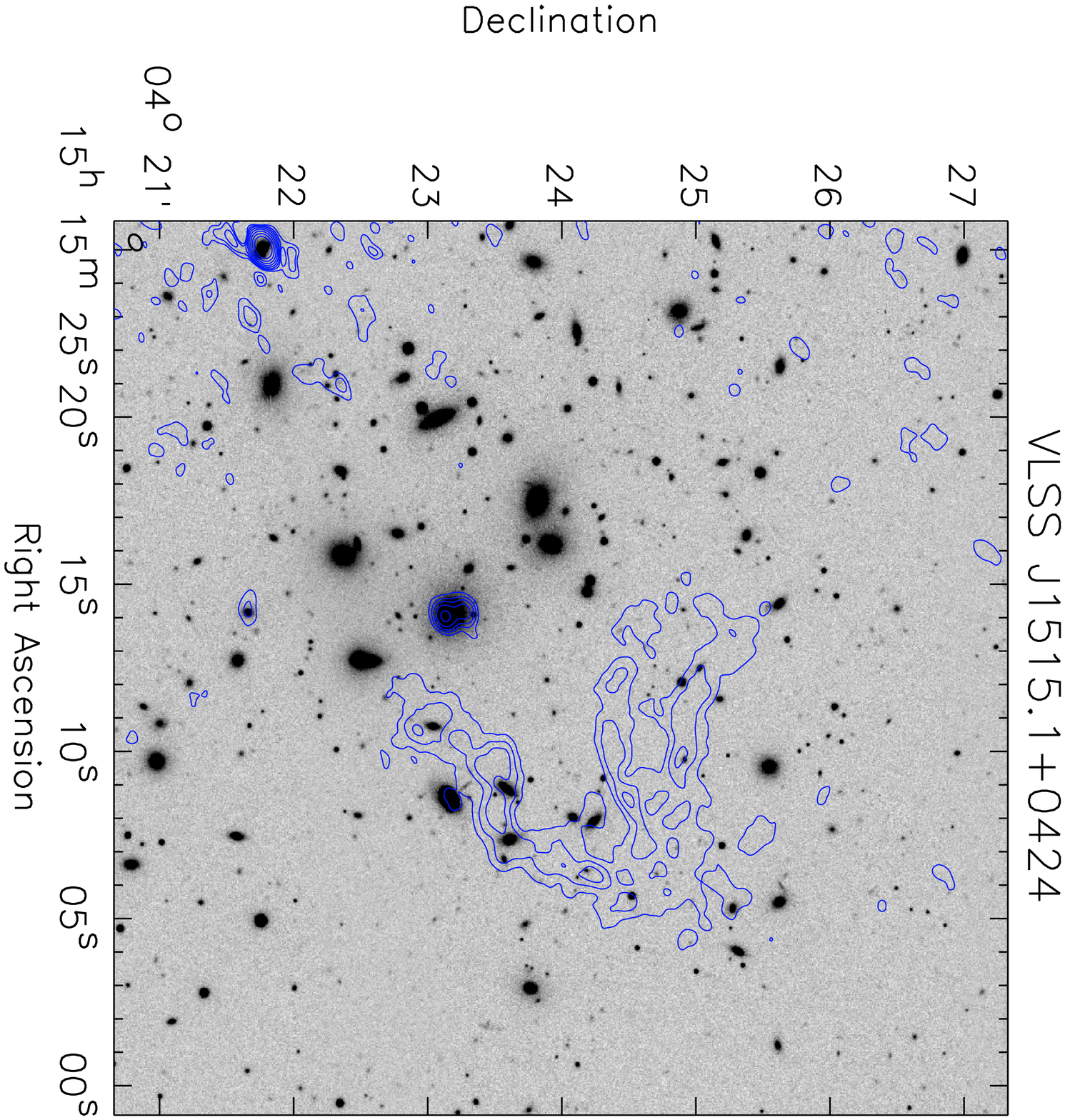}
       \end{center}
      \caption{SDSS r band image overlaid with GMRT 610~MHz contours. Contour levels are drawn as in Fig.~\ref{fig:radiomap5_o}.}
       \label{fig:radiomap25_o}
 \end{figure}

\section{Spectral index modeling}
\label{sec:spectra}

Radio spectra can be an important tool to understand the origin of the relativistic electrons and to determine the age of the radio emitting plasma. We have combined our flux measurements at 610~MHz with literature values to determine the radio spectra for the sources in our sample. We have included flux measurements from the following surveys: 38~MHz~8C \citep{1990MNRAS.244..233R,1995MNRAS.274..447H}, 74~MHz~VLSS, 151~MHz~7C  \citep{1996MNRAS.282..779W, 2007MNRAS.382.1639H}, 232~MHz~Miyun \citep{1997A&AS..121...59Z},  325~MHz~WENSS \citep{1997A&AS..124..259R}, 352~MHz~WISH \citep{2002A&A...394...59D}, 365~MHz~TEXAS \citep{1996AJ....111.1945D} and 1400~MHz~NVSS.

We model the integrated radio spectra using the Jaffe-Perola (JP) model \citep{1973A&A....26..423J} described by \cite{1994A&A...285...27K}, see also \cite{2001AJ....122.1172S}. We assume that (i) synchrotron self-absorption is negligible as this only occurs in compact sources, (ii) radiative energy losses dominate over other losses such as adiabatic ones, (iii) the magnetic field is spatially uniform and constant in time, (iv) radiative electrons do not escape from the region, (v) the emission in the region is uniform, (iv) relativistic electrons were injected at a single point with a power-law distribution of energy, and (vii) the pitch angles of the synchrotron emitting electrons are assumed to be continuously isotropized on a timescale shorter than the radiative timescale. Relativistic electrons lose their energy by synchrotron emission and IC scattering off the cosmic microwave background (CMB).

In the Kardashev-Pacholczyk (KP) model \citep{1962SvA.....6..317K, 1970ranp.book.....P}, also used by \cite{1994A&A...285...27K}, the pitch angle of the electrons remains in its original orientation with respect to the magnetic field. This introduces one more free parameter in the spectral index modeling (the ratio between the source magnetic field ($B$) and the effective magnetic field for IC losses ($B_{\mathrm{IC}}$). We have chosen not to fit this model to limit the number of free parameters. Furthermore, the JP model is more realistic from a physical point of view, as an anisotropic pitch angle distribution will become more isotropic due to changes in the magnetic field strength between different regions and scattering by self-induced Alfv\'en waves \citep[e.g.,][]{1991ApJ...383..554C, 2001AJ....122.1172S}. 

\begin{table*}
\begin{center}
\caption{JP model fits with $\alpha_{\mathrm{inj}} = -0.75$}
\begin{tabular}{lllll}
\hline
\hline
Source Name& $t_{\mathrm{CI}}$ &  $t_{\mathrm{RE}}$  & $B^{\prime}_{\mathrm{eq}}$ \\
                          &   $10^8$ yr                & $10^8$ yr  & $\mu$Gauss \\
\hline
VLSS~J0646.8$-$0722  &  $0.227 \pm 0.0775$  & $0.0869 \pm 0.00624$  & 24.5&\\  
VLSS~J1133.7+2324 & $0.121 \pm 0.034$ &$0.0870 \pm 0.00634$&23.3&  \\ 
VLSS~J1710.5+6844 & $0.421 \pm 0.259$&  $0.0946 \pm 0.00812$ &23.1&\\ 
24P73 (VLSS~J2217.5+5943), $z=0.05$ &$0.156\pm 0.0464$  & $0.120 \pm 0.00406$ &29.1&\\ 
24P73 (VLSS~J2217.5+5943), $z=0.10$ &$0.173\pm 0.0534$  & $0.131 \pm 0.00503$ &27.0&\\ 
24P73 (VLSS~J2217.5+5943), $z=0.15$ &$0.176\pm 0.0557$  & $0.131 \pm 0.00443$ &26.4&\\ 
24P73 (VLSS~J2217.5+5943), $z=0.20$ &$0.169\pm 0.0500$  & $0.129 \pm 0.00497$ &26.3&\\ 
\hline
\hline
\end{tabular}
\label{tab:jp}
\end{center}
\end{table*}

Our adopted scenario is as follows: When the source starts, we assume that it is fueled at a constant rate for a certain time $t_{\mathrm{CI}}$ (the time of the continuous injection (CI) of relativistic electrons, with an injection spectral index $\alpha_{\mathrm{inj}}$). This is followed by a relic phase (RE) where the injection of electrons is switched off ($t_{\mathrm{RE}}$). During both of these phases the electrons lose energy by synchrotron and IC losses. The total source age is $t_{\mathrm{age}}= t_{\mathrm{RE}} + t_{\mathrm{CI}}$. 

In the spectrum two break frequencies occur both related to spectral aging. The first is the break frequency $\nu_{\mathrm{b}}$ of the first injected electron population (at the beginning of the CI phase)
\begin{equation}
\nu_{\mathrm{b}} \propto \frac{B}{\left([B^2 + B_{\mathrm{IC}}^2]t\right)^2} \mbox{ .}
\end{equation}
A second higher frequency break $\nu\prime_{\mathrm{b}}$ is caused by the last electron population injected, at the end of the CI phase
\begin{equation}
\nu\prime_{\mathrm{b}} = \nu_{\mathrm{b}}  \left(1 + \frac{t_{\mathrm{CI}}}{t_{\mathrm{RE}}} \right)^2   \mbox{ .} 
\label{eq:mach-alpha}
\end{equation}

The injection spectral index ( $\alpha_{\mathrm{inj}}$) is determined by the AGN in the case of (relic) radio lobes or radio phoenices. For relics where the particles are accelerated by DSA the injection spectral index is related to the Mach number ($\mathcal{M}$) of the shock. In linear theory the relevant expression is
\begin{equation}
\alpha_{\mathrm{inj}} = -\frac{3{\mathcal M}^{-2}+1}{2 - 2{\mathcal M}^{-2}} \mbox{ ,}
\end{equation}
see \cite{2009arXiv0908.0728V}. The injection time must be comparable with the crossing time of the relic region by the shock front.  Since ``DSA-relics'' are mostly located in the cluster periphery, where the pressure of the ICM gas is lower, adiabatic energy losses may become important. These adiabatic energy losses affect the radio spectrum, depending on the expansion rate of the relic and evolution of the magnetic field strength ($B(t)$) with time \citep{1962SvA.....6..317K, 1994ApJ...431..586G, 2002NewAR..46..307M}. However, we only have a limited number of flux measurements for our sources and only one proposed DSA-relic with enough flux measurements to model the spectrum. Furthermore, the location of this relic in the cluster as well as the identification of the cluster itself are uncertain. We therefore chose to ignore the effects adiabatic expansion losses.

The JP model is thus characterized by four free parameters: (1) the injection spectral index $\alpha_{\mathrm{inj}}$, (2) the length of the CI phase $t_{\mathrm{CI}}$, (3) the length of the RE phase $t_{\mathrm{RE}}$, and (4) a flux normalization constant. To reduce the number of free parameters we have chosen to keep $\alpha_{\mathrm{inj}}$ fixed to a value of $-0.75$ \citep{2007A&A...470..875P}.

The spectra are fitted by minimizing the $\chi$-squared value of the fit in a two-step process. We first determine the shape of the spectrum for different values of  $t_{\mathrm{RE}}$ and $t_{\mathrm{CI}}$, both ranging from 0 to $10^{9}$ yrs in 25 equal logarithmically spaced steps. Then an overall flux scaling (normalization) constant is determined by multiplying the spectrum with a constant until the $\chi$-squared value is minimized. In this way, a 2-dimensional ($25\times25$) array of $\chi$-squared values is created. We continue the fitting using the same process but now centering the  $t_{\mathrm{RE}}$ and $t_{\mathrm{CI}}$ values around the minimum $\chi$-squared value in the array and increasing the time resolution by a factor of $1.5$. The process is repeated until both $t_{\mathrm{RE}}$ and $t_{\mathrm{CI}}$ converge to a constant value, i.e., do not change by more than 1\% between subsequent iterations. The formal errors in the fitted parameters are determined by the corresponding distribution of the $\chi$-squared values.  For the sources where we successfully fitted the JP-model we should take in mind that some of the simplifying assumptions we made may not be valid and could have affected our results.

For the magnetic field strength we use the revised equipartition magnetic field strength $B^{\prime}_{\mathrm{eq}}$ \citep{1997A&A...325..898B, 2005AN....326..414B}. We use the same procedure as in \cite{2009arXiv0908.0728V} and take for the depth of the source ($d$) the average of the major and minor axis. The ratio of energy in relativistic protons to that in electrons ($k$) is set to $100$. For the low-energy cutoff ($\gamma_{\mathrm{min}}$, the energy boundary indicated by the Lorentz factor) we take $100$. For other values of $k$ and $\gamma_{\mathrm{min}}$, $B^{\prime}_{\mathrm{eq}}$ scales with $ \gamma_{\mathrm{min}}^{(1+2\alpha)/(3-\alpha)} (1+k)^{1/(3-\alpha)}$, with $\alpha$ the spectral index.

We have only fitted the radio spectra for sources with a redshift (because the energy loss rate due to IC scattering is proportional to $(1+z)^4$) and at least four flux measurements available. An exception is \object{24P73} (or VLSS~J2217.5+5943) for which we have no redshift. If 24P73 has a similar size as the relic in A85 \citep[150~kpc, ][]{2001AJ....122.1172S}, then its redshift would be around $0.1$. Since this is a very rough estimate we have also fitted the radio spectrum using redshifts of $0.05$, $0.15$, and $0.2$.

The fluxes for the sources could be contaminated by the presence of field sources within the beam of the low-resolution surveys. In the case of 24P73, there is a 5.0~mJy NVSS source (\object{NVSS~J221736+594403}) nearby which is blended with the diffuse source in the 8C, VLSS, and WENSS surveys. We fitted a second order polynomial in log-log space to  the 610 MHz GMRT, NVSS, and the 1.4 and 8.4~GHz fluxes from \cite{1994A&AS..104..481G} for this source. We then extrapolated the flux to 325, 74, and 38~MHz and subtracted off this flux to recover the uncontaminated flux for the diffuse source.  For the three other diffuse sources no unrelated sources were found that could have significantly contributed to the flux.

\begin{figure}
    \begin{center}
      \includegraphics[angle = 0, trim =0cm 0cm 0cm 0cm,width=0.5\textwidth]{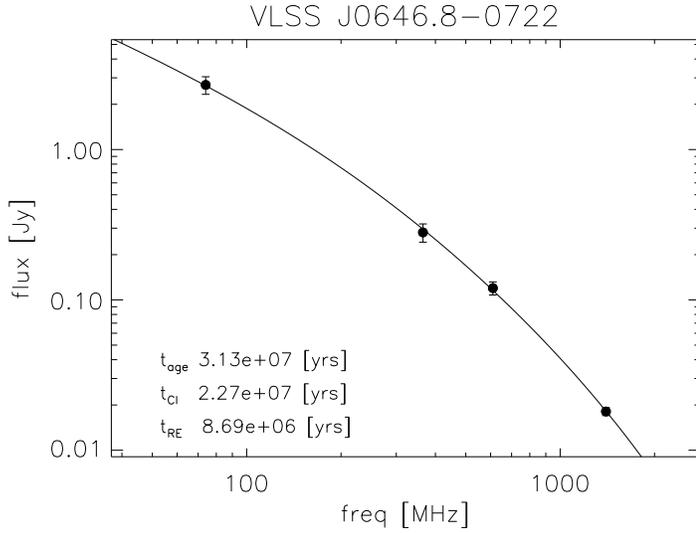}
       \end{center}
      \caption{Jaffe-Perola fit to the flux measurements. The duration of the CI and RE phases are indicated in the figure.}
                  \label{fig:spectrajp_s3}
 \end{figure}

\begin{figure}
    \begin{center}
      \includegraphics[angle = 0, trim =0cm 0cm 0cm 0cm,width=0.5\textwidth]{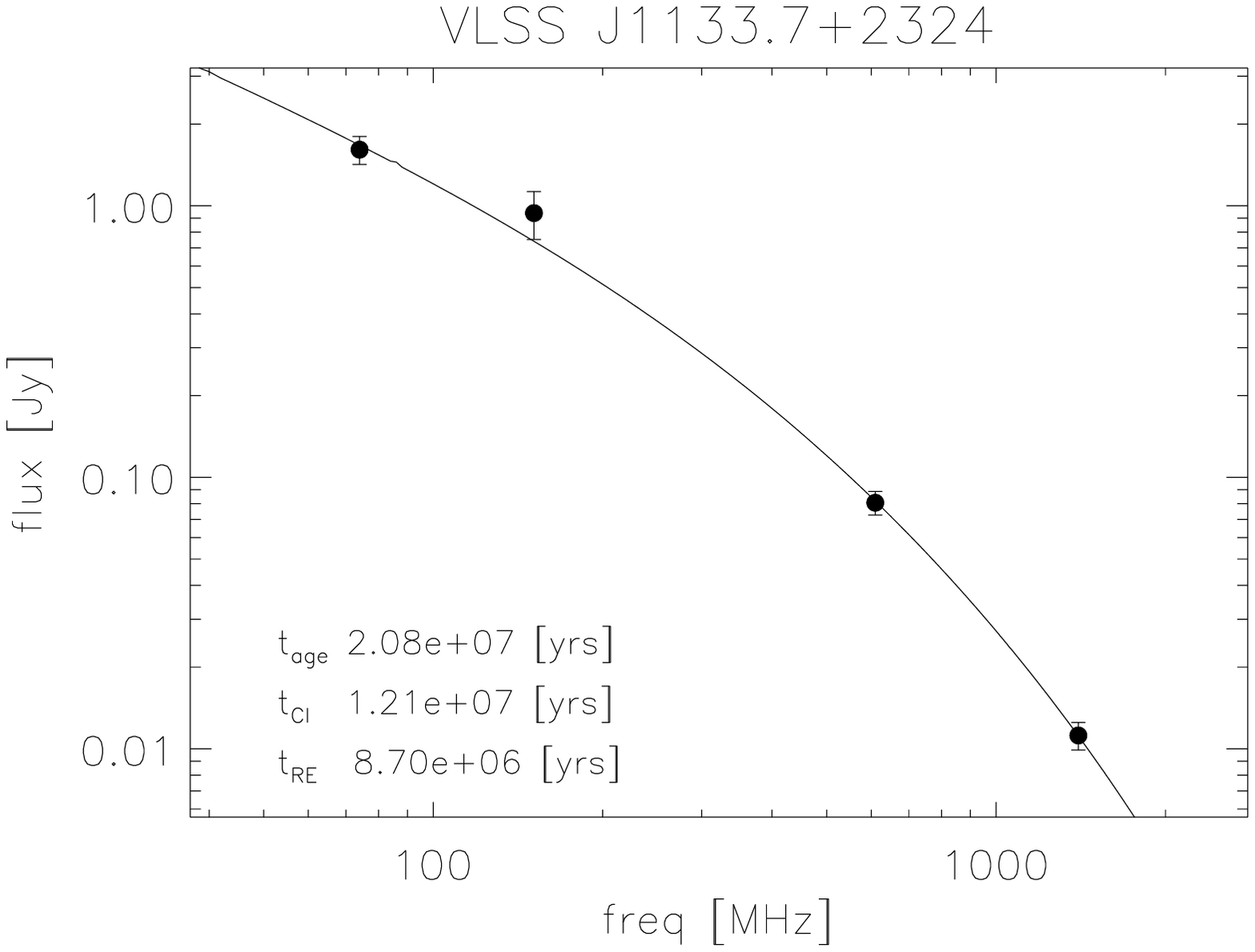}
       \end{center}
      \caption{Same as Fig.~\ref{fig:spectrajp_s3} but for VLSS~J1133.7+2324.}
                  \label{fig:spectrajp_s5}
 \end{figure}

\begin{figure}
    \begin{center}
      \includegraphics[angle = 0, trim =0cm 0cm 0cm 0cm,width=0.5\textwidth]{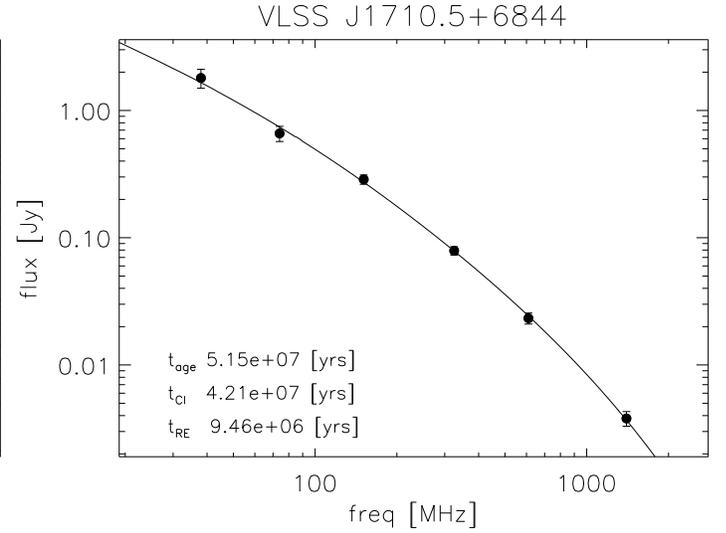}
       \end{center}
      \caption{Same as Fig.~\ref{fig:spectrajp_s3} but for VLSS~J1710.5+6844.}
                  \label{fig:spectrajp_s8}
 \end{figure}

\begin{figure}
    \begin{center}
      \includegraphics[angle = 0, trim =0cm 0cm 0cm 0cm,width=0.5\textwidth]{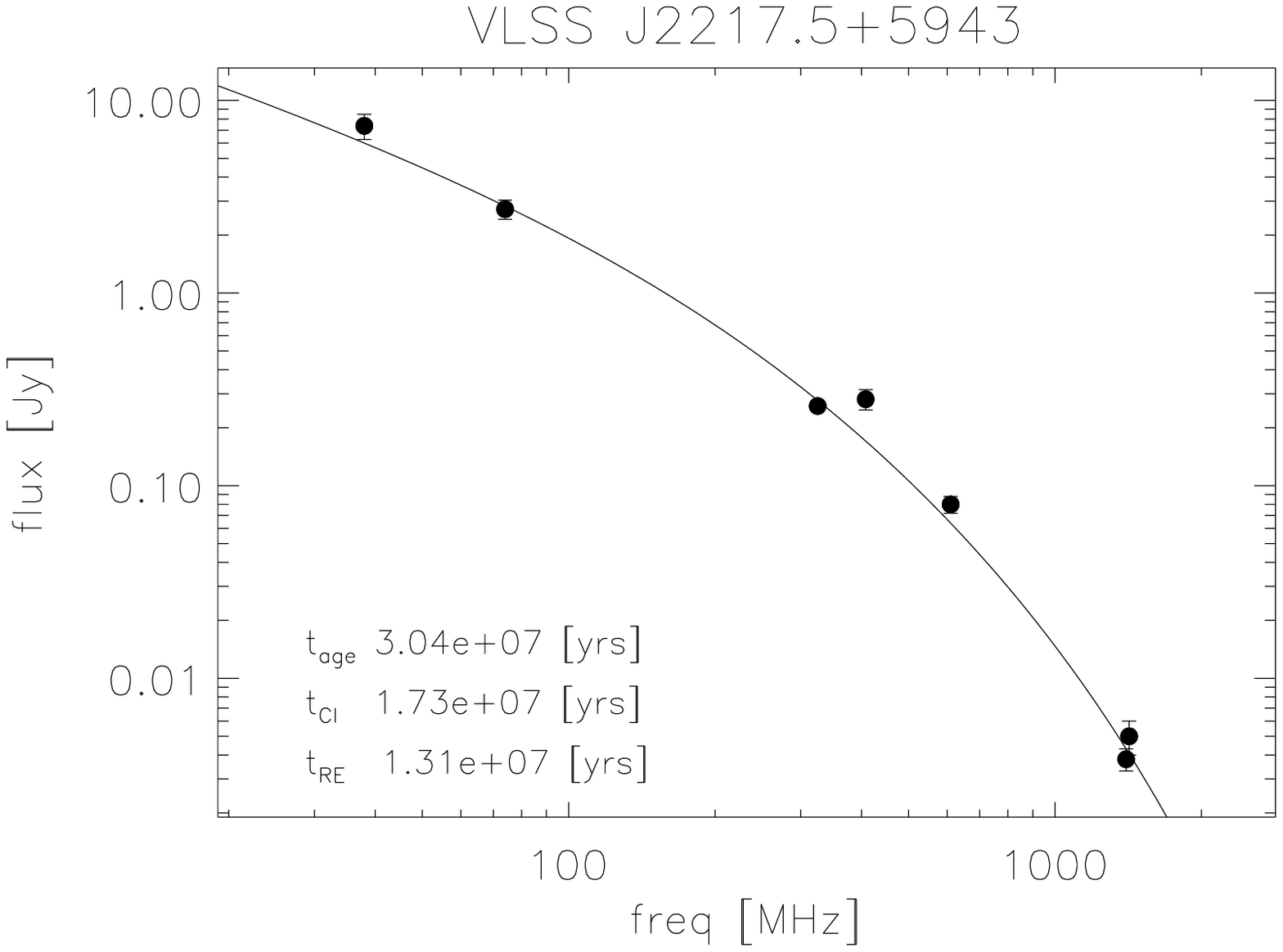}
       \end{center}
      \caption{Same as Fig.~\ref{fig:spectrajp_s3} but for 24P73 (VLSS~J2217.5+5943) using $z=0.1$.}
                  \label{fig:spectrajp_s14}
 \end{figure}

The fitted radio spectra are show in Figs.~\ref{fig:spectrajp_s3} to \ref{fig:spectrajp_s14}. The values for $t_{\mathrm{CI}}$, $t_{\mathrm{RE}}$, and $B^{\prime}_{\mathrm{eq}}$ are reported in Table~\ref{tab:jp}. The fitting process for the sources VLSS~J1636.5+3326, VLSS~J2213.2+3411, VLSS~J0004.9$-$3457 did not converge.  For VLSS~J1636.5+3326 the 1.4~GHz flux measurement is relatively high, in this case the source may have restarted its activity causing the high 1.4~GHz flux value. This is also seen for several radio sources by \cite{2007A&A...470..875P}. For the other sources the flux measurements are too closely spaced in frequency to provide enough constraints for the fitting process.

For the sources where we successfully modeled the synchrotron spectrum $t_{\mathrm{CI}} > t_{\mathrm{RE}}$. All of these sources show steepening of the radio spectrum at higher frequencies which is expected in case of spectral aging. For 24P73 the spectral index modeling is consistent with the source being a radio phoenix with a total source age of $3 \times 10^{7}$~yrs.  We find that the derived synchrotron age does not critically depend on the adopted redshift of 0.1 for this source (see Table~\ref{tab:jp}).

\section{Discussion}
\label{sec:discussion}

Based on their toroidal or filamentary morphologies and curved radio spectra two of our relics could be classified as radio phoenices: VLSS~J1515.1+0424 and VLSS J2217.5+5943. Contrary,  VLSS J1133.7+2324  and the twisted structure in MACS~J0717.5+3745 are probably the result of DSA from structure formation shocks. In the case of VLSS J0717.5+3745, the distorted morphology and high temperature of the ICM are clear evidence for a cluster undergoing a merger. For the other relics more observations are needed to determine their origin.

Most of the sources in our sample are associated with galaxies in clusters and show irregular morphologies. This can be caused by the interaction of the radio plasma with the ICM. The steep spectrum of the sources is caused by spectral aging, e.g., in some cases the central AGN activity may have stopped, producing a so called ``dying'' radio source. Confinement of the radio plasma by the ICM could also have contributed to the steep radio spectra, as in this case the radio plasma has ``the time'' to display the effects of spectral aging. 
\label{sec:correlation}

 \begin{table*}
\begin{center}
\caption{Radio relic properties}
\begin{tabular}{llllll}
\hline
\hline
Source Name& $\alpha_{74 \mathrm{MHz}}^{1400 \mathrm{MHz}}$& spectral curvature   & $P_{1.4}$& size$^{a}$& $R_{\mathrm{projected}}$\\
                          &         &$\alpha_{74 \mathrm{MHz}}^{610 \mathrm{MHz}} - \alpha_{610 \mathrm{MHz}}^{1400 \mathrm{MHz}}$        & $10^{24}$ W Hz$^{-1}$  & kpc  & kpc\\
\hline
VLSS J1133.7+2324 & $-1.69$& 0.96&24.1&570 \\ 
VLSS~J1431.8+1331$^{d}$ &  \ldots& \ldots & \ldots & 125 & 175\\ 
 24P73 (VLSS J2217.5+5943)  & $-2.20$ &  2.0& \ldots& \ldots &\ldots\\ 
 \object{MACS J0717.5+3745} & $-1.09^{b}$ & $-0.02$ & 138$^{b}$ & 700 & 320\\ 
Abell 2048 (VLSS~J1515.1+0424) &$-1.50$& 1.60& 0.566 & 310& 310 \\
Abell 2256 &$-1.2$&\ldots & 3.95 &1100 & 300\\
Abell 1240-N &$-1.2$&\ldots &0.427&650& 700\\
Abell 1240-S &$-1.3$& \ldots&0.730&1250& 1100\\
Abell 2345-E &$-1.3$& \ldots  &2.62 &1500& 890\\
Abell 2345-W  &$-1.5$&\ldots &2.83 & 1150&1000\\
Abell 13 &$-1.79$& 0.65& 0.853& 260 & 170\\
Abell 85 &$-2.30$&1.36&0.322&150&430 \\
Abell 133 &$-1.70$&1.39 &1.07&55 & 40\\ 
Abell 4038 &$-1.88$& 1.03& 0.0983 &55 & 35\\
Abell 3667 &$-1.1$&\ldots &17.4&1920 &1950 \\
Abell 548b A&$\sim-2$&\ldots & 0.260& 310 & 500 \\ 
Abell 548b B&$<-2.0$& \ldots & 0.250 & 370 & 430\\
Abell 521 &$-1.48$& \ldots &2.90 &1000 &930\\
Coma cluster$^{c}$ & $-1.18$&\ldots & 0.284& 850& 1940\\
Abell 2163 &$-1.02$& \ldots & 2.23& 450 & 1550 \\ 
Abell 2744 &$-1.1$& \ldots& 6.20 & 1620 & 1560 \\
Abell S753 &$-2.0$&\ldots & 0.205&350  & 410  \\
Abell 115 &$-1.1$& \ldots & 16.7 & 1960 & 1510\\  
Abell 610 &$-1.4$& \ldots& 0.444 & 330 & 310\\
RXCJ1314.4-2515-E & $-1.41$ & \ldots&2.08 & 920& 685\\
RXCJ1314.4-2515-W& $-1.40$ & \ldots& 4.81 & 920& 685\\
\hline
\hline
\end{tabular}
\label{tab:correlation}
\end{center}
$^{a}$ largest linear scale\\
$^{b}$ excluding the flux of the halo and central head-tail source \citep[see also][]{2009arXiv0905.3650V, 2009A&A...503..707B}\\
$^{c}$ 1253+275\\
$^{d}$ blended with a nearby AGN (except in the GMRT 610~MHz image), therefore no spectral index or spectral curvature could be calculated
\end{table*}

We have calculated the physical size (largest extent), projected distance from the cluster center ($R_{\mathrm{projected}}$), and 1.4~GHz radio power ($P_{1.4}$) for the radio relics in our sample. We have complemented this with values from the literature for radio relics with measured spectral indices: \object{A13}, \object{A85}, \object{A133}, and \object{A4038} \citep{2001AJ....122.1172S}, \object{A1240} and \object{A2345}   \citep{2009A&A...494..429B}, \object{A3667} \citep{1997MNRAS.290..577R}, \object{A548b} \citep{2006MNRAS.368..544F}, \object{A2256} \citep{2006AJ....131.2900C}, \object{A521} \citep{2008A&A...486..347G}, \object{1253+275} \citep{1991A&A...252..528G}, \object{A2163} \citep{2004A&A...423..111F}, \object{A2744} \citep{2007A&A...467..943O}, \object{AS753} \citep{2003AJ....125.1095S}, \object{A115}  \citep{2001A&A...376..803G} \object{A610} \citep{2000NewA....5..335G} and \object{RXC J1314.4-2515} \citep{2007A&A...463..937V}.  The spectral indices for our newly found relics were calculated between 74 and 1400~MHz. The spectral indices for the relics taken from the literature were usually measured between 325 and 1400~MHz, but for some relics the frequency range is somewhat different (for more details the reader is referred the references given above).

\begin{figure}
    \begin{center}
      \includegraphics[angle = 90, trim =0cm 0cm 0cm 0cm,width=0.5\textwidth]{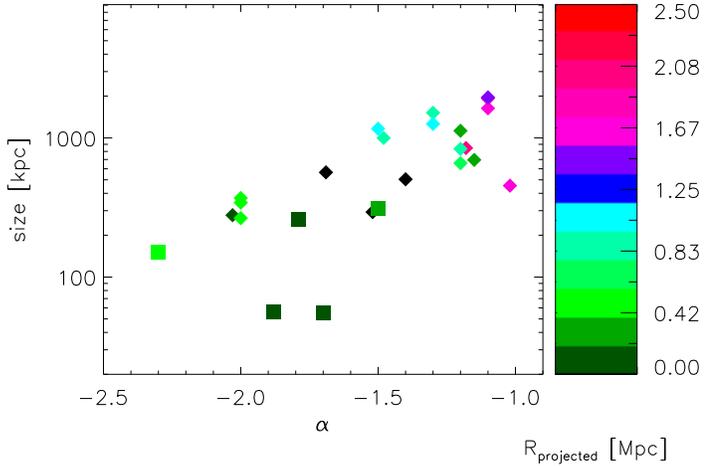}
       \end{center}
      \caption{Spectral index of radio relics versus their size. Squares are the proposed radio phoenices, i.e., the sources from \cite{2001AJ....122.1172S}, 24P73 and the relic in Abell~2048. Diamonds represent the radio relics tracing merger shocks where particles are being accelerated by the DSA mechanism. The color coding is according to the projected distance from the cluster center. For the relics represented by black symbols we could not obtain a reliable projected distance to the cluster center.}
                  \label{fig:alpha_size}
                   \end{figure}

The spectral index of the radio relics, versus the physical size is shown in Fig.~\ref{fig:alpha_size}. The projected distance from the cluster center is color coded. We find that on average the smaller relics have steeper spectra. There are two different explanations for this correlation. The found correlation between physical size and spectral index may (partly) be the result of the proposed radio phoenices occupying the lower left region in Fig.~\ref{fig:alpha_size}. Ignoring the proposed radio phoenices the trend remains, although the scatter is large so this result is less significant. Such a trend though, is in line with predictions from shock statistics derived from cosmological simulations \citep{2008MNRAS.391.1511H, 2008ApJ...689.1063S}. They find that larger shock waves occur mainly in lower-density regions and have larger Mach numbers, and consequently shallower spectra. Conversely, smaller shock waves are more likely to be found in cluster centers and have lower Mach numbers and steeper spectra. We note that more spectral index measurements of high quality are needed to confirm the correlation between physical size and spectral index (leaving out the proposed radio phoenices).

\begin{figure}
    \begin{center}
      \includegraphics[angle = 90, trim =0cm 0cm 0cm 0cm,width=0.5\textwidth]{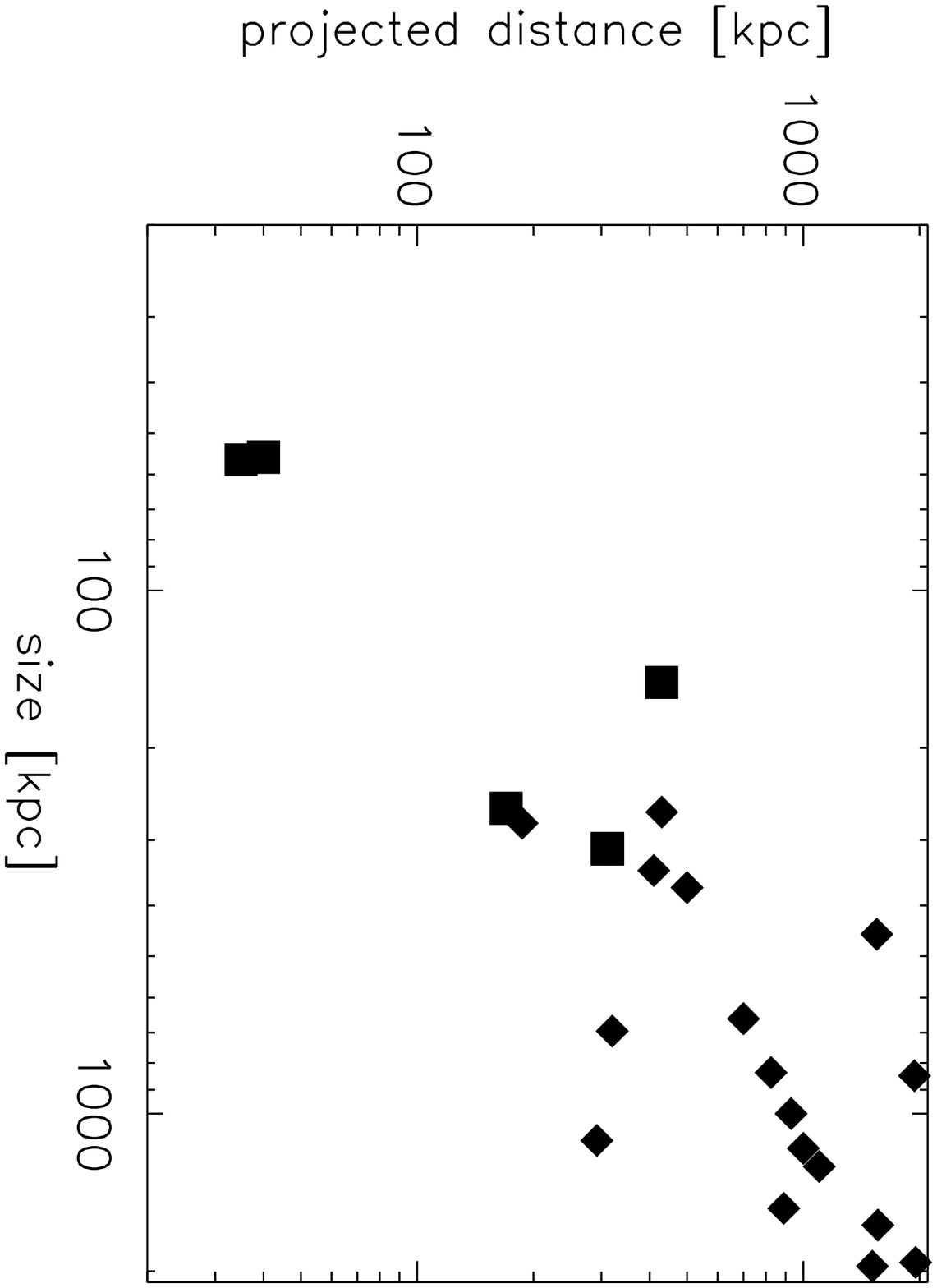}
       \end{center}
      \caption{Projected distance from the cluster center of radio relics versus their size. Symbols are defined in Fig.~\ref{fig:alpha_size}.}
                  \label{fig:proj_size}
                   \end{figure}

The size of the relics versus $R_{\mathrm{projected}}$ is shown in Fig.~\ref{fig:proj_size}. The projected distance from the cluster center for the shallow spectra relics is indeed on average larger than for the steep spectrum relics, indicating that they are mostly located in the periphery of clusters where the Mach numbers of the shock are higher.  We contribute this to the fact that in the periphery the shock surfaces are larger and the density/pressure of the ICM is lower. Therefore large radio relics are mainly located in the cluster periphery. Because we are using the projected distance from the cluster center, and not the (unknown) de-projected distance, some additional scatter is introduced in the plot. Furthermore, the spectral indices for the various radio relics are calculated using data from radio telescopes with different array configurations, sensitivities, and/or frequencies ranges, adding to the scatter.

Giant Mpc-scale radio relics are probably caused by DSA in an outwards traveling shock front. It is unlikely that they are the result of compression and reignition of fossil radio plasma as the time to compress such a large radio ``ghost'' would remove most of the electrons responsible for the radio emission by radiative energy losses \citep{2006AJ....131.2900C}. In the case of giant peripheral relics shock-acceleration is ongoing resulting in relatively flat spectral indices of about $-1$, i.e., a balance between electron cooling in the post-shock regions and continuous acceleration at the shock front. 
Behind the shock front, the spectral index is indeed observed to steepen for some giant relics \citep[e.g.,][]{1997MNRAS.290..577R}. 
After a few times $10^{8}$~yrs the electrons behind the shock front have lost most of their energy and cause little synchrotron emission. It takes of the order of $1$ Gyr for a shock wave to travel from the center of the cluster to about the virial radius. 

As has been mentioned, an alternative explanation for the found size-spectral index correlation could be a possible different origin of the smaller radio relics. With the smaller radio relics  originating from the adiabatic compression of fossil radio plasma. In fossil radio plasma, the high-frequency synchrotron emitting electrons have lost most of their energy. Due to compression and the increase in magnetic field strength the radio plasma becomes visible again. Therefore these sources are characterized by (very) steep and curved spectra. Proposed examples of such sources are 24P73 and those found by \cite{2001AJ....122.1172S}. If spectral aging occurs, the steeper radio spectra should be more curved.

It would be interesting to determine the injection spectral indices ( $\alpha_{\mathrm{inj}}$) for the radio relics.  This requires reliable flux measurements over a wide range of frequencies, especially below the break frequency $\nu_{\mathrm{b}}$. With enough flux measurements it should be possible to separate the effect of spectral aging and a steep $\alpha_{\mathrm{inj}}$. In particular, if the correlation is explained by the different Mach numbers of shocks the injection spectral indices for smaller relics located close to the cluster center should be steeper than that of larger relics. If this is indeed the case this is evidence for DSA. For radio phoenices the injection spectral index should be around $-0.5$ to $-0.75$ because the fossil radio plasma originated from AGN. Their steep spectral indices should then solely be the result of spectral aging and not a steep injection spectral index. Currently, with a limited number of flux measurements, mainly at frequencies where electrons have already lost some of their energy, there is a degeneracy between the injection spectral index and the amount of spectral aging. New low-frequency radio facilities, such as LOFAR, will be needed to determine $\alpha_{\mathrm{inj}}$ for radio relics.

\begin{figure}
    \begin{center}
      \includegraphics[angle = 90, trim =0cm 0cm 0cm 0cm,width=0.5\textwidth]{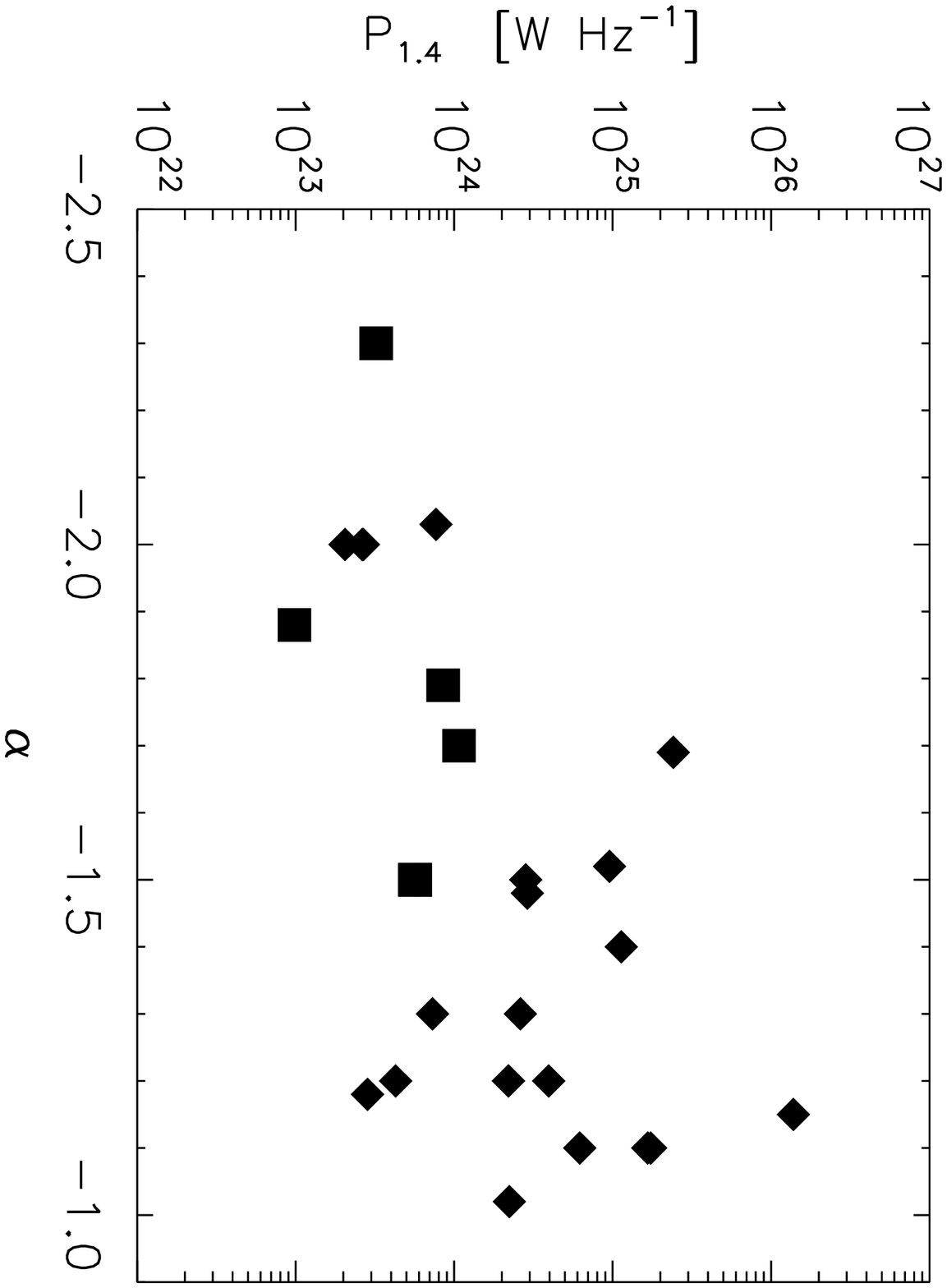}
       \end{center}
      \caption{Radio power at 1.4~GHz of radio relics versus their spectral index. Symbols are defined in Fig.~\ref{fig:alpha_size}.}
                  \label{fig:alpha_P}
 \end{figure}

The spectral index versus the 1.4~GHz radio power for the relics is shown in Fig.~\ref{fig:alpha_P}. From the figure we can see that there is a lack of high power sources with steep radio spectra. This is not surprising since relics with less steep spectra are larger (Fig.~\ref{fig:alpha_size}) and therefore should have a higher radio power.

\begin{figure}
    \begin{center}
      \includegraphics[angle = 90, trim =0cm 0cm 0cm 0cm,width=0.5\textwidth]{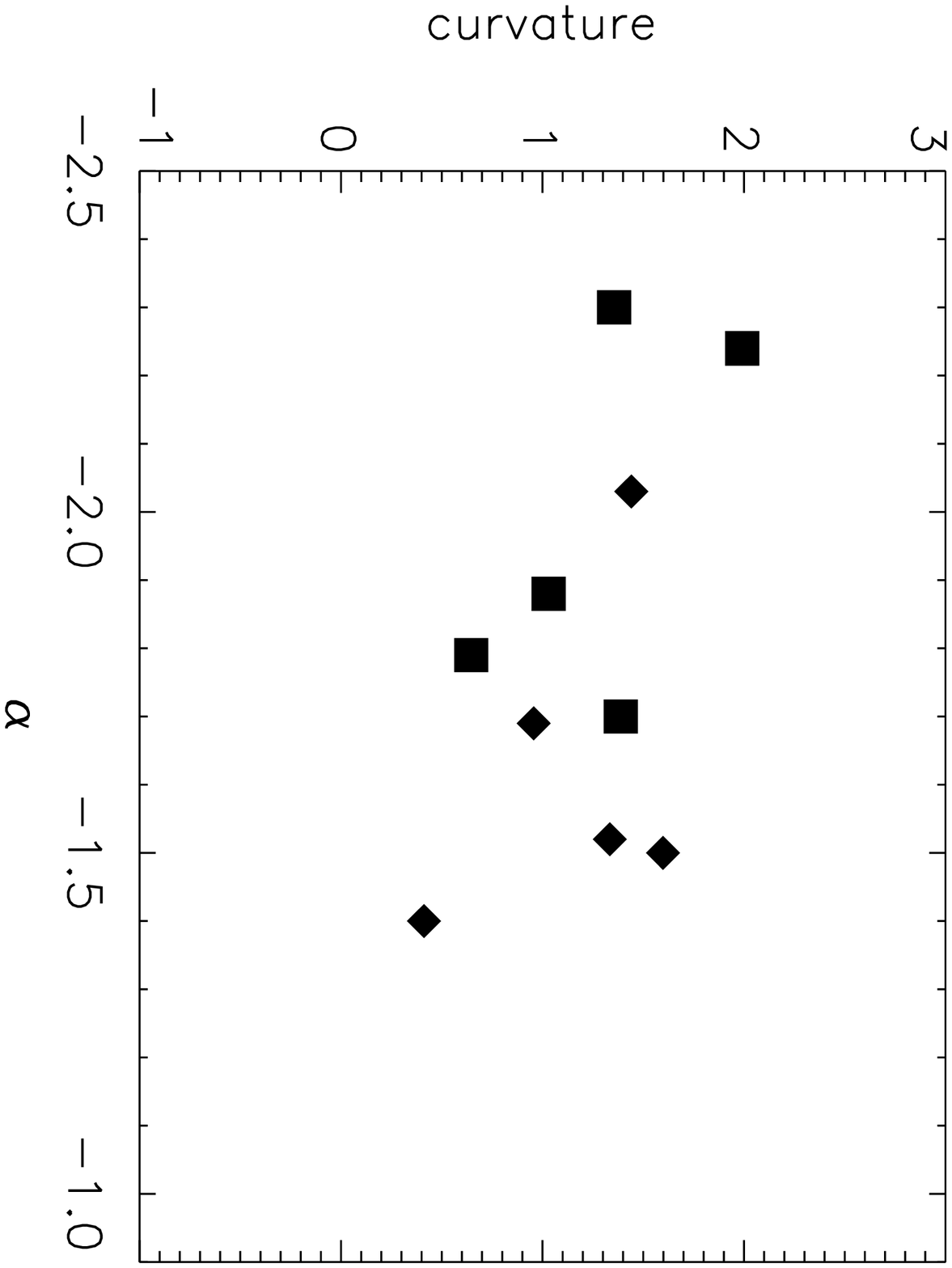}
       \end{center}
      \caption{Curvature of the radio spectra versus their spectral index between 74 and 1400 MHz. The curvature is defined as $\alpha_{74-610}~-~\alpha_{610-1400}$. Symbols are defined in Fig.~\ref{fig:alpha_size}.}
                  \label{fig:alpha_curvature}
 \end{figure}

 We have plotted the amount of spectral curvature for our sources versus their spectral index, between 74 and 1400~MHz, in Fig.~\ref{fig:alpha_curvature}. The curvature is defined as $\alpha_{74-610} - \alpha_{610-1400}$. A higher positive value for the curvature implies that the spectra are steeper at higher frequencies. We have only added the four sources from \cite{2001AJ....122.1172S} for which flux measurements at both 1.4~GHz and 74~MHz (VLSS) are available. We calculated the 610~MHz fluxes by interpolating the fluxes (linearly in log-log space) between 408 and 834~MHz. For other radio relics, flux measurements are less reliable or not available at 74 and/or 610~MHz.  As shown in Fig.~\ref{fig:alpha_curvature} there is an indication that the curvature is higher for the sources with a steeper spectral index, although the number of sources in the plot is rather low. If the spectral index of relics steepens by spectral aging such a trend is expected, but more measurements of spectral curvature will be needed to show that this correlation with the spectral index really exists.

\section{ Conclusions}
\label{sec:conclusion}
We carried out 610~MHz radio continuum observations with the GMRT of 26 diffuse steep spectrum sources. All of these sources were resolved out by 1.4~GHz~VLA~B-array snapshot observations or in the 1.4~GHz~FIRST survey. Of these 26 sources 25 were detected with the GMRT. 
The radio observations show a wide range of sources: radio relics, a giant radio halo, a mini-halo, FR-I, head-tail, and USS sources. Here we shortly summarize the properties of the relics and halos in our sample.

-The radio relic 24P73 (VLSS~J2217.5+5943) is located close to the galactic plane. This source was previously known to have a steep spectrum but remained unclassified due to the limited sensitivity and resolution of previous observations. Due to the crowed Milky Way field and high extinction no cluster is found in POSS-II images. Deep NIR imaging will be needed to reveal the presence of a cluster. The sources has an extremely steep curved spectrum, likely the result of adiabatic compression of fossil radio plasma and spectral aging.

-VLSS~J1133.7+2324 is a filamentary relic located next to a nearby spiral galaxy. The radio source breaks up into two parallel filamentary structures. This relic source is probably associated with a structure of galaxies located at a distance of $z \sim 0.6$. The nearby spiral galaxy itself is also detected in the radio images, the flux density consistent with that predicated by the FIR-radio correlation (using IRAS fluxes for this galaxy). The galaxy is partly blocking our view of the distant structure of galaxies, making follow-up observations more difficult. 

-VLSS~J1431.8+1331 consists of two components. The eastern part is probably associated with a cD galaxy in the center of the cluster \object{MaxBCG~J217.95869+13.53470} at $z=0.16$. The radio morphology suggests we are seeing signs of interaction between the radio plasma and the ICM. The other part is probably a relic source. These sources makes an interesting target for follow-up X-ray observations.

-VLSS ~J0004.9$-$3457 is associated with an elliptical galaxy in a small cluster at $z\sim0.3$. The diffuse radio emission surrounds the galaxy in the form of a mini-halo. The source has an arc-like extension to the east. 

- VLSS~J0717.5+3745 is located in the cluster MACS~J0717.5+3745, a massive merging system at a redshift of 0.5548. The radio emission is complex, consisting of a large elongated radio relic and a giant 1.2~Mpc radio halo. The halo has a radio power ($P_{1.4}$) of $5\times 10^{25}$~W~Hz$^{-1}$, the highest one known to date. The relic might trace a giant shock wave related to the cluster merger. This is consistent with the very hot ICM and relatively flat spectral index. 
          
-VLSS~J1515.1+0424 is a radio relic consisting of three filamentary structures in the periphery of the cluster Abell~2048.  Adiabatic compression of fossil radio plasma probably resulted in the complex morphology.\\

We find that larger relics are mostly located in the cluster periphery, while smaller relics are found closer to the cluster center. We also discovered a correlation between the spectral index and the physical size of the relics. A likely explanation for this correlation is that the larger shock waves occur mainly in lower-density regions and have larger Mach numbers. As a consequence they have shallower spectra.
However, the correlation can also (partly) be explained by invoking different origins for relics. Relics formed by diffusive shock acceleration are direct tracers of large shock fronts in clusters, while the compression of fossil radio plasma is thought to produce relatively small relics with steep curved spectra.

If the correlation is explained by the fact that the larger shock waves occur mainly in lower-density regions and have larger Mach numbers then the injection spectral indices should flatten away from the cluster center for the various relics. To measure this requires a sufficient number of flux measurements, especially below the break frequency in the spectrum. With upcoming new radio facilities operating at lower frequencies it should therefore be possible to break the degeneracy between the injection spectral index and spectral aging.

Follow-up observations to measure the polarization properties are currently underway. We also plan to create spectral index maps for our sources which should give more information on the life-times of the various synchrotron emitting regions. Optical observations have been taken to characterize the large-scale environment around the radio sources. For two of our diffuse sources these observations should confirm the presence of a cluster. X-ray observations will be needed to study the dynamical state of the clusters and reveal the presence of shock fronts in the ICM.

\begin{acknowledgements}

We would like to thank the anonymous referee for useful comments. We thank the staff of the GMRT who have made these observations possible. The GMRT is run by the National Centre for Radio Astrophysics of the Tata Institute of Fundamental Research. RJvW would like to thank J.B.R.~Oonk and A.~Omar for helping with the observations, D.~A.~Green for the discussion about source 24P73, and F.~Govoni for pointing out the correct flux of A115. RJvW acknowledges funding from the Royal Netherlands Academy of Arts and Sciences. 

This publication makes use of data products from the Two Micron All Sky Survey, which is a joint project of the University of Massachusetts and the Infrared Processing and Analysis Center/California Institute of Technology, funded by the National Aeronautics and Space Administration and the National Science Foundation. This research has made use of the VizieR catalogue access tool, CDS, Strasbourg, France. 

The Digitized Sky Surveys were produced at the Space Telescope Science Institute under U.S. Government grant NAG W-2166. The images of these surveys are based on photographic data obtained using the Oschin Schmidt Telescope on Palomar Mountain and the UK Schmidt Telescope. The plates were processed into the present compressed digital form with the permission of these institutions. The Second Palomar Observatory Sky Survey (POSS-II) was made by the California Institute of Technology with funds from the National Science Foundation, the National Geographic Society, the Sloan Foundation, the Samuel Oschin Foundation, and the Eastman Kodak Corporation.

\end{acknowledgements}

\bibliographystyle{aa}
\bibliography{12501b}

\begin{thebibliography}{119}
\expandafter\ifx\csname natexlab\endcsname\relax\def\natexlab#1{#1}\fi

\bibitem[{{Abazajian} {et~al.}(2009){Abazajian}, {Adelman-McCarthy},
  {Ag{\"u}eros}, {Allam}, {Allende Prieto}, {An}, {Anderson}, {Anderson},
  {Annis}, {Bahcall}, {Bailer-Jones}, {Barentine}, {Bassett}, {Becker},
  {Beers}, {Bell}, {Belokurov}, {Berlind}, {Berman}, {Bernardi}, {Bickerton},
  {Bizyaev}, {Blakeslee}, {Blanton}, {Bochanski}, {Boroski}, {Brewington},
  {Brinchmann}, {Brinkmann}, {Brunner}, {Budav{\'a}ri}, {Carey}, {Carliles},
  {Carr}, {Castander}, {Cinabro}, {Connolly}, {Csabai}, {Cunha}, {Czarapata},
  {Davenport}, {de Haas}, {Dilday}, {Doi}, {Eisenstein}, {Evans}, {Evans},
  {Fan}, {Friedman}, {Frieman}, {Fukugita}, {G{\"a}nsicke}, {Gates},
  {Gillespie}, {Gilmore}, {Gonzalez}, {Gonzalez}, {Grebel}, {Gunn},
  {Gy{\"o}ry}, {Hall}, {Harding}, {Harris}, {Harvanek}, {Hawley}, {Hayes},
  {Heckman}, {Hendry}, {Hennessy}, {Hindsley}, {Hoblitt}, {Hogan}, {Hogg},
  {Holtzman}, {Hyde}, {Ichikawa}, {Ichikawa}, {Im}, {Ivezi{\'c}}, {Jester},
  {Jiang}, {Johnson}, {Jorgensen}, {Juri{\'c}}, {Kent}, {Kessler}, {Kleinman},
  {Knapp}, {Konishi}, {Kron}, {Krzesinski}, {Kuropatkin}, {Lampeitl},
  {Lebedeva}, {Lee}, {Lee}, {Leger}, {L{\'e}pine}, {Li}, {Lima}, {Lin}, {Long},
  {Loomis}, {Loveday}, {Lupton}, {Magnier}, {Malanushenko}, {Malanushenko},
  {Mandelbaum}, {Margon}, {Marriner}, {Mart{\'{\i}}nez-Delgado}, {Matsubara},
  {McGehee}, {McKay}, {Meiksin}, {Morrison}, {Mullally}, {Munn}, {Murphy},
  {Nash}, {Nebot}, {Neilsen}, {Newberg}, {Newman}, {Nichol}, {Nicinski},
  {Nieto-Santisteban}, {Nitta}, {Okamura}, {Oravetz}, {Ostriker}, {Owen},
  {Padmanabhan}, {Pan}, {Park}, {Pauls}, {Peoples}, {Percival}, {Pier}, {Pope},
  {Pourbaix}, {Price}, {Purger}, {Quinn}, {Raddick}, {Fiorentin}, {Richards},
  {Richmond}, {Riess}, {Rix}, {Rockosi}, {Sako}, {Schlegel}, {Schneider},
  {Scholz}, {Schreiber}, {Schwope}, {Seljak}, {Sesar}, {Sheldon}, {Shimasaku},
  {Sibley}, {Simmons}, {Sivarani}, {Smith}, {Smith}, {Smol{\v c}i{\'c}},
  {Snedden}, {Stebbins}, {Steinmetz}, {Stoughton}, {Strauss}, {Subba Rao},
  {Suto}, {Szalay}, {Szapudi}, {Szkody}, {Tanaka}, {Tegmark}, {Teodoro},
  {Thakar}, {Tremonti}, {Tucker}, {Uomoto}, {Vanden Berk}, {Vandenberg},
  {Vidrih}, {Vogeley}, {Voges}, {Vogt}, {Wadadekar}, {Watters}, {Weinberg},
  {West}, {White}, {Wilhite}, {Wonders}, {Yanny}, {Yocum}, {York}, {Zehavi},
  {Zibetti}, \& {Zucker}}]{2009ApJS..182..543A}
{Abazajian}, K.~N., {Adelman-McCarthy}, J.~K., {Ag{\"u}eros}, M.~A., {et~al.}
  2009, \apjs, 182, 543

\bibitem[{{Baars} {et~al.}(1977){Baars}, {Genzel}, {Pauliny-Toth}, \&
  {Witzel}}]{1977A&A....61...99B}
{Baars}, J.~W.~M., {Genzel}, R., {Pauliny-Toth}, I.~I.~K., \& {Witzel}, A.
  1977, \aap, 61, 99

\bibitem[{{Bade} {et~al.}(1998){Bade}, {Engels}, {Voges}, {Beckmann}, {Boller},
  {Cordis}, {Dahlem}, {Englhauser}, {Molthagen}, {Nass}, {Studt}, \&
  {Reimers}}]{1998A&AS..127..145B}
{Bade}, N., {Engels}, D., {Voges}, W., {et~al.} 1998, \aaps, 127, 145

\bibitem[{{Bagchi} {et~al.}(2009){Bagchi}, {Jacob}, {Gopal-Krishna}, {Werner},
  {Wadnerkar}, {Belapure}, \& {Kumbharkhane}}]{2009MNRAS.tmp.1144B}
{Bagchi}, J., {Jacob}, J., {Gopal-Krishna}, {et~al.} 2009, \mnras, 1144

\bibitem[{{Beck} \& {Krause}(2005)}]{2005AN....326..414B}
{Beck}, R. \& {Krause}, M. 2005, Astronomische Nachrichten, 326, 414

\bibitem[{{Becker} {et~al.}(1995){Becker}, {White}, \&
  {Helfand}}]{1995ApJ...450..559B}
{Becker}, R.~H., {White}, R.~L., \& {Helfand}, D.~J. 1995, \apj, 450, 559

\bibitem[{{Blandford} \& {Eichler}(1987)}]{1987PhR...154....1B}
{Blandford}, R. \& {Eichler}, D. 1987, \physrep, 154, 1

\bibitem[{{Bonafede} {et~al.}(2009{\natexlab{a}}){Bonafede}, {Feretti},
  {Giovannini}, {Govoni}, {Murgia}, {Taylor}, {Ebeling}, {Allen}, {Gentile}, \&
  {Pihlstr{\"o}m}}]{2009A&A...503..707B}
{Bonafede}, A., {Feretti}, L., {Giovannini}, G., {et~al.} 2009{\natexlab{a}},
  \aap, 503, 707

\bibitem[{{Bonafede} {et~al.}(2009{\natexlab{b}}){Bonafede}, {Giovannini},
  {Feretti}, {Govoni}, \& {Murgia}}]{2009A&A...494..429B}
{Bonafede}, A., {Giovannini}, G., {Feretti}, L., {Govoni}, F., \& {Murgia}, M.
  2009{\natexlab{b}}, \aap, 494, 429

\bibitem[{{Briggs}(1995)}]{briggs_phd}
{Briggs}, D.~S. 1995, PhD thesis, New Mexico Institute of Mining Technology,
  Socorro, New Mexico, USA

\bibitem[{{Brunetti} {et~al.}(2008){Brunetti}, {Giacintucci}, {Cassano},
  {Lane}, {Dallacasa}, {Venturi}, {Kassim}, {Setti}, {Cotton}, \&
  {Markevitch}}]{2008Natur.455..944B}
{Brunetti}, G., {Giacintucci}, S., {Cassano}, R., {et~al.} 2008, \nat, 455, 944

\bibitem[{{Brunetti} {et~al.}(1997){Brunetti}, {Setti}, \&
  {Comastri}}]{1997A&A...325..898B}
{Brunetti}, G., {Setti}, G., \& {Comastri}, A. 1997, \aap, 325, 898

\bibitem[{{Carilli} {et~al.}(1991){Carilli}, {Perley}, {Dreher}, \&
  {Leahy}}]{1991ApJ...383..554C}
{Carilli}, C.~L., {Perley}, R.~A., {Dreher}, J.~W., \& {Leahy}, J.~P. 1991,
  \apj, 383, 554

\bibitem[{{Cassano} {et~al.}(2006){Cassano}, {Brunetti}, \&
  {Setti}}]{2006MNRAS.369.1577C}
{Cassano}, R., {Brunetti}, G., \& {Setti}, G. 2006, \mnras, 369, 1577

\bibitem[{{Cassano} {et~al.}(2007){Cassano}, {Brunetti}, {Setti}, {Govoni}, \&
  {Dolag}}]{2007MNRAS.378.1565C}
{Cassano}, R., {Brunetti}, G., {Setti}, G., {Govoni}, F., \& {Dolag}, K. 2007,
  \mnras, 378, 1565

\bibitem[{{Cassano} {et~al.}(2008){Cassano}, {Brunetti}, {Venturi}, {Setti},
  {Dallacasa}, {Giacintucci}, \& {Bardelli}}]{2008A&A...480..687C}
{Cassano}, R., {Brunetti}, G., {Venturi}, T., {et~al.} 2008, \aap, 480, 687

\bibitem[{{Cen} \& {Ostriker}(1999)}]{1999ApJ...514....1C}
{Cen}, R. \& {Ostriker}, J.~P. 1999, \apj, 514, 1

\bibitem[{{Chandra} {et~al.}(2004){Chandra}, {Ray}, \&
  {Bhatnagar}}]{2004ApJ...612..974C}
{Chandra}, P., {Ray}, A., \& {Bhatnagar}, S. 2004, \apj, 612, 974

\bibitem[{{Clarke} \& {Ensslin}(2006)}]{2006AJ....131.2900C}
{Clarke}, T.~E. \& {Ensslin}, T.~A. 2006, \aj, 131, 2900

\bibitem[{{Cohen} {et~al.}(2007){Cohen}, {Lane}, {Cotton}, {Kassim}, {Lazio},
  {Perley}, {Condon}, \& {Erickson}}]{2007AJ....134.1245C}
{Cohen}, A.~S., {Lane}, W.~M., {Cotton}, W.~D., {et~al.} 2007, \aj, 134, 1245

\bibitem[{{Condon} {et~al.}(1991){Condon}, {Anderson}, \&
  {Helou}}]{1991ApJ...376...95C}
{Condon}, J.~J., {Anderson}, M.~L., \& {Helou}, G. 1991, \apj, 376, 95

\bibitem[{{Condon} {et~al.}(1998){Condon}, {Cotton}, {Greisen}, {Yin},
  {Perley}, {Taylor}, \& {Broderick}}]{1998AJ....115.1693C}
{Condon}, J.~J., {Cotton}, W.~D., {Greisen}, E.~W., {et~al.} 1998, \aj, 115,
  1693

\bibitem[{{Cornwell} \& {Perley}(1992)}]{1992A&A...261..353C}
{Cornwell}, T.~J. \& {Perley}, R.~A. 1992, \aap, 261, 353

\bibitem[{{Cotton} {et~al.}(2004){Cotton}, {Condon}, {Perley}, {Kassim},
  {Lazio}, {Cohen}, {Lane}, \& {Erickson}}]{2004SPIE.5489..180C}
{Cotton}, W.~D., {Condon}, J.~J., {Perley}, R.~A., {et~al.} 2004, in Presented
  at the Society of Photo-Optical Instrumentation Engineers (SPIE) Conference,
  Vol. 5489, Society of Photo-Optical Instrumentation Engineers (SPIE)
  Conference Series, ed. J.~M. {Oschmann}, Jr., 180--189

\bibitem[{{De Breuck} {et~al.}(2002{\natexlab{a}}){De Breuck}, {Tang}, {de
  Bruyn}, {R{\"o}ttgering}, \& {van Breugel}}]{2002A&A...394...59D}
{De Breuck}, C., {Tang}, Y., {de Bruyn}, A.~G., {R{\"o}ttgering}, H., \& {van
  Breugel}, W. 2002{\natexlab{a}}, \aap, 394, 59

\bibitem[{{De Breuck} {et~al.}(2000){De Breuck}, {van Breugel},
  {R{\"o}ttgering}, \& {Miley}}]{2000A&AS..143..303D}
{De Breuck}, C., {van Breugel}, W., {R{\"o}ttgering}, H.~J.~A., \& {Miley}, G.
  2000, \aaps, 143, 303

\bibitem[{{De Breuck} {et~al.}(2002{\natexlab{b}}){De Breuck}, {van Breugel},
  {Stanford}, {R{\"o}ttgering}, {Miley}, \& {Stern}}]{2002AJ....123..637D}
{De Breuck}, C., {van Breugel}, W., {Stanford}, S.~A., {et~al.}
  2002{\natexlab{b}}, \aj, 123, 637

\bibitem[{{de Vries} {et~al.}(2007){de Vries}, {Snellen}, {Schilizzi},
  {Lehnert}, \& {Bremer}}]{2007A&A...464..879D}
{de Vries}, N., {Snellen}, I.~A.~G., {Schilizzi}, R.~T., {Lehnert}, M.~D., \&
  {Bremer}, M.~N. 2007, \aap, 464, 879

\bibitem[{{Dolag} {et~al.}(2008){Dolag}, {Bykov}, \&
  {Diaferio}}]{2008SSRv..134..311D}
{Dolag}, K., {Bykov}, A.~M., \& {Diaferio}, A. 2008, Space Science Reviews,
  134, 311

\bibitem[{{Douglas} {et~al.}(1996){Douglas}, {Bash}, {Bozyan}, {Torrence}, \&
  {Wolfe}}]{1996AJ....111.1945D}
{Douglas}, J.~N., {Bash}, F.~N., {Bozyan}, F.~A., {Torrence}, G.~W., \&
  {Wolfe}, C. 1996, \aj, 111, 1945

\bibitem[{{Drury}(1983)}]{1983RPPh...46..973D}
{Drury}, L.~O. 1983, Reports on Progress in Physics, 46, 973

\bibitem[{{Ebeling} {et~al.}(2004){Ebeling}, {Barrett}, \&
  {Donovan}}]{2004ApJ...609L..49E}
{Ebeling}, H., {Barrett}, E., \& {Donovan}, D. 2004, \apjl, 609, L49

\bibitem[{{Ebeling} {et~al.}(2007){Ebeling}, {Barrett}, {Donovan}, {Ma},
  {Edge}, \& {van Speybroeck}}]{2007ApJ...661L..33E}
{Ebeling}, H., {Barrett}, E., {Donovan}, D., {et~al.} 2007, \apjl, 661, L33

\bibitem[{{Ebeling} {et~al.}(2001){Ebeling}, {Edge}, \&
  {Henry}}]{2001ApJ...553..668E}
{Ebeling}, H., {Edge}, A.~C., \& {Henry}, J.~P. 2001, \apj, 553, 668

\bibitem[{{Edge} {et~al.}(2003){Edge}, {Ebeling}, {Bremer}, {R{\"o}ttgering},
  {van Haarlem}, {Rengelink}, \& {Courtney}}]{2003MNRAS.339..913E}
{Edge}, A.~C., {Ebeling}, H., {Bremer}, M., {et~al.} 2003, \mnras, 339, 913

\bibitem[{{Ensslin} {et~al.}(1998){Ensslin}, {Biermann}, {Klein}, \&
  {Kohle}}]{1998A&A...332..395E}
{Ensslin}, T.~A., {Biermann}, P.~L., {Klein}, U., \& {Kohle}, S. 1998, \aap,
  332, 395

\bibitem[{{En{\ss}lin} \& {Br{\"u}ggen}(2002)}]{2002MNRAS.331.1011E}
{En{\ss}lin}, T.~A. \& {Br{\"u}ggen}, M. 2002, \mnras, 331, 1011

\bibitem[{{En{\ss}lin} \& {Gopal-Krishna}(2001)}]{2001A&A...366...26E}
{En{\ss}lin}, T.~A. \& {Gopal-Krishna}. 2001, \aap, 366, 26

\bibitem[{{En{\ss}lin} \& {R{\"o}ttgering}(2002)}]{2002A&A...396...83E}
{En{\ss}lin}, T.~A. \& {R{\"o}ttgering}, H. 2002, \aap, 396, 83

\bibitem[{{Fabian} {et~al.}(1991){Fabian}, {Nulsen}, \&
  {Canizares}}]{1991A&ARv...2..191F}
{Fabian}, A.~C., {Nulsen}, P.~E.~J., \& {Canizares}, C.~R. 1991, \aapr, 2, 191

\bibitem[{{Fanaroff} \& {Riley}(1974)}]{1974MNRAS.167P..31F}
{Fanaroff}, B.~L. \& {Riley}, J.~M. 1974, \mnras, 167, 31P

\bibitem[{{Feretti} {et~al.}(2006){Feretti}, {Bacchi}, {Slee}, {Giovannini},
  {Govoni}, {Andernach}, \& {Tsarevsky}}]{2006MNRAS.368..544F}
{Feretti}, L., {Bacchi}, M., {Slee}, O.~B., {et~al.} 2006, \mnras, 368, 544

\bibitem[{{Feretti} \& {Giovannini}(1996)}]{1996IAUS..175..333F}
{Feretti}, L. \& {Giovannini}, G. 1996, in IAU Symposium, Vol. 175,
  Extragalactic Radio Sources, ed. R.~D. {Ekers}, C.~{Fanti}, \&
  L.~{Padrielli}, 333--+

\bibitem[{{Feretti} {et~al.}(2004){Feretti}, {Orr{\`u}}, {Brunetti},
  {Giovannini}, {Kassim}, \& {Setti}}]{2004A&A...423..111F}
{Feretti}, L., {Orr{\`u}}, E., {Brunetti}, G., {et~al.} 2004, \aap, 423, 111

\bibitem[{{Ferrari} {et~al.}(2008){Ferrari}, {Govoni}, {Schindler}, {Bykov}, \&
  {Rephaeli}}]{2008SSRv..134...93F}
{Ferrari}, C., {Govoni}, F., {Schindler}, S., {Bykov}, A.~M., \& {Rephaeli}, Y.
  2008, Space Science Reviews, 134, 93

\bibitem[{{Giacintucci} {et~al.}(2008){Giacintucci}, {Venturi}, {Macario},
  {Dallacasa}, {Brunetti}, {Markevitch}, {Cassano}, {Bardelli}, \&
  {Athreya}}]{2008A&A...486..347G}
{Giacintucci}, S., {Venturi}, T., {Macario}, G., {et~al.} 2008, \aap, 486, 347

\bibitem[{{Giovannini} \& {Feretti}(2000)}]{2000NewA....5..335G}
{Giovannini}, G. \& {Feretti}, L. 2000, New Astronomy, 5, 335

\bibitem[{{Giovannini} {et~al.}(1991){Giovannini}, {Feretti}, \&
  {Stanghellini}}]{1991A&A...252..528G}
{Giovannini}, G., {Feretti}, L., \& {Stanghellini}, C. 1991, \aap, 252, 528

\bibitem[{{Goldshmidt} \& {Rephaeli}(1994)}]{1994ApJ...431..586G}
{Goldshmidt}, O. \& {Rephaeli}, Y. 1994, \apj, 431, 586

\bibitem[{{Gopal-Krishna} {et~al.}(2002){Gopal-Krishna}, {Kulkarni}, {Bagchi},
  \& {Melnick}}]{2002IAUS..199..159G}
{Gopal-Krishna}, {Kulkarni}, V.~K., {Bagchi}, J., \& {Melnick}, J. 2002, in IAU
  Symposium, Vol. 199, The Universe at Low Radio Frequencies, ed. A.~{Pramesh
  Rao}, G.~{Swarup}, \& {Gopal-Krishna}, 159--+

\bibitem[{{Govoni} {et~al.}(2001){Govoni}, {Feretti}, {Giovannini},
  {B{\"o}hringer}, {Reiprich}, \& {Murgia}}]{2001A&A...376..803G}
{Govoni}, F., {Feretti}, L., {Giovannini}, G., {et~al.} 2001, \aap, 376, 803

\bibitem[{{Govoni} {et~al.}(2009){Govoni}, {Murgia}, {Markevitch}, {Feretti},
  {Giovannini}, {Taylor}, \& {Carretti}}]{2009A&A...499..371G}
{Govoni}, F., {Murgia}, M., {Markevitch}, M., {et~al.} 2009, \aap, 499, 371

\bibitem[{{Green} \& {Joncas}(1994)}]{1994A&AS..104..481G}
{Green}, D.~A. \& {Joncas}, G. 1994, \aaps, 104, 481

\bibitem[{{Hales} {et~al.}(2007){Hales}, {Riley}, {Waldram}, {Warner}, \&
  {Baldwin}}]{2007MNRAS.382.1639H}
{Hales}, S.~E.~G., {Riley}, J.~M., {Waldram}, E.~M., {Warner}, P.~J., \&
  {Baldwin}, J.~E. 2007, \mnras, 382, 1639

\bibitem[{{Hales} {et~al.}(1995){Hales}, {Waldram}, {Rees}, \&
  {Warner}}]{1995MNRAS.274..447H}
{Hales}, S.~E.~G., {Waldram}, E.~M., {Rees}, N., \& {Warner}, P.~J. 1995,
  \mnras, 274, 447

\bibitem[{{Harris} {et~al.}(1996){Harris}, {Forman}, {Gioa}, {Hale}, {Harnden},
  {Jones}, {Karakashian}, {Maccacaro}, {McSweeney}, {Primnini}, {Schwarz},
  {Tananbaum}, \& {Thurman}}]{1996yCat.9013....0H}
{Harris}, D.~E., {Forman}, W., {Gioa}, I.~M., {et~al.} 1996, VizieR Online Data
  Catalog, 9013, 0

\bibitem[{{Haynes} {et~al.}(1997){Haynes}, {Giovanelli}, {Herter}, {Vogt},
  {Freudling}, {Maia}, {Salzer}, \& {Wegner}}]{1997AJ....113.1197H}
{Haynes}, M.~P., {Giovanelli}, R., {Herter}, T., {et~al.} 1997, \aj, 113, 1197

\bibitem[{{Higgs}(1989)}]{1989JRASC..83..105H}
{Higgs}, L.~A. 1989, \jrasc, 83, 105

\bibitem[{{Hoeft} \& {Br{\"u}ggen}(2007)}]{2007MNRAS.375...77H}
{Hoeft}, M. \& {Br{\"u}ggen}, M. 2007, \mnras, 375, 77

\bibitem[{{Hoeft} {et~al.}(2008){Hoeft}, {Br{\"u}ggen}, {Yepes},
  {Gottl{\"o}ber}, \& {Schwope}}]{2008MNRAS.391.1511H}
{Hoeft}, M., {Br{\"u}ggen}, M., {Yepes}, G., {Gottl{\"o}ber}, S., \& {Schwope},
  A. 2008, \mnras, 391, 1511

\bibitem[{{Inoue} {et~al.}(2008){Inoue}, {Sigl}, {Miniati}, \& {et
  al.}}]{2008ICRC....4..555I}
{Inoue}, S., {Sigl}, G., {Miniati}, F., \& {et al.} 2008, in International
  Cosmic Ray Conference, Vol.~4, International Cosmic Ray Conference, 555--558

\bibitem[{{Jaffe} \& {Perola}(1973)}]{1973A&A....26..423J}
{Jaffe}, W.~J. \& {Perola}, G.~C. 1973, \aap, 26, 423

\bibitem[{{Joncas} \& {Higgs}(1990)}]{1990A&AS...82..113J}
{Joncas}, G. \& {Higgs}, L.~A. 1990, \aaps, 82, 113

\bibitem[{{Jones} \& {Ellison}(1991)}]{1991SSRv...58..259J}
{Jones}, F.~C. \& {Ellison}, D.~C. 1991, Space Science Reviews, 58, 259

\bibitem[{{Kang} {et~al.}(1996){Kang}, {Ryu}, \& {Jones}}]{1996ApJ...456..422K}
{Kang}, H., {Ryu}, D., \& {Jones}, T.~W. 1996, \apj, 456, 422

\bibitem[{{Kardashev}(1962)}]{1962SvA.....6..317K}
{Kardashev}, N.~S. 1962, Soviet Astronomy, 6, 317

\bibitem[{{Kempner} {et~al.}(2004){Kempner}, {Blanton}, {Clarke}, {En{\ss}lin},
  {Johnston-Hollitt}, \& {Rudnick}}]{2004rcfg.proc..335K}
{Kempner}, J.~C., {Blanton}, E.~L., {Clarke}, T.~E., {et~al.} 2004, in The
  Riddle of Cooling Flows in Galaxies and Clusters of galaxies, ed.
  T.~{Reiprich}, J.~{Kempner}, \& N.~{Soker}, 335--+

\bibitem[{{Kennicutt}(1998)}]{1998ApJ...498..541K}
{Kennicutt}, Jr., R.~C. 1998, \apj, 498, 541

\bibitem[{{Keshet} {et~al.}(2004){Keshet}, {Waxman}, \&
  {Loeb}}]{2004ApJ...617..281K}
{Keshet}, U., {Waxman}, E., \& {Loeb}, A. 2004, \apj, 617, 281

\bibitem[{{Koester} {et~al.}(2007){Koester}, {McKay}, {Annis}, {Wechsler},
  {Evrard}, {Bleem}, {Becker}, {Johnston}, {Sheldon}, {Nichol}, {Miller},
  {Scranton}, {Bahcall}, {Barentine}, {Brewington}, {Brinkmann}, {Harvanek},
  {Kleinman}, {Krzesinski}, {Long}, {Nitta}, {Schneider}, {Sneddin}, {Voges},
  \& {York}}]{2007ApJ...660..239K}
{Koester}, B.~P., {McKay}, T.~A., {Annis}, J., {et~al.} 2007, \apj, 660, 239

\bibitem[{{Komissarov} \& {Gubanov}(1994)}]{1994A&A...285...27K}
{Komissarov}, S.~S. \& {Gubanov}, A.~G. 1994, \aap, 285, 27

\bibitem[{{Liang} {et~al.}(2000){Liang}, {Hunstead}, {Birkinshaw}, \&
  {Andreani}}]{2000ApJ...544..686L}
{Liang}, H., {Hunstead}, R.~W., {Birkinshaw}, M., \& {Andreani}, P. 2000, \apj,
  544, 686

\bibitem[{{Lonsdale} \& {Helou}(1985)}]{1985cgqo.book.....L}
{Lonsdale}, C.~J. \& {Helou}, G. 1985, {Cataloged galaxies and quasars observed
  in the IRAS survey} (Pasadena: Jet Propulsion Laboratory (JPL), 1985)

\bibitem[{{Ma} {et~al.}(2009){Ma}, {Ebeling}, \&
  {Barrett}}]{2009ApJ...693L..56M}
{Ma}, C., {Ebeling}, H., \& {Barrett}, E. 2009, \apjl, 693, L56

\bibitem[{{Malkov} \& {O'C Drury}(2001)}]{2001RPPh...64..429M}
{Malkov}, M.~A. \& {O'C Drury}, L. 2001, Reports on Progress in Physics, 64,
  429

\bibitem[{{Miley} \& {De Breuck}(2008)}]{2008A&ARv..15...67M}
{Miley}, G. \& {De Breuck}, C. 2008, \aapr, 15, 67

\bibitem[{{Miniati}(2002)}]{2002MNRAS.337..199M}
{Miniati}, F. 2002, \mnras, 337, 199

\bibitem[{{Miniati} {et~al.}(2001){Miniati}, {Jones}, {Kang}, \&
  {Ryu}}]{2001ApJ...562..233M}
{Miniati}, F., {Jones}, T.~W., {Kang}, H., \& {Ryu}, D. 2001, \apj, 562, 233

\bibitem[{{Miniati} {et~al.}(2000){Miniati}, {Ryu}, {Kang}, {Jones}, {Cen}, \&
  {Ostriker}}]{2000ApJ...542..608M}
{Miniati}, F., {Ryu}, D., {Kang}, H., {et~al.} 2000, \apj, 542, 608

\bibitem[{{Miyauchi-Isobe} \& {Maehara}(2002)}]{2002PNAOJ...6..107M}
{Miyauchi-Isobe}, N. \& {Maehara}, H. 2002, Publications of the National
  Astronomical Observatory of Japan, 6, 107

\bibitem[{{Moshir} \& {et al.}(1990)}]{1990IRASF.C......0M}
{Moshir}, M. \& {et al.} 1990, in IRAS Faint Source Catalogue, version 2.0
  (1990), 0--+

\bibitem[{{Murgia} {et~al.}(2002){Murgia}, {Fanti}, {Fanti}, {Gregorini},
  {Klein}, {Mack}, \& {Vigotti}}]{2002NewAR..46..307M}
{Murgia}, M., {Fanti}, C., {Fanti}, R., {et~al.} 2002, New Astronomy Review,
  46, 307

\bibitem[{{Nicastro} {et~al.}(2005){Nicastro}, {Mathur}, {Elvis}, {Drake},
  {Fang}, {Fruscione}, {Krongold}, {Marshall}, {Williams}, \&
  {Zezas}}]{2005Natur.433..495N}
{Nicastro}, F., {Mathur}, S., {Elvis}, M., {et~al.} 2005, \nat, 433, 495

\bibitem[{{Nilson}(1973)}]{1973UGC...C...0000N}
{Nilson}, P. 1973, Nova Acta Regiae Soc.~Sci.~Upsaliensis Ser.~V, 0

\bibitem[{{Norman} {et~al.}(1995){Norman}, {Melrose}, \&
  {Achterberg}}]{1995ApJ...454...60N}
{Norman}, C.~A., {Melrose}, D.~B., \& {Achterberg}, A. 1995, \apj, 454, 60

\bibitem[{{Orr{\'u}} {et~al.}(2007){Orr{\'u}}, {Murgia}, {Feretti}, {Govoni},
  {Brunetti}, {Giovannini}, {Girardi}, \& {Setti}}]{2007A&A...467..943O}
{Orr{\'u}}, E., {Murgia}, M., {Feretti}, L., {et~al.} 2007, \aap, 467, 943

\bibitem[{{Pacholczyk}(1970)}]{1970ranp.book.....P}
{Pacholczyk}, A.~G. 1970, {Radio astrophysics. Nonthermal processes in galactic
  and extragalactic sources} (Series of Books in Astronomy and Astrophysics,
  San Francisco: Freeman, 1970)

\bibitem[{{Parma} {et~al.}(2007){Parma}, {Murgia}, {de Ruiter}, {Fanti},
  {Mack}, \& {Govoni}}]{2007A&A...470..875P}
{Parma}, P., {Murgia}, M., {de Ruiter}, H.~R., {et~al.} 2007, \aap, 470, 875

\bibitem[{{Paturel} {et~al.}(2003){Paturel}, {Petit}, {Prugniel}, {Theureau},
  {Rousseau}, {Brouty}, {Dubois}, \& {Cambr{\'e}sy}}]{2003A&A...412...45P}
{Paturel}, G., {Petit}, C., {Prugniel}, P., {et~al.} 2003, \aap, 412, 45

\bibitem[{{Perley}(1989)}]{1989ASPC....6..259P}
{Perley}, R.~A. 1989, in Astronomical Society of the Pacific Conference Series,
  Vol.~6, Synthesis Imaging in Radio Astronomy, ed. R.~A. {Perley}, F.~R.
  {Schwab}, \& A.~H. {Bridle}, 259--+

\bibitem[{{Perley} \& {Taylor}(1999)}]{perleyandtaylor}
{Perley}, R.~T. \& {Taylor}, G.~B. 1999, {VLA Calibrator Manual}, Tech. rep.,
  NRAO

\bibitem[{{Peterson} \& {Fabian}(2006)}]{2006PhR...427....1P}
{Peterson}, J.~R. \& {Fabian}, A.~C. 2006, \physrep, 427, 1

\bibitem[{{Pfrommer}(2008)}]{2008MNRAS.385.1242P}
{Pfrommer}, C. 2008, \mnras, 385, 1242

\bibitem[{{Pfrommer} {et~al.}(2008){Pfrommer}, {En{\ss}lin}, \&
  {Springel}}]{2008MNRAS.385.1211P}
{Pfrommer}, C., {En{\ss}lin}, T.~A., \& {Springel}, V. 2008, \mnras, 385, 1211

\bibitem[{{Pfrommer} {et~al.}(2006){Pfrommer}, {Springel}, {En{\ss}lin}, \&
  {Jubelgas}}]{2006MNRAS.367..113P}
{Pfrommer}, C., {Springel}, V., {En{\ss}lin}, T.~A., \& {Jubelgas}, M. 2006,
  \mnras, 367, 113

\bibitem[{{Rees}(1990)}]{1990MNRAS.244..233R}
{Rees}, N. 1990, \mnras, 244, 233

\bibitem[{{Rengelink} {et~al.}(1997){Rengelink}, {Tang}, {de Bruyn}, {Miley},
  {Bremer}, {Roettgering}, \& {Bremer}}]{1997A&AS..124..259R}
{Rengelink}, R.~B., {Tang}, Y., {de Bruyn}, A.~G., {et~al.} 1997, \aaps, 124,
  259

\bibitem[{{Roettiger} {et~al.}(1999){Roettiger}, {Burns}, \&
  {Stone}}]{1999ApJ...518..603R}
{Roettiger}, K., {Burns}, J.~O., \& {Stone}, J.~M. 1999, \apj, 518, 603

\bibitem[{{R\"ottgering} {et~al.}(1997){R\"ottgering}, {Wieringa}, {Hunstead},
  \& {Ekers}}]{1997MNRAS.290..577R}
{R\"ottgering}, H.~J.~A., {Wieringa}, M.~H., {Hunstead}, R.~W., \& {Ekers},
  R.~D. 1997, \mnras, 290, 577

\bibitem[{{Ryu} {et~al.}(2003){Ryu}, {Kang}, {Hallman}, \&
  {Jones}}]{2003ApJ...593..599R}
{Ryu}, D., {Kang}, H., {Hallman}, E., \& {Jones}, T.~W. 2003, \apj, 593, 599

\bibitem[{{Saito} {et~al.}(1990){Saito}, {Ohtani}, {Asonuma}, {Kashikawa},
  {Maki}, {Nishida}, \& {Watanabe}}]{1990PASJ...42..603S}
{Saito}, M., {Ohtani}, H., {Asonuma}, A., {et~al.} 1990, \pasj, 42, 603

\bibitem[{{Schlegel} {et~al.}(1998){Schlegel}, {Finkbeiner}, \&
  {Davis}}]{1998ApJ...500..525S}
{Schlegel}, D.~J., {Finkbeiner}, D.~P., \& {Davis}, M. 1998, \apj, 500, 525

\bibitem[{{Skillman} {et~al.}(2008){Skillman}, {O'Shea}, {Hallman}, {Burns}, \&
  {Norman}}]{2008ApJ...689.1063S}
{Skillman}, S.~W., {O'Shea}, B.~W., {Hallman}, E.~J., {Burns}, J.~O., \&
  {Norman}, M.~L. 2008, \apj, 689, 1063

\bibitem[{{Slee} {et~al.}(2001){Slee}, {Roy}, {Murgia}, {Andernach}, \&
  {Ehle}}]{2001AJ....122.1172S}
{Slee}, O.~B., {Roy}, A.~L., {Murgia}, M., {Andernach}, H., \& {Ehle}, M. 2001,
  \aj, 122, 1172

\bibitem[{{Solomon} {et~al.}(1997){Solomon}, {Downes}, {Radford}, \&
  {Barrett}}]{1997ApJ...478..144S}
{Solomon}, P.~M., {Downes}, D., {Radford}, S.~J.~E., \& {Barrett}, J.~W. 1997,
  \apj, 478, 144

\bibitem[{{Springob} {et~al.}(2005){Springob}, {Haynes}, {Giovanelli}, \&
  {Kent}}]{2005ApJS..160..149S}
{Springob}, C.~M., {Haynes}, M.~P., {Giovanelli}, R., \& {Kent}, B.~R. 2005,
  \apjs, 160, 149

\bibitem[{{Struble} \& {Rood}(1999)}]{1999ApJS..125...35S}
{Struble}, M.~F. \& {Rood}, H.~J. 1999, \apjs, 125, 35

\bibitem[{{Subrahmanyan} {et~al.}(2003){Subrahmanyan}, {Beasley}, {Goss},
  {Golap}, \& {Hunstead}}]{2003AJ....125.1095S}
{Subrahmanyan}, R., {Beasley}, A.~J., {Goss}, W.~M., {Golap}, K., \&
  {Hunstead}, R.~W. 2003, \aj, 125, 1095

\bibitem[{{van Weeren} {et~al.}(2009{\natexlab{a}}){van Weeren}, {Rottgering},
  {Bagchi}, {Raychaudhury}, {Intema}, {Miniati}, {Ensslin}, {Markevitch}, \&
  {Erben}}]{2009arXiv0908.0728V}
{van Weeren}, R.~J., {Rottgering}, H.~J.~A., {Bagchi}, J., {et~al.}
  2009{\natexlab{a}}, ArXiv e-prints (0908.0728), \aap~accepted

\bibitem[{{van Weeren} {et~al.}(2009{\natexlab{b}}){van Weeren},
  {R\"ottgering}, {Br\"uggen}, \& {Cohen}}]{2009arXiv0905.3650V}
{van Weeren}, R.~J., {R\"ottgering}, H.~J.~A., {Br\"uggen}, M., \& {Cohen}, A.
  2009{\natexlab{b}}, ArXiv e-prints (0905.3650), \aap~accepted

\bibitem[{{Vazza} {et~al.}(2009){Vazza}, {Brunetti}, \&
  {Gheller}}]{2009MNRAS.395.1333V}
{Vazza}, F., {Brunetti}, G., \& {Gheller}, C. 2009, \mnras, 395, 1333

\bibitem[{{Venturi} {et~al.}(2007){Venturi}, {Giacintucci}, {Brunetti},
  {Cassano}, {Bardelli}, {Dallacasa}, \& {Setti}}]{2007A&A...463..937V}
{Venturi}, T., {Giacintucci}, S., {Brunetti}, G., {et~al.} 2007, \aap, 463, 937

\bibitem[{{Voges} {et~al.}(1999){Voges}, {Aschenbach}, {Boller},
  {Br{\"a}uninger}, {Briel}, {Burkert}, {Dennerl}, {Englhauser}, {Gruber},
  {Haberl}, {Hartner}, {Hasinger}, {K{\"u}rster}, {Pfeffermann}, {Pietsch},
  {Predehl}, {Rosso}, {Schmitt}, {Tr{\"u}mper}, \&
  {Zimmermann}}]{1999A&A...349..389V}
{Voges}, W., {Aschenbach}, B., {Boller}, T., {et~al.} 1999, \aap, 349, 389

\bibitem[{{Voges} {et~al.}(2000){Voges}, {Aschenbach}, {Boller}, {Brauninger},
  {Briel}, {Burkert}, {Dennerl}, {Englhauser}, {Gruber}, {Haberl}, {Hartner},
  {Hasinger}, {Pfeffermann}, {Pietsch}, {Predehl}, {Schmitt}, {Trumper}, \&
  {Zimmermann}}]{2000IAUC.7432R...1V}
{Voges}, W., {Aschenbach}, B., {Boller}, T., {et~al.} 2000, \iaucirc, 7432, 1

\bibitem[{{Waldram} {et~al.}(1996){Waldram}, {Yates}, {Riley}, \&
  {Warner}}]{1996MNRAS.282..779W}
{Waldram}, E.~M., {Yates}, J.~A., {Riley}, J.~M., \& {Warner}, P.~J. 1996,
  \mnras, 282, 779

\bibitem[{{Weinberger} {et~al.}(1995){Weinberger}, {Saurer}, \&
  {Seeberger}}]{1995A&AS..110..269W}
{Weinberger}, R., {Saurer}, W., \& {Seeberger}, R. 1995, \aaps, 110, 269

\bibitem[{{Willott} {et~al.}(2003){Willott}, {Rawlings}, {Jarvis}, \&
  {Blundell}}]{2003MNRAS.339..173W}
{Willott}, C.~J., {Rawlings}, S., {Jarvis}, M.~J., \& {Blundell}, K.~M. 2003,
  \mnras, 339, 173

\bibitem[{{Zanichelli} {et~al.}(2001){Zanichelli}, {Vigotti}, {Scaramella},
  {Grueff}, \& {Vettolani}}]{2001A&A...379...21Z}
{Zanichelli}, A., {Vigotti}, M., {Scaramella}, R., {Grueff}, G., \&
  {Vettolani}, G. 2001, \aap, 379, 21

\bibitem[{{Zhang} {et~al.}(1997){Zhang}, {Zheng}, {Chen}, {Wang}, {Cao},
  {Peng}, \& {Nan}}]{1997A&AS..121...59Z}
{Zhang}, X., {Zheng}, Y., {Chen}, H., {et~al.} 1997, \aaps, 121, 59

\end{thebibliography}
 \newpage
 \begin{appendix} 
 
\section{Other sources in the sample} 
 \label{sec:fr-i}

 \subsection{VLSS J0227.4$-$1642 } 

The radio map shows two resolved elongated patches of emission. The brighter eastern component has a flux of $25.8$~mJy and the weaker western component $11.2$~mJy.  Both components are resolved, 28.3\arcsec~by 10.3\arcsec~(eastern component) and  22.3\arcsec~by 9.0\arcsec~(western component), and show structure at the limit of the $\sim 5\arcsec$ beam. The position angles of the longest axes are roughly the same, which suggests a link between the two components. No optical counterpart is detected at the limit of the POSS-II survey. The steep spectral index of about $-1.5$ could be explained if this source is a FR-I type AGN, where the central engine has stopped and the synchrotron emission in the two lobes has steepened by spectral aging. In this case both components should roughly have the same spectral index. 

\subsection{VLSS J0250.5$-$1247} 
The radio observations show a double source. The two components are slightly resolved, have a similar flux and seem to be connected by a faint bridge. The separation between the two radio components is 15\arcsec. No optical counterpart is detected. 

\subsection{VLSS J0646.8$-$0722} 
The radio map shows a source with an angular size of 1.0\arcmin~by 0.5\arcmin. An optical counterpart (\object{PCG~76039}, mag$_{R}$=18.4) is detected in both the POSS-II and 2MASS surveys. This galaxy was also listed in the catalog of Galaxies Behind the Milky Way \citep[\object{CGMW~1-0379},][]{1990PASJ...42..603S} and in the catalog Galaxies in the ``zone of avoidance'' \citep[\object{ZOAG~G218.99-04.37},][]{1995A&AS..110..269W} and was in both cases classified as an elliptical galaxy. The amount of extinction in this region is A$_{B} = 2.2$ mag. 
The radio map shows a one-sided tail originating from the optical counterpart. This might indicate that the source is located in a cluster and we see the effects of the interaction between radio plasma and the surrounding ICM. 

\subsection{VLSS J1117.1+7003} 
This source has a remarkable steep and straight radio spectrum without any indication of a flux turnover at lower frequencies. The source is resolved into a smooth featureless roughly spherical blob ( 26\arcsec~by 23\arcsec ). No optical counterpart is detected. Given the steep straight radio spectrum it could be a radio halo, however the small size of the source argues against such an identification. The source could be a mini-radio halo surrounding a central cD galaxy of a distant (proto) cluster.

\subsection{VLSS J1636.5+3326} 
The radio map shows two compact components, with the northern component extending to the west. SDSS images show several faint mostly red galaxies around the radio source. Photometric redshifts (SDSS DR7) indicate the presence of a cluster at  $z=0.65 \pm 0.19$. Several galaxies can be identified as possible optical counterparts. 

\subsection{VLSS J1710.5+6844} 
This  ``head-tail''  radio source has a length of 0.73\arcmin~and a width of 0.27\arcmin. An optical counterpart is detected, a galaxy with a K-magnitude of 14.75. An optical $i$-band image from the 2.5m Isaac Newton Telescope (INT)  shows that this source is located within a cluster at a redshift of $z\sim 0.3$. An east-west elongated  X-ray source \object{1RXS J171034.0+684403} follows roughly the galaxy distribution confirming the presence of a cluster. 

\subsection{VLSS J1930.4+1048} 
The radio map shows an elongated source with a LAS of 101\arcsec. The radio morphology is consistent with a FR-I radio source. No optical counterpart is detected at the position of the core, but this is not surprising given the amount of extinction ($A_{B}=3.759$) at  this galactic latitude of $-3.6$ degrees and the high star density in the POSS-II images. A  $69.6$~mJy (probably unrelated) radio source is located 70\arcsec to the north.

\subsection{VLSS J2043.9$-$1118} 
An optical counterpart for this  source is detected in POSS-II images. The faint optical source is located on the peak of the radio emission. The radio emission surrounds the galaxy and there is a hint of a faint extension to the east. 
\subsection{VLSS J2044.7+0447} 
The source is elongated in the east-west direction and has a LAS of 101\arcsec. A R-magnitude 19.56 galaxy corresponds to a peak in the radio emission, consistent with this being a FR-I radio source. The high rms noise is the radio map is caused by dynamic range limitations from \object{4C 04.71}, a $1.24$~Jy source (610~MHz), located 3.3\arcmin~to the south. 

\subsection{VLSS 2122.9+0012} 
The radio source is compact, 11.6\arcsec~by 10.7\arcsec. This is probably a USS source associated with a HzRG. The lack of an optical counterpart in SDSS images in consistent with this explanation, and implies the source is located at $z \gtrsim1.4$.

\subsection{VLSS J2213.2+3411} 
This source was also included in the USS sample of \cite{2000A&AS..143..303D}, \object{WN J2213+3411}. A K$_{s}$ band image from \cite{2002AJ....123..637D} showed a possible counterpart. However, because no high-resolution ($\lesssim 10\arcsec$) VLA images were available the identification was uncertain on the basis of position alone. Our high-resolution $5\arcsec$~image shows a distorted double-lobe source. The central radio core is not detected. The NIR counterpart is located in between the two lobes, confirming that this is indeed the host galaxy. The K-band magnitude of $18.3$ implies a redshift of $1.6 \pm 0.5$ for this source using the K-z relation.

\subsection{VLSS J2341.1+1231} 
The radio map shows a FR-I type radio source. The northwestern and southeastern lobes are listed as \object{NVSS J234104+123203} and \object{NVSS J234107+123126}, respectively. An optical counterpart is detected on POSS-II images at roughly the expected position based on the location of the lobes. This source may be a so called ``dying'' radio source. 
 
\subsection{VLSS J2345.2+2157} 
The radio source shows an S-shape symmetry. An optical counterpart is detected in 2MASS and POSS-II images being the central cD galaxy of a cluster. This is confirmed by the presence of the X-ray source \object{1RXS~J234518.4+215753}. 

\subsection{VLSS J0511.6+0254}   
The radio map shows a FR-I  type radio source. Two possible optical counterparts are detected which are the central elliptical galaxies in a cluster at a redshift of about 0.2.

\subsection{VLSS J0516.2+0103}   
VLSS~J0516.2+0103 is a relatively compact source, without a double lobe structure. The source does not have an optical counterpart in POSS-II images. This implies the source is probably located at $z \gtrsim 0.7$. 

\subsection{VLSS J2209.5+1546}   
The radio map shows an elongated source. No optical counterpart is detected at the limit of the POSS-II images. The source is probably a distant FR-I source. Optical imaging will be needed to confirm this classification.

\subsection{VLSS J2241.3$-$1626}
This source could be a disturbed FR-I source, no optical counterpart is detected. Optical imaging will be needed to confirm this classification.

\subsection{VLSS J2357.0+0441}   
The source is found to be a blend of three compact sources, all without optical counterparts. By convolving the GMRT image to the NVSS resolution (45\arcsec) it seems that the two outer components are responsible for the steep spectrum. These are probably distant FR-II sources.

\begin{figure*}
    \begin{center}
      \includegraphics[angle = 90, trim =0cm 0cm 0cm 0cm,width=0.3\textwidth]{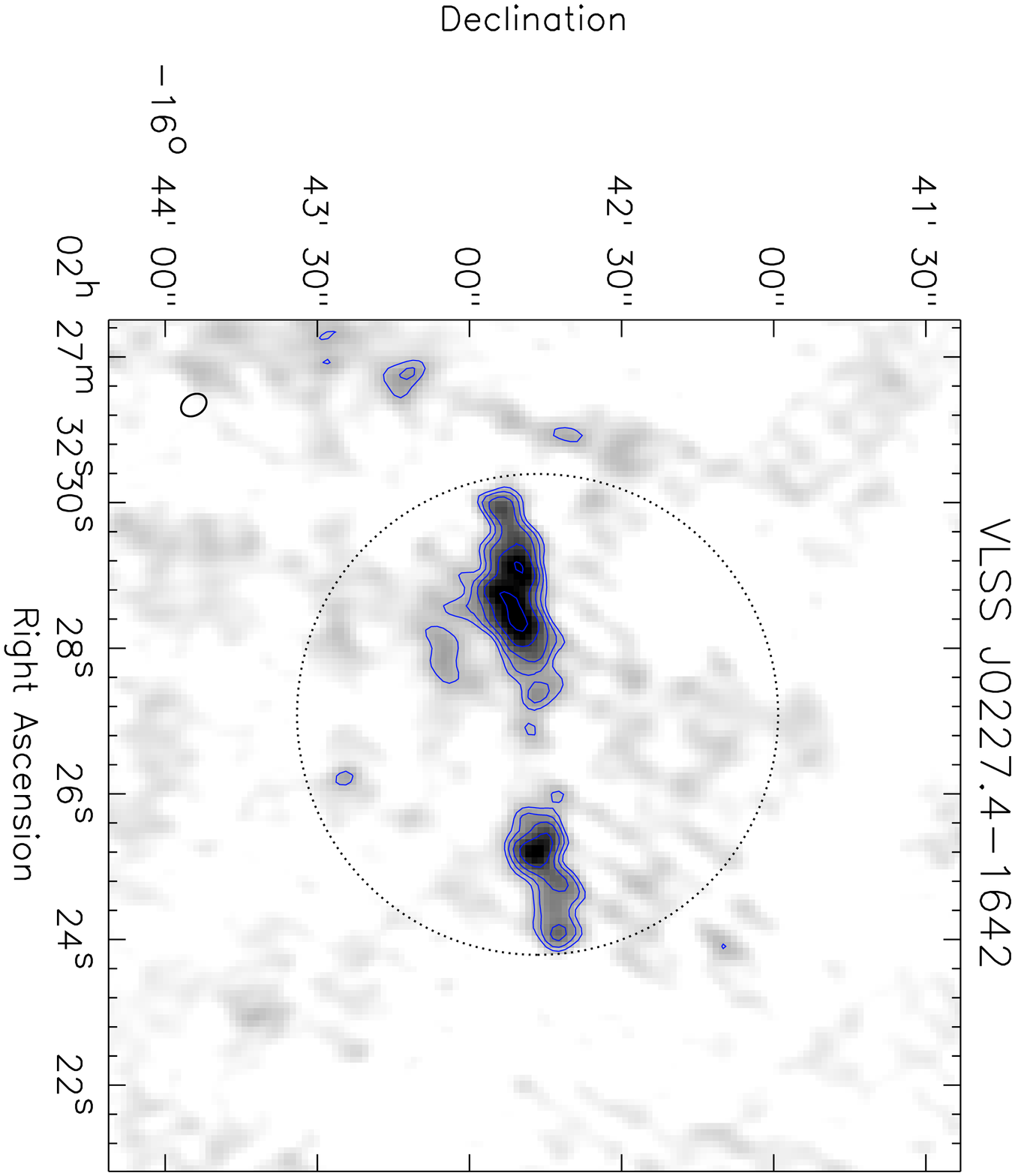}
      \includegraphics[angle = 90, trim =0cm 0cm 0cm 0cm,width=0.3\textwidth]{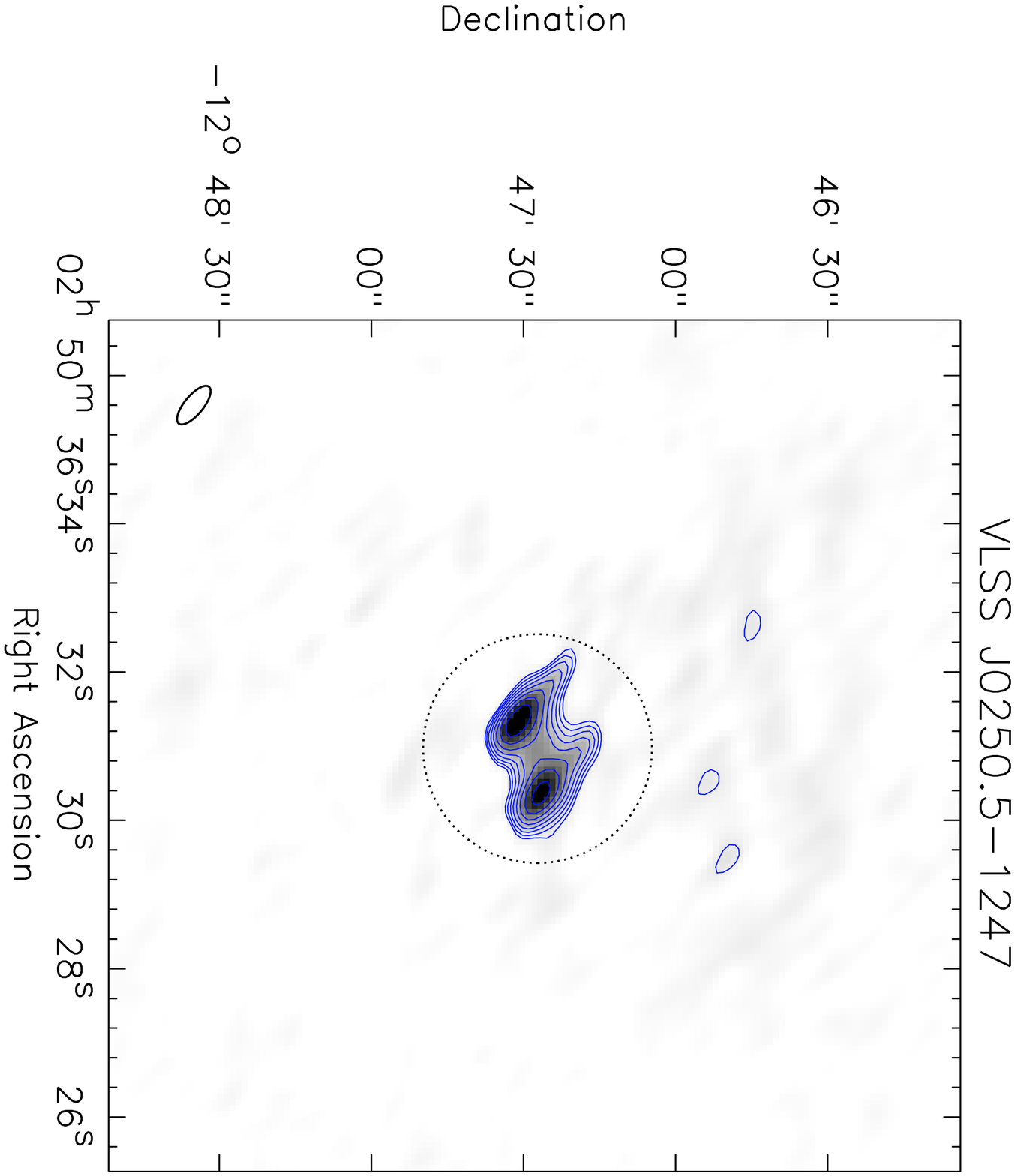}
      \includegraphics[angle = 90, trim =0cm 0cm 0cm 0cm,width=0.3\textwidth]{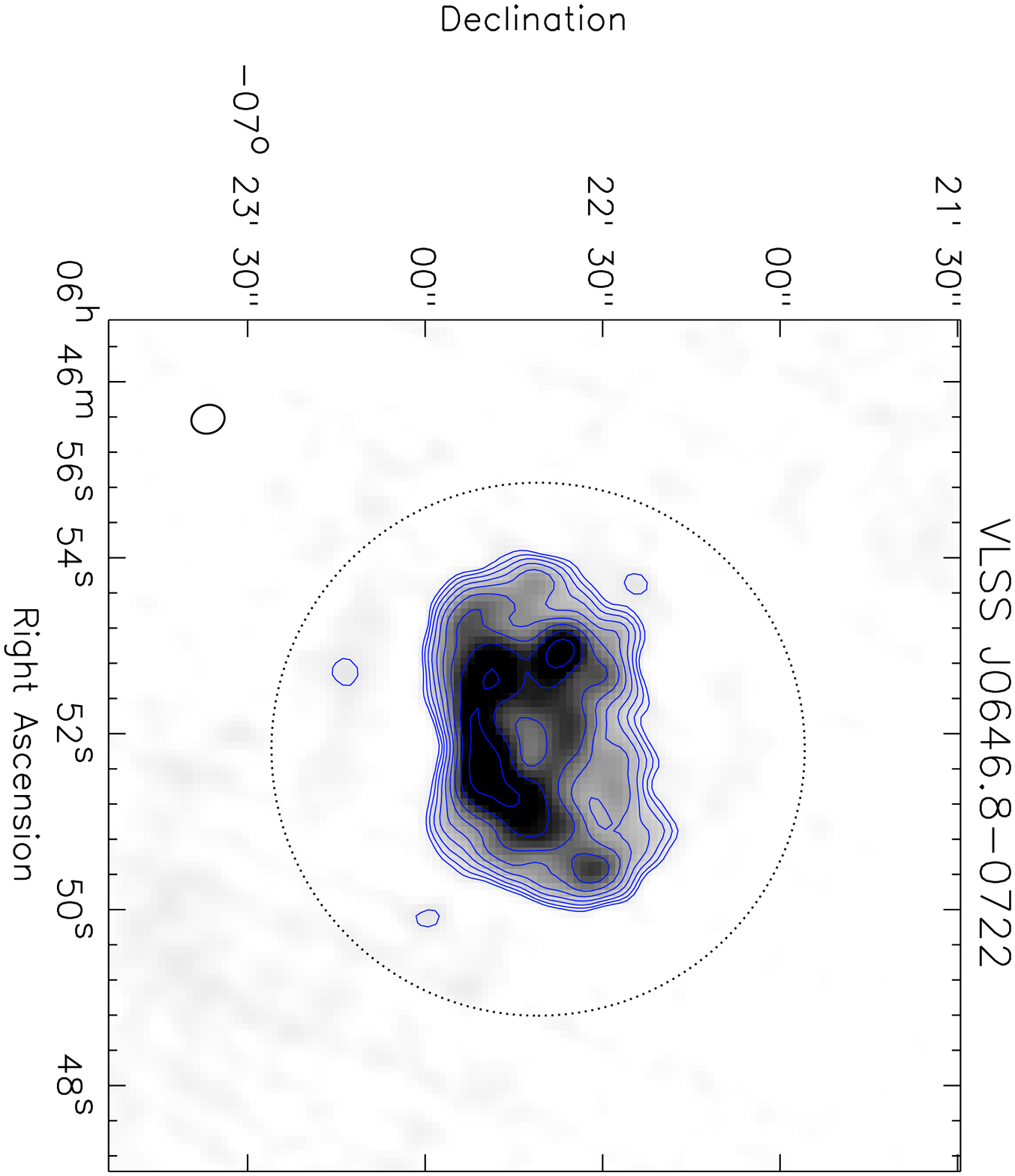}
      \includegraphics[angle = 90, trim =0cm 0cm 0cm 0cm,width=0.3\textwidth]{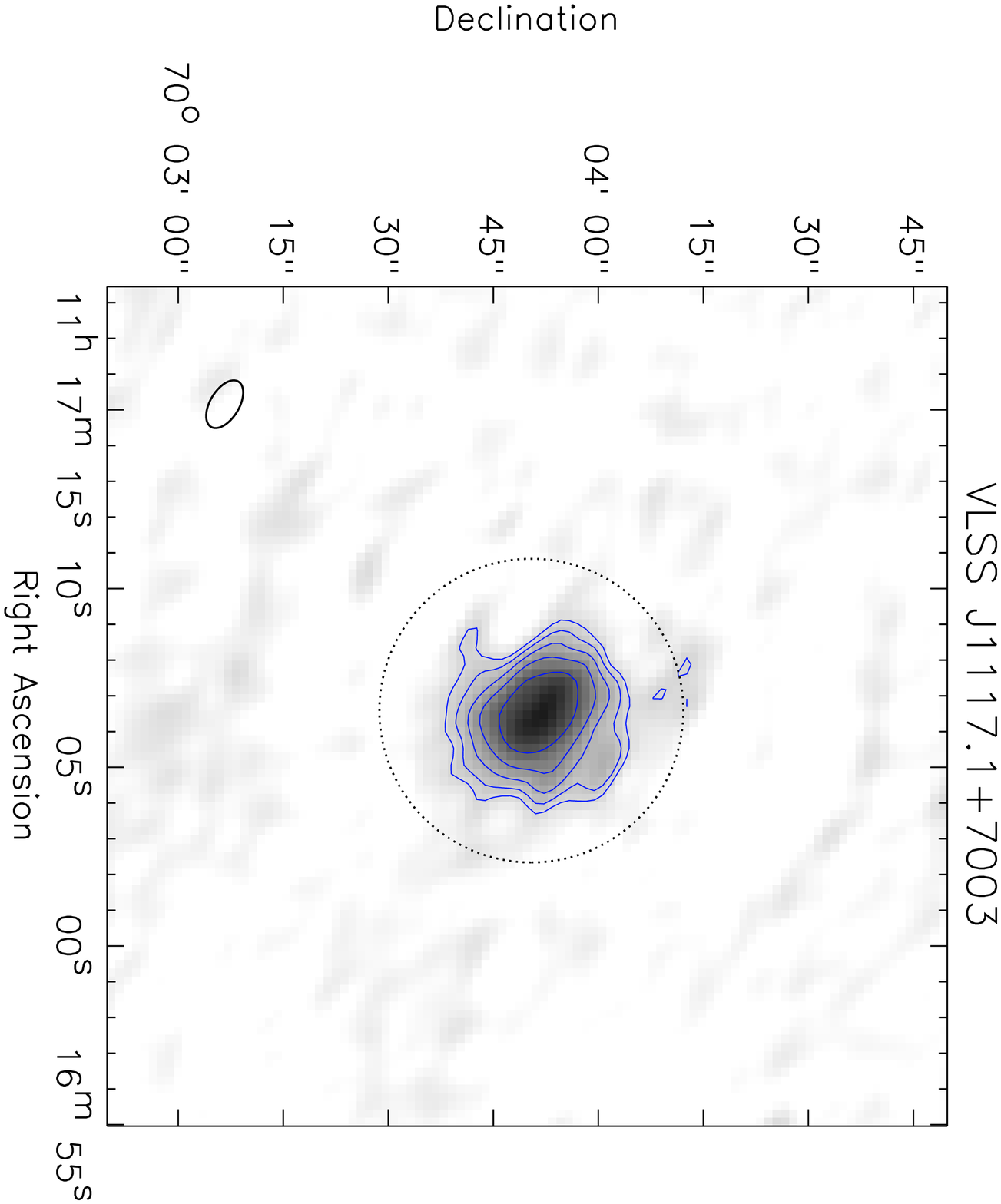}
      \includegraphics[angle = 90, trim =0cm 0cm 0cm 0cm,width=0.3\textwidth]{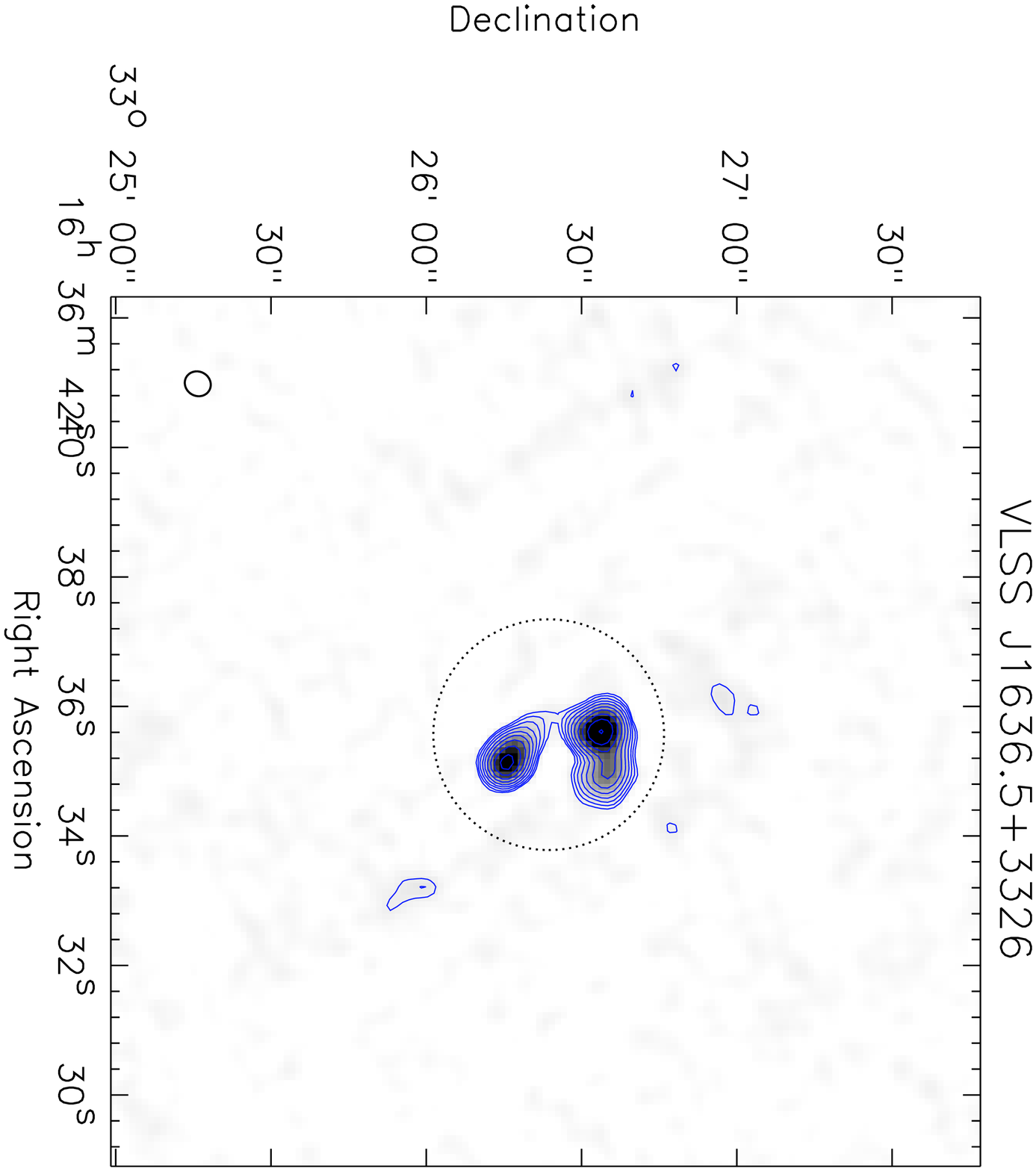}
      \includegraphics[angle = 90, trim =0cm 0cm 0cm 0cm,width=0.3\textwidth]{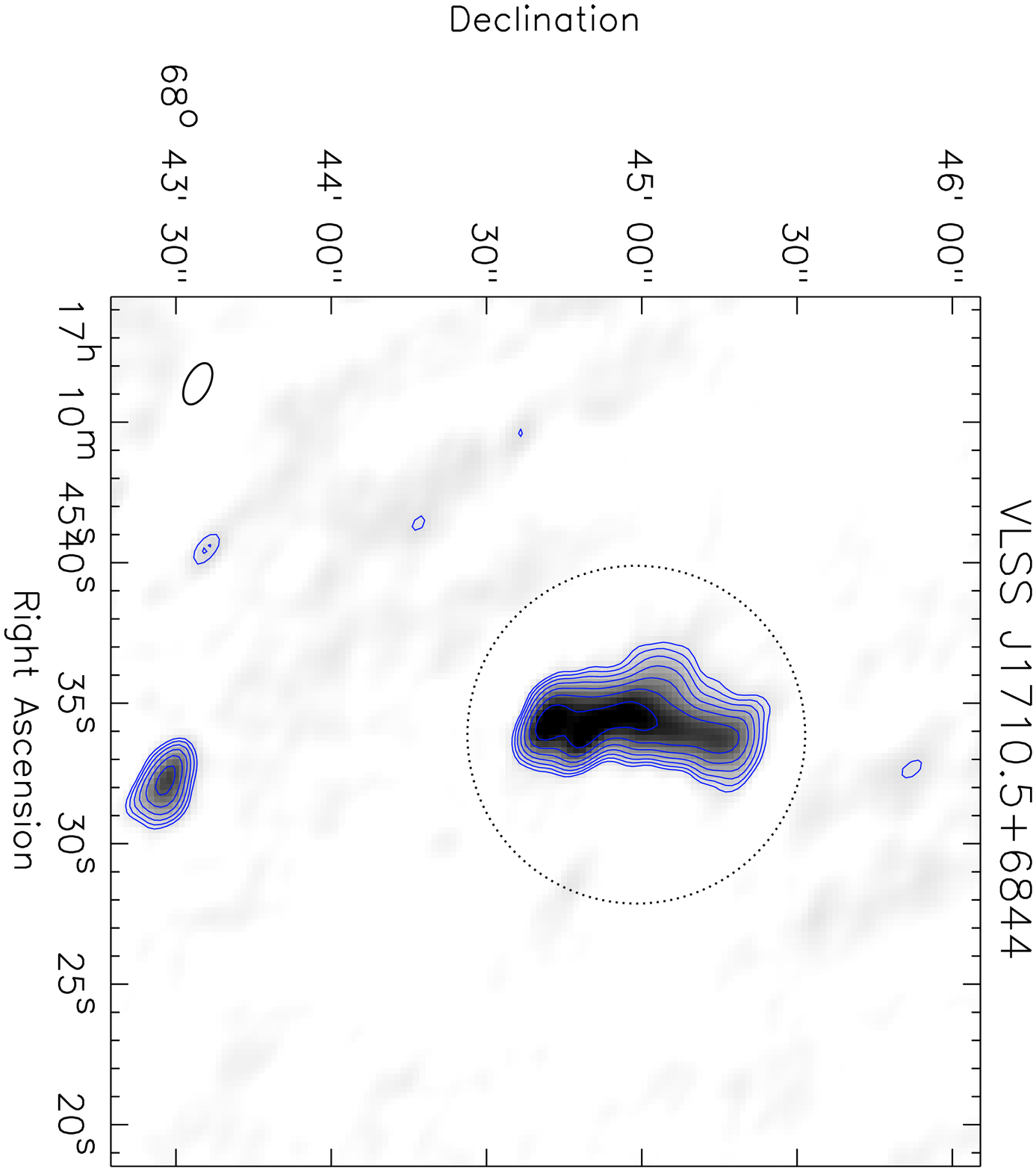}
      \includegraphics[angle = 90, trim =0cm 0cm 0cm 0cm,width=0.3\textwidth]{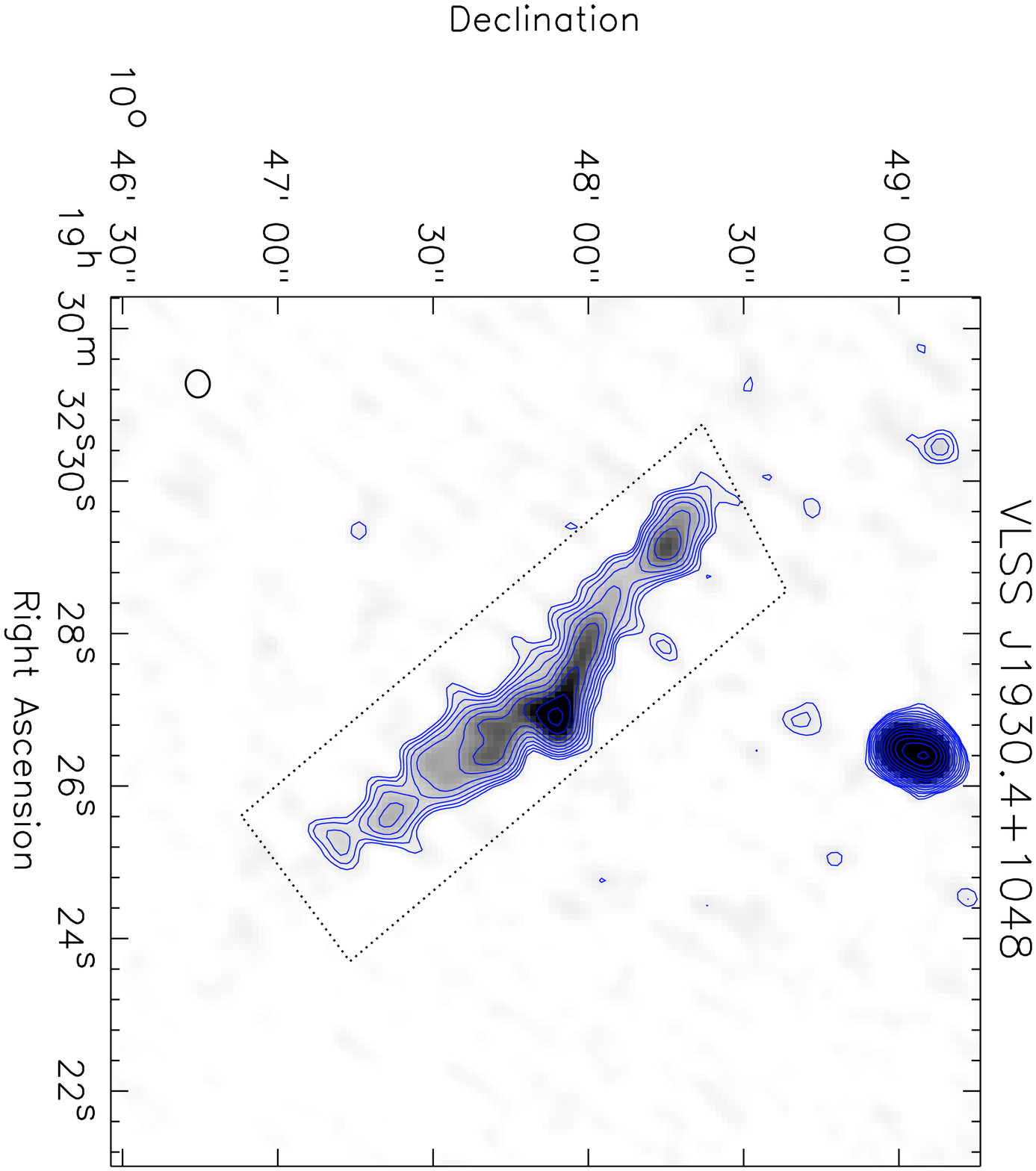}
      \includegraphics[angle = 90, trim =0cm 0cm 0cm 0cm,width=0.3\textwidth]{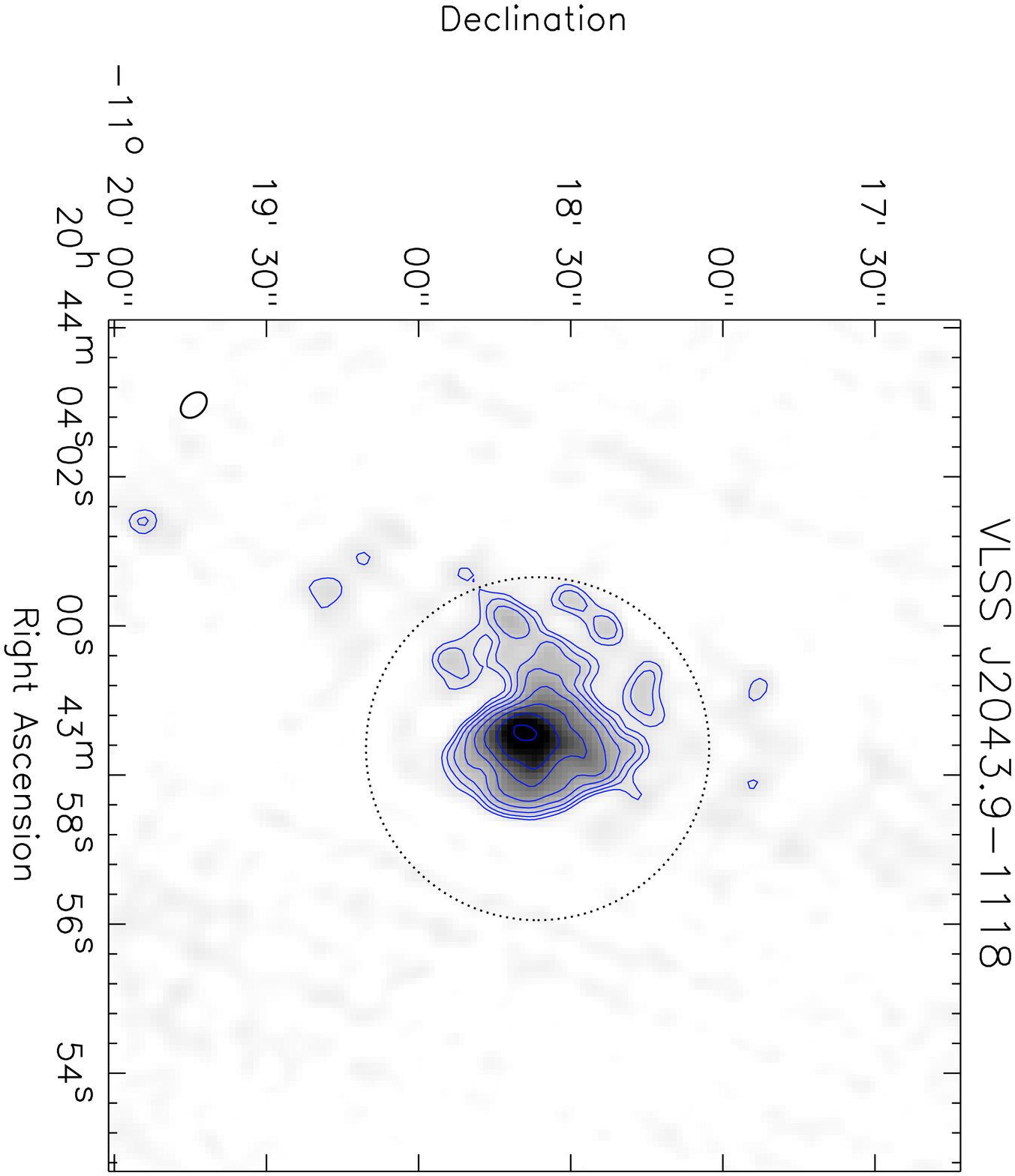}
      \includegraphics[angle = 90, trim =0cm 0cm 0cm 0cm,width=0.3\textwidth]{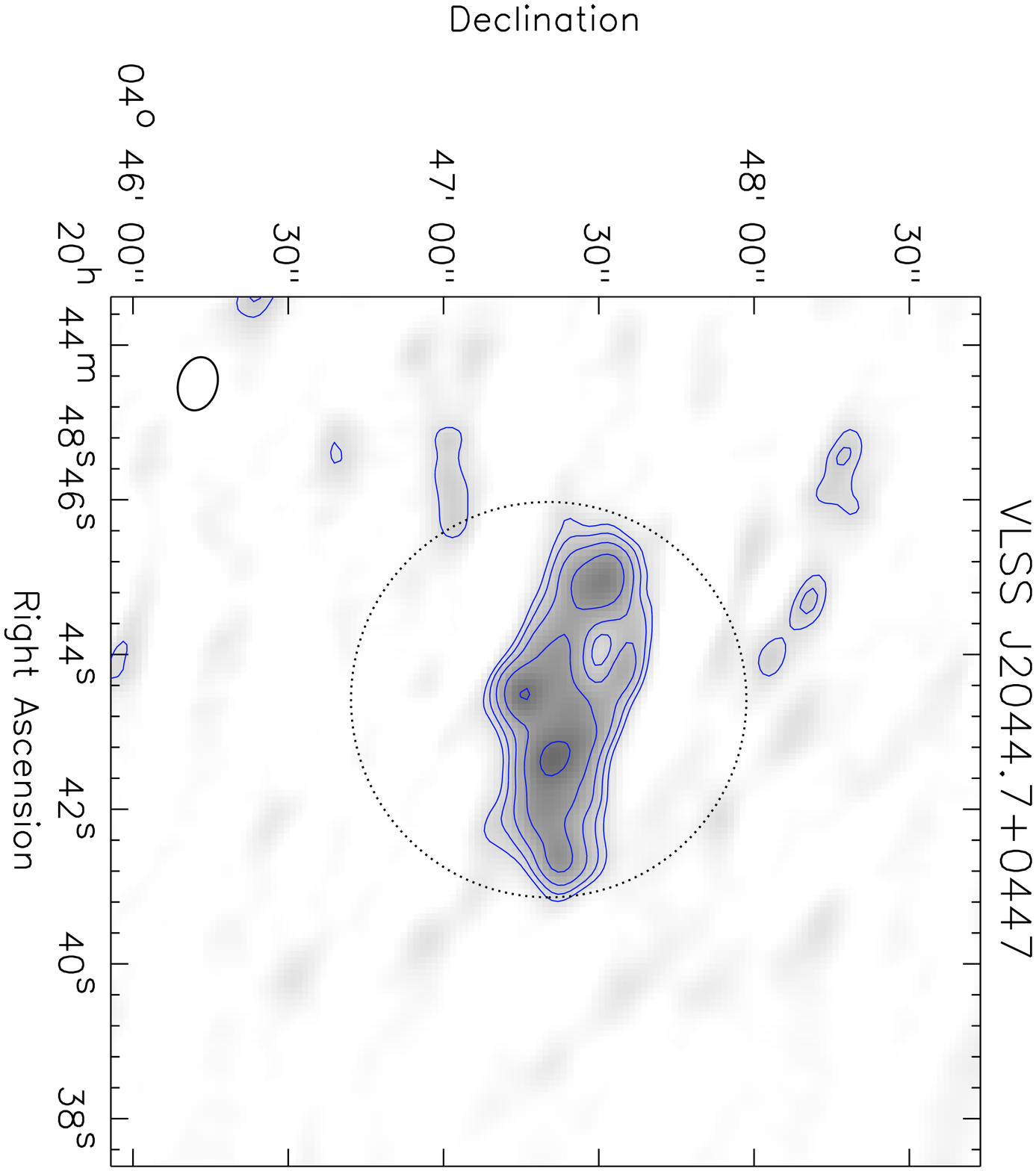}
      \includegraphics[angle = 90, trim =0cm 0cm 0cm 0cm,width=0.3\textwidth]{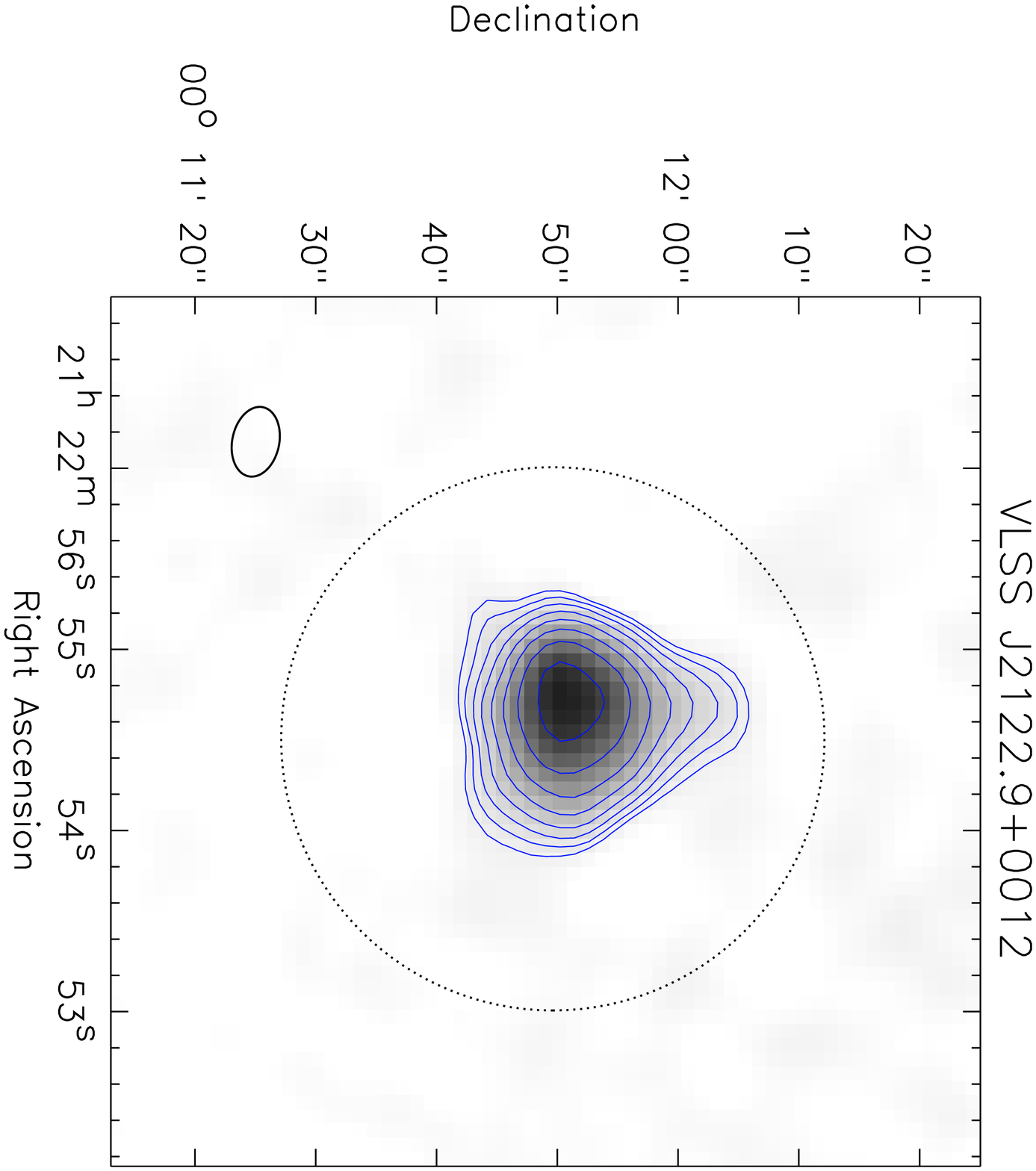}
      \includegraphics[angle = 90, trim =0cm 0cm 0cm 0cm,width=0.3\textwidth]{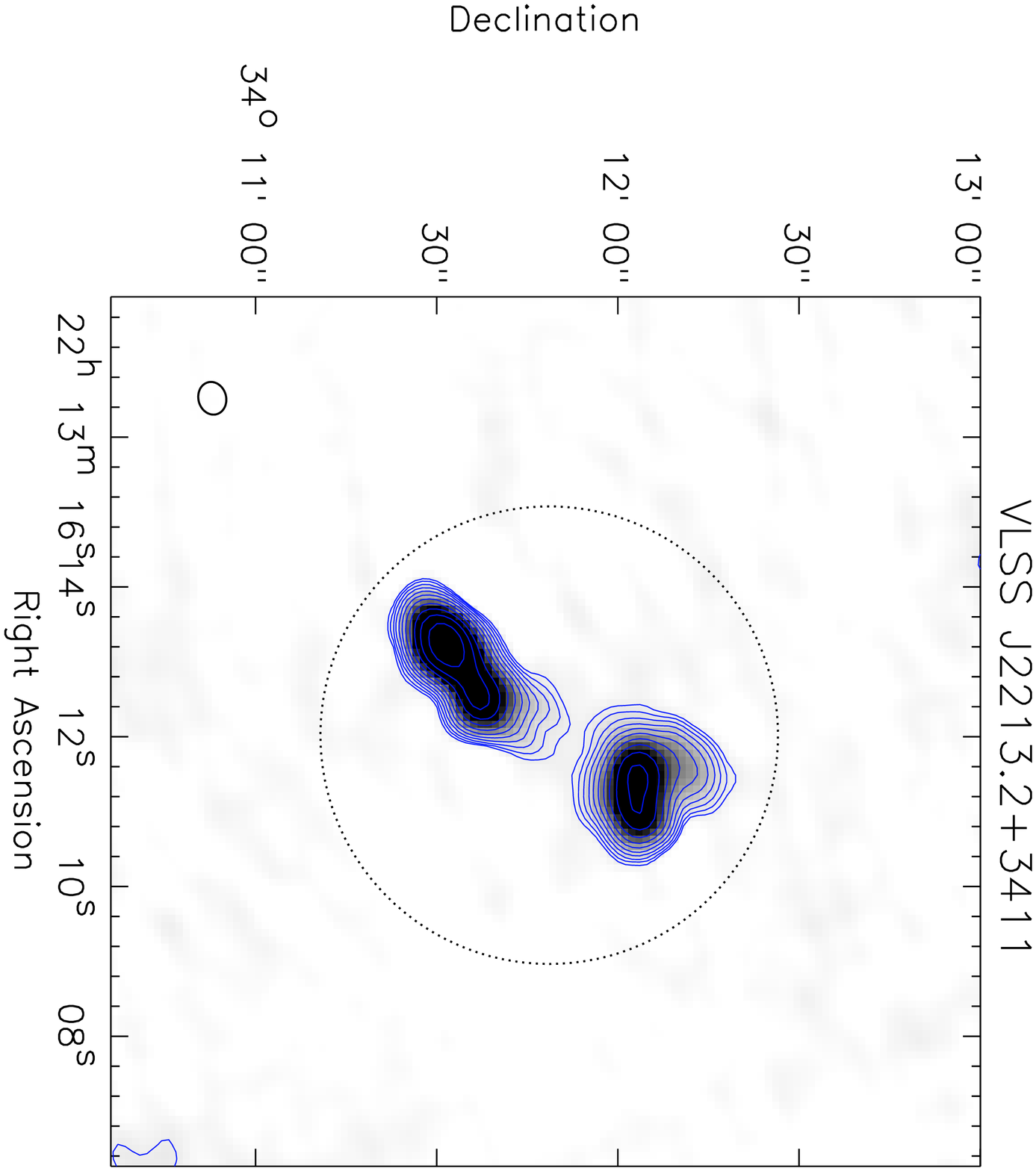}
      \includegraphics[angle = 90, trim =0cm 0cm 0cm 0cm,width=0.3\textwidth]{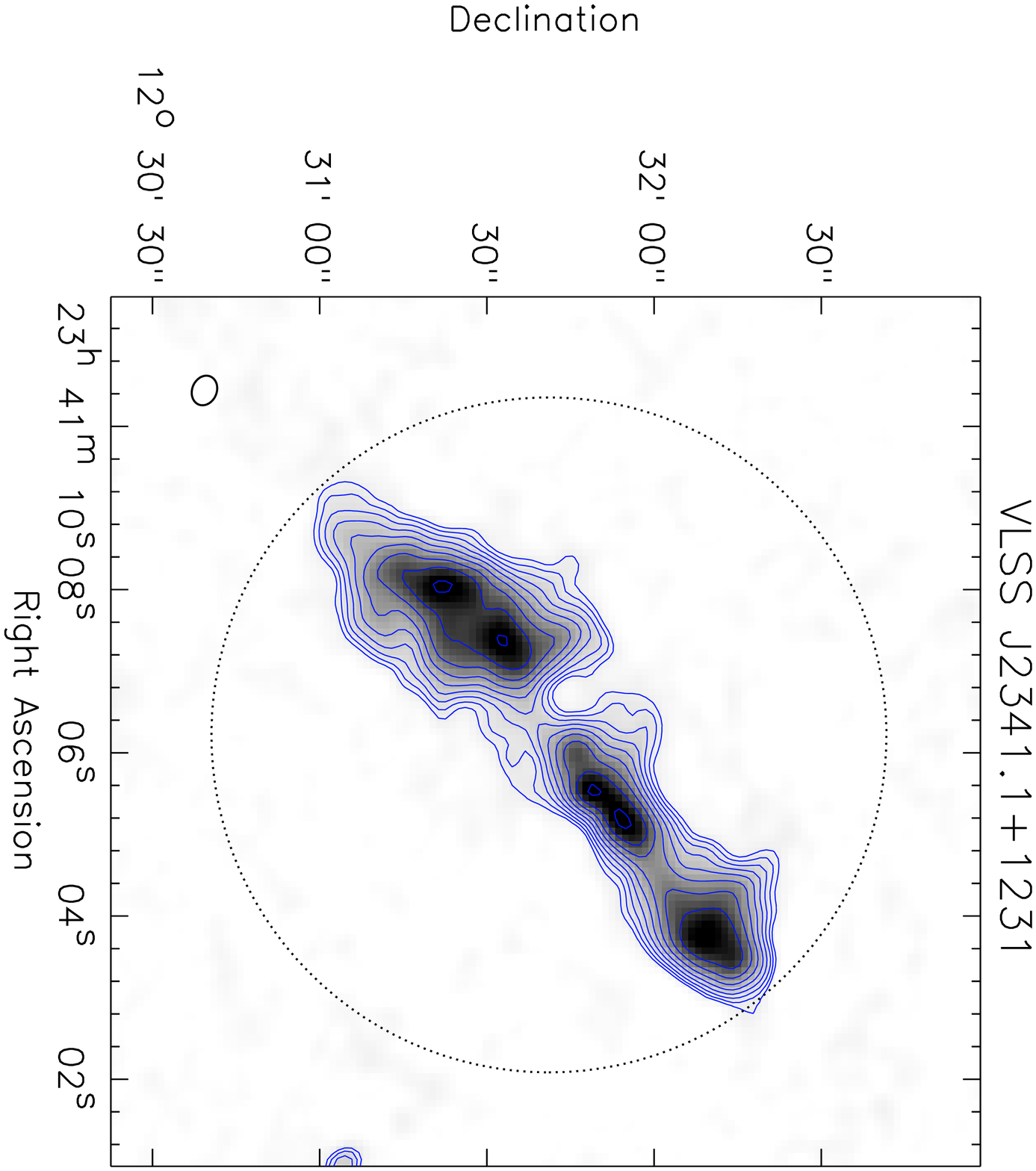}
       \end{center}
      \caption{GMRT 610 MHz radio map. Contour levels are drawn as in Fig.~\ref{fig:radiomap5}. }
              \label{fig:radiomap1}
 \end{figure*}
 
  \begin{figure*}
    \begin{center}
      \includegraphics[angle = 90, trim =0cm 0cm 0cm 0cm,width=0.3\textwidth]{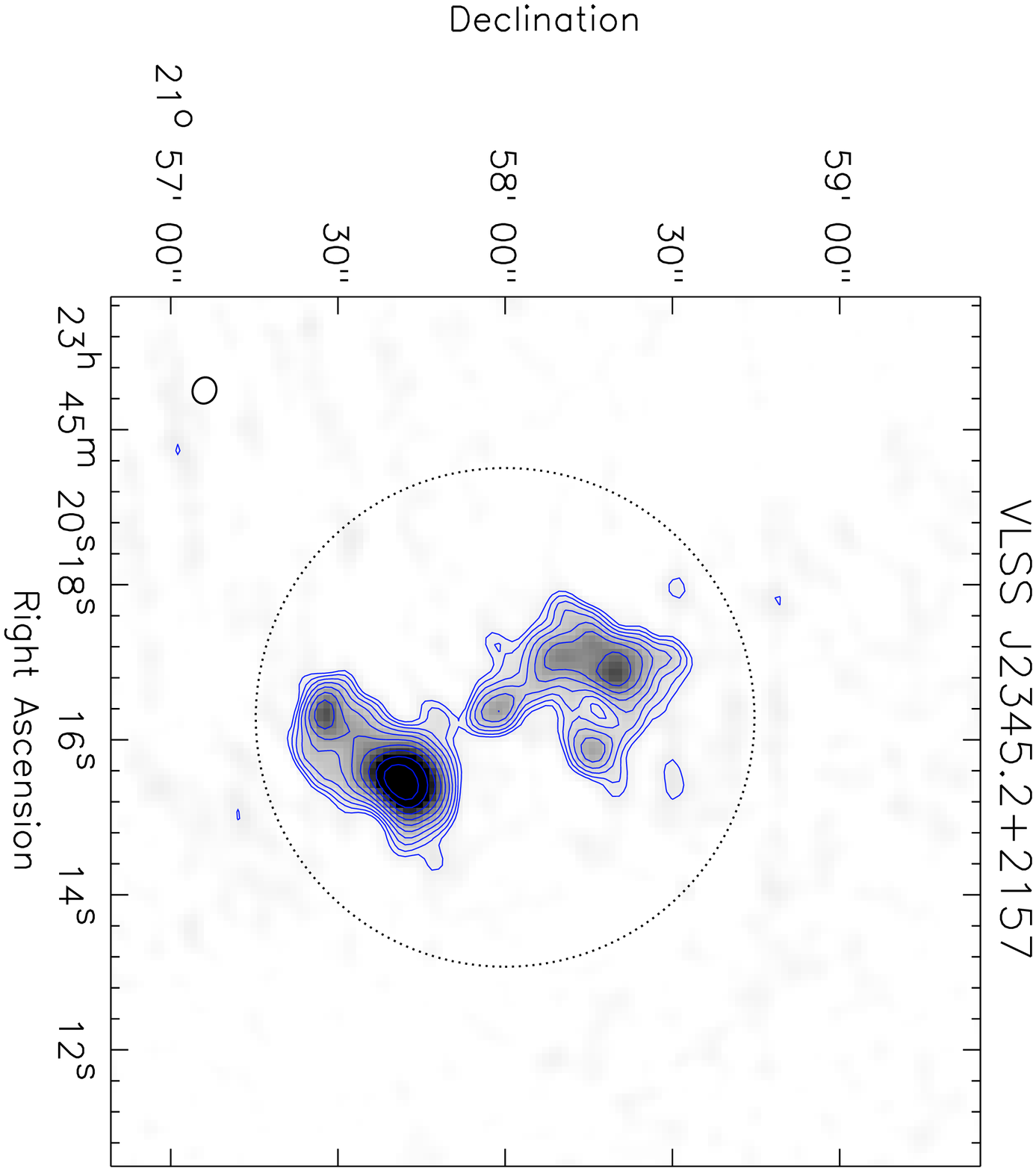}
      \includegraphics[angle = 90, trim =0cm 0cm 0cm 0cm,width=0.3\textwidth]{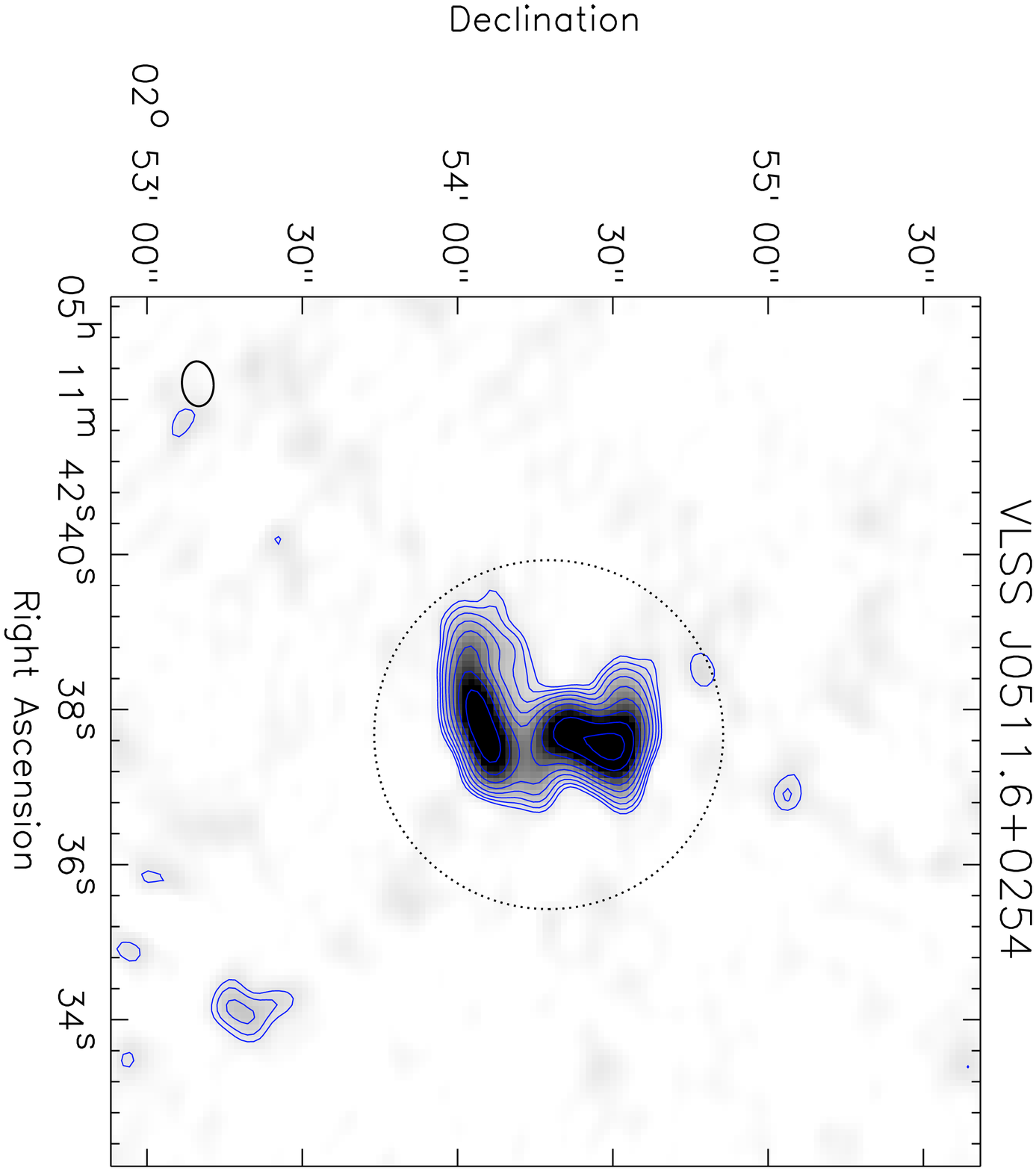}
      \includegraphics[angle = 90, trim =0cm 0cm 0cm 0cm,width=0.3\textwidth]{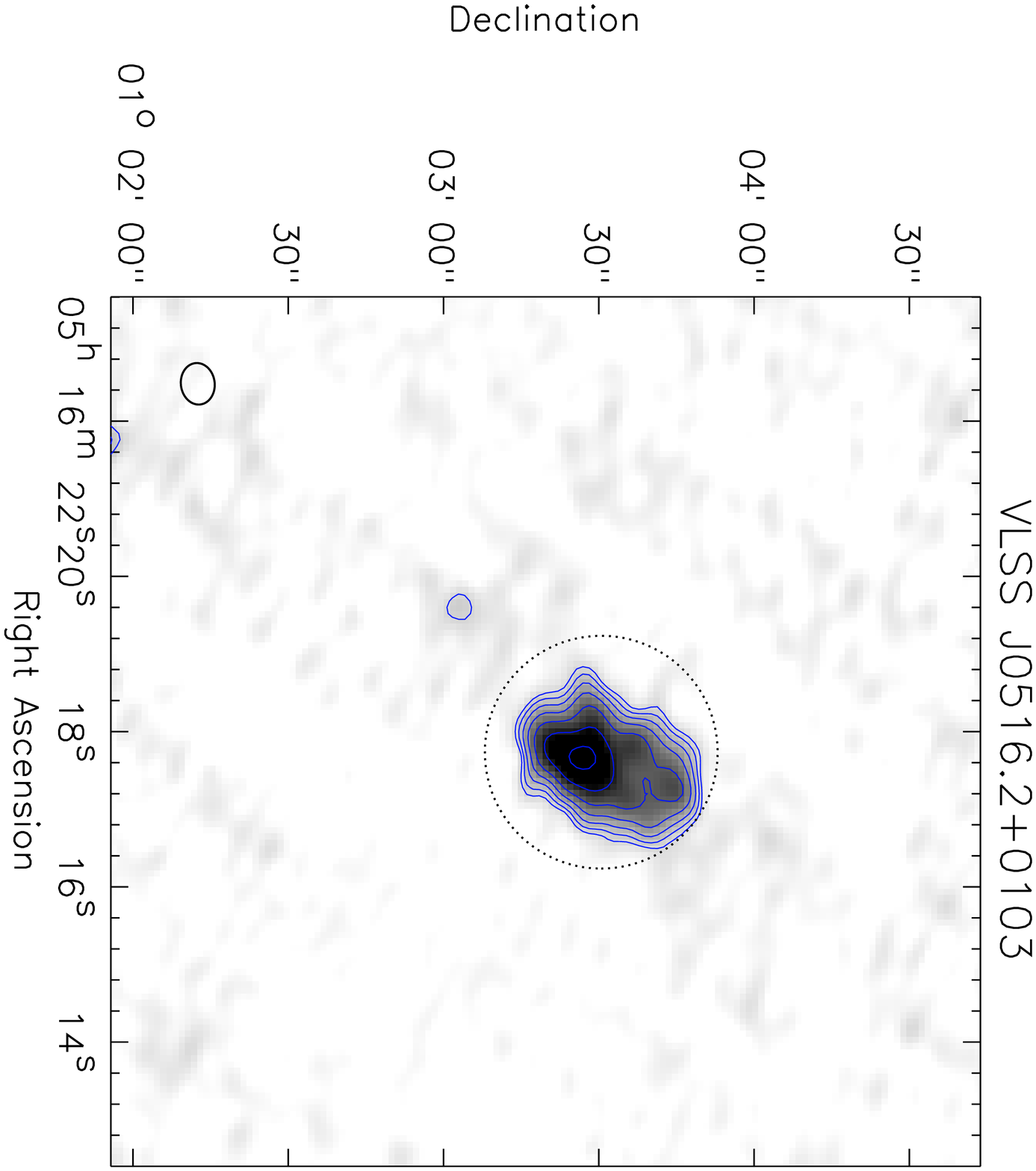}
      \includegraphics[angle = 90, trim =0cm 0cm 0cm 0cm,width=0.3\textwidth]{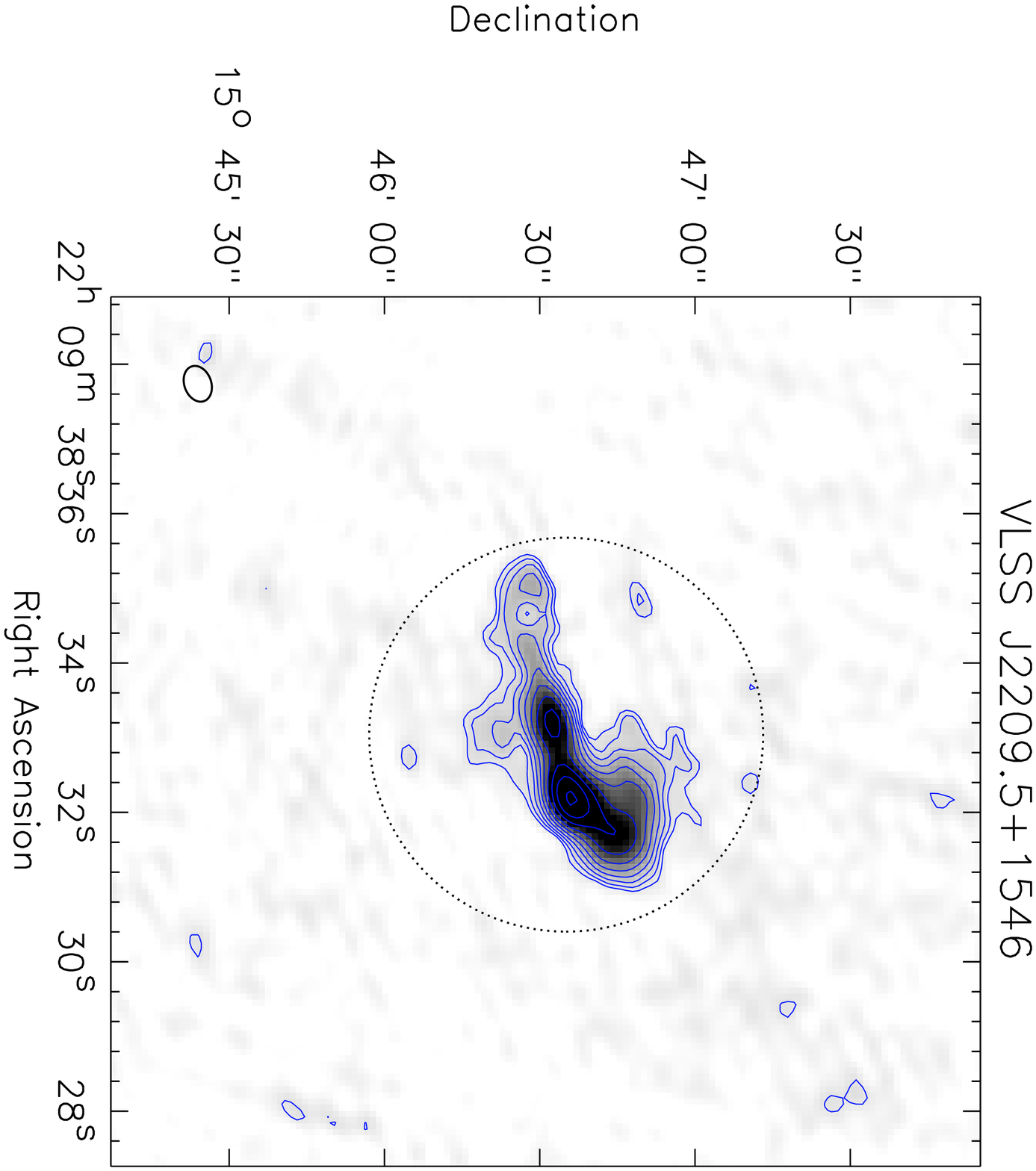}
      \includegraphics[angle = 90, trim =0cm 0cm 0cm 0cm,width=0.3\textwidth]{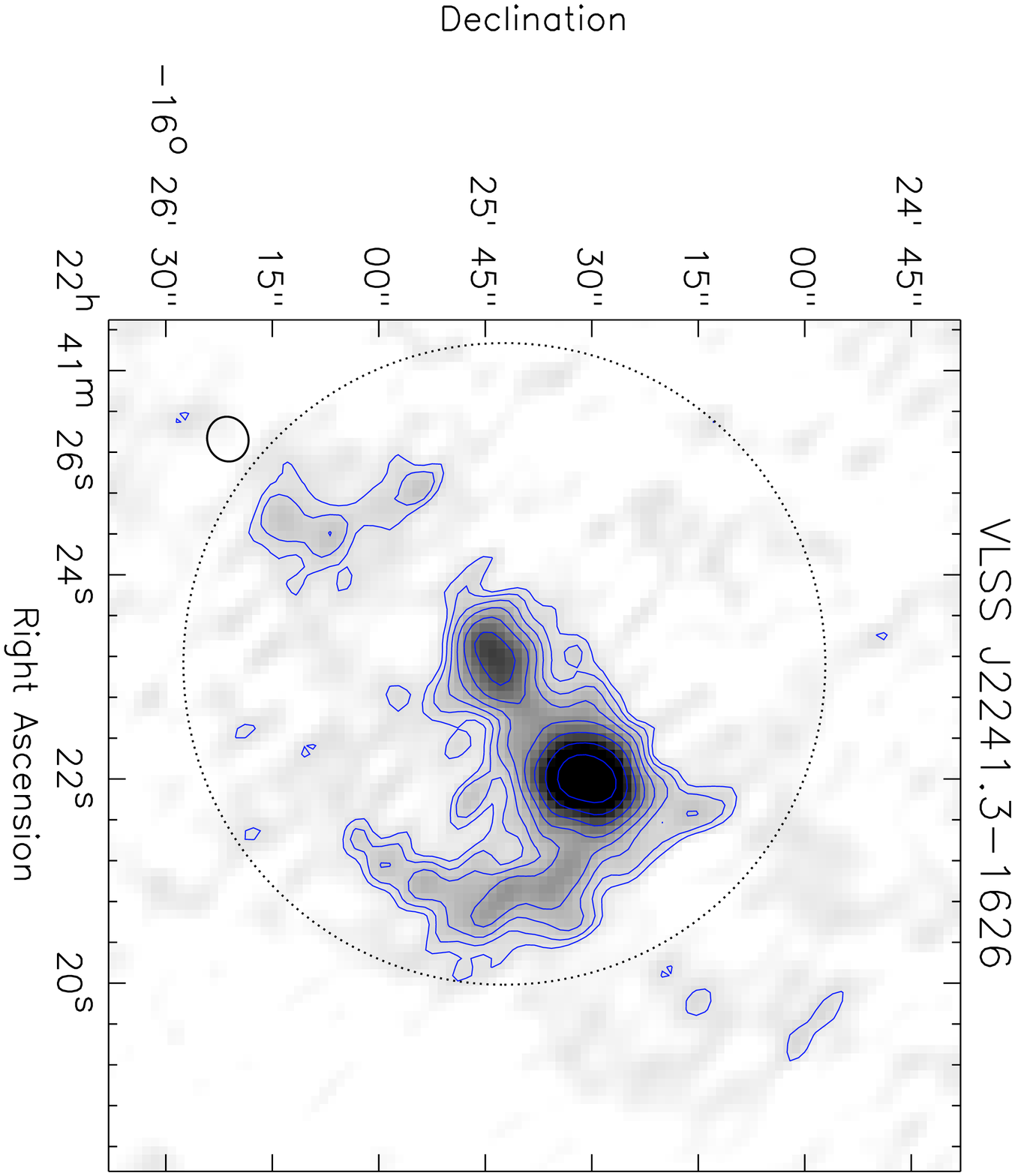}
      \includegraphics[angle = 90, trim =0cm 0cm 0cm 0cm,width=0.3\textwidth]{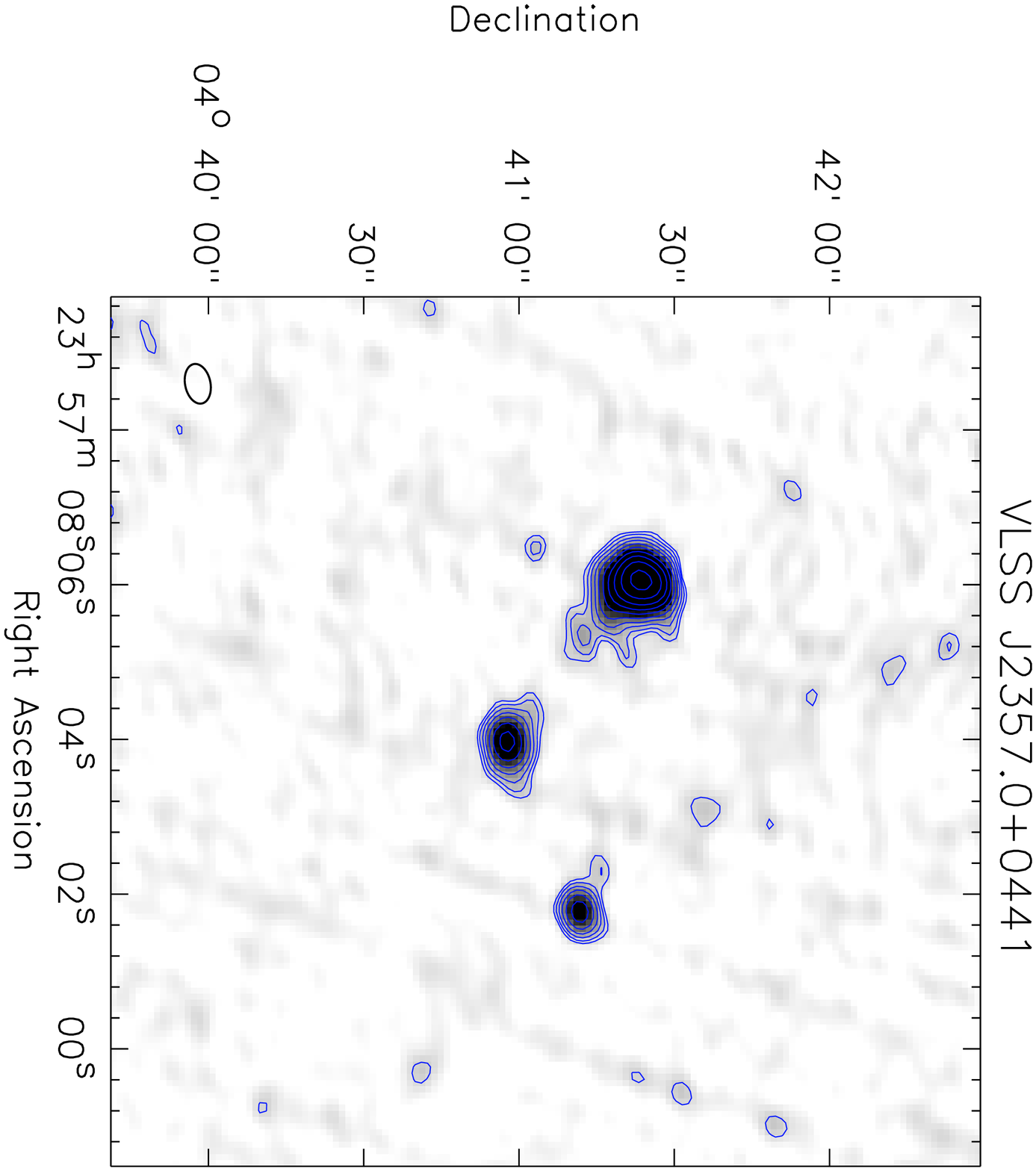}
       \end{center}
      \caption{GMRT 610~MHz radio map. Contour levels are drawn as in Fig.~\ref{fig:radiomap5}. }
              \label{fig:radiomap2}
 \end{figure*}

\clearpage

\begin{figure*}
    \begin{center}
      \includegraphics[angle = 90, trim =0cm 0cm 0cm 0cm,width=0.3\textwidth]{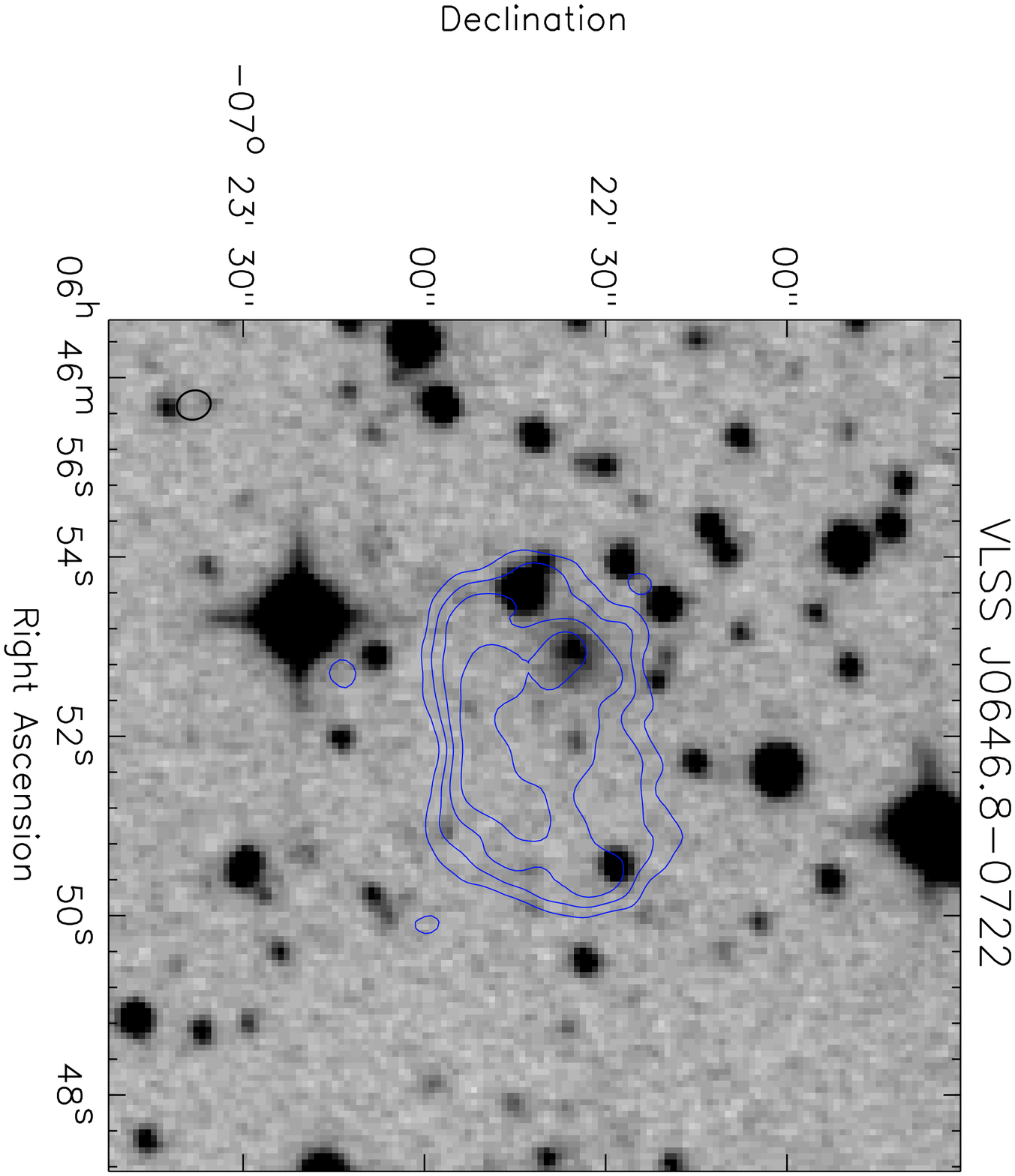}
      \includegraphics[angle = 90, trim =0cm 0cm 0cm 0cm,width=0.3\textwidth]{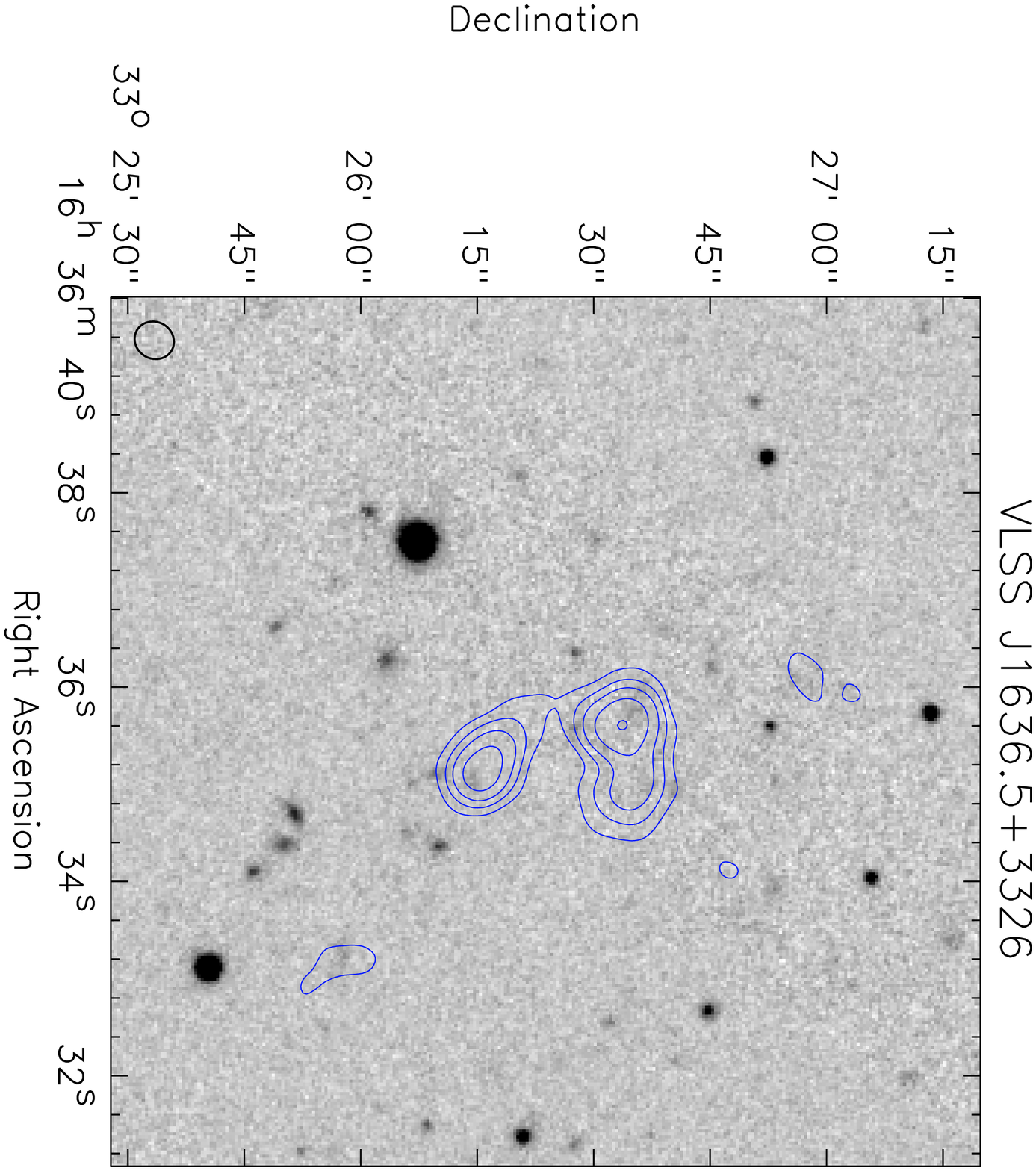}
      \includegraphics[angle = 90, trim =0cm 0cm 0cm 0cm,width=0.3\textwidth]{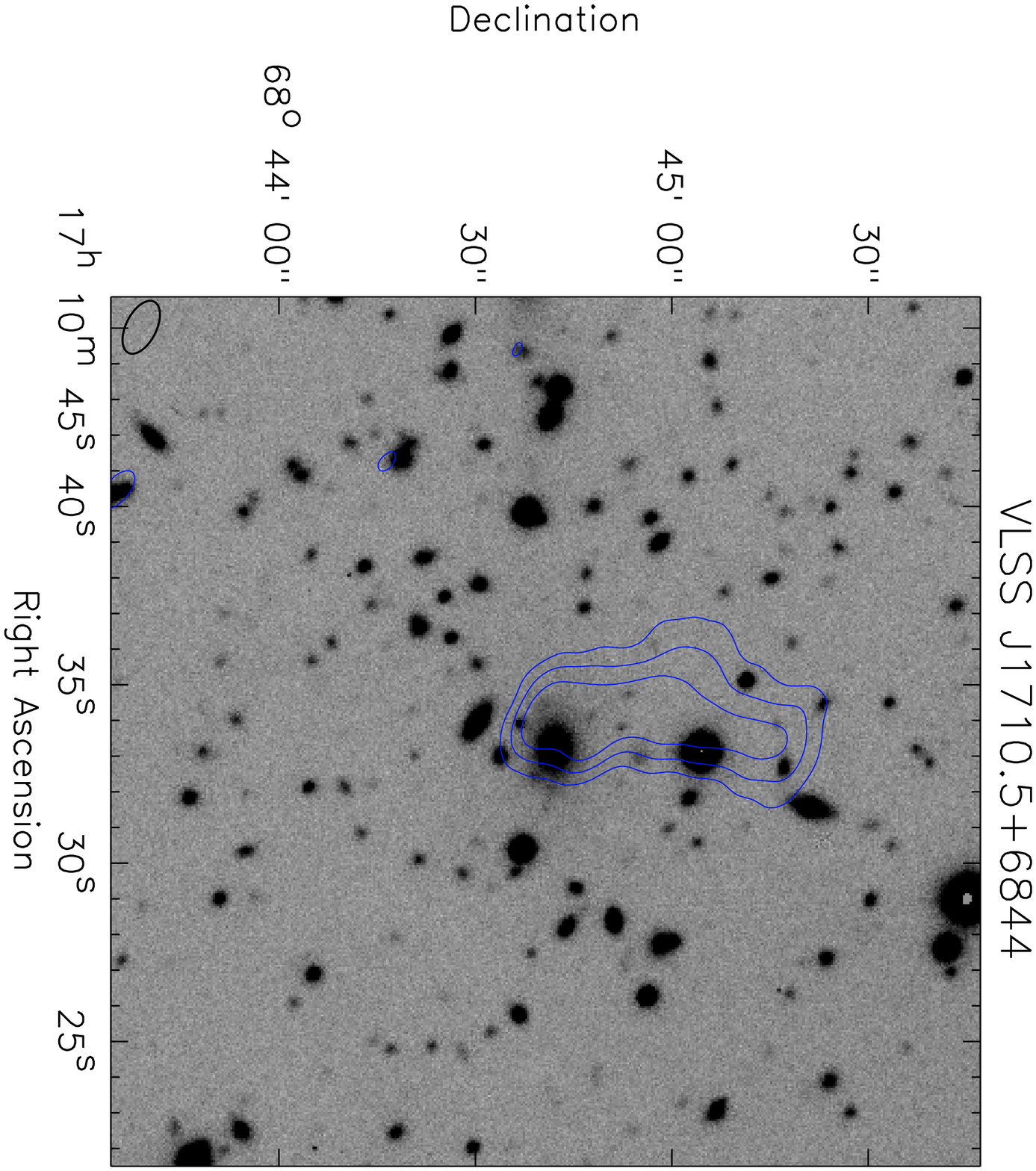}
      \includegraphics[angle = 90, trim =0cm 0cm 0cm 0cm,width=0.3\textwidth]{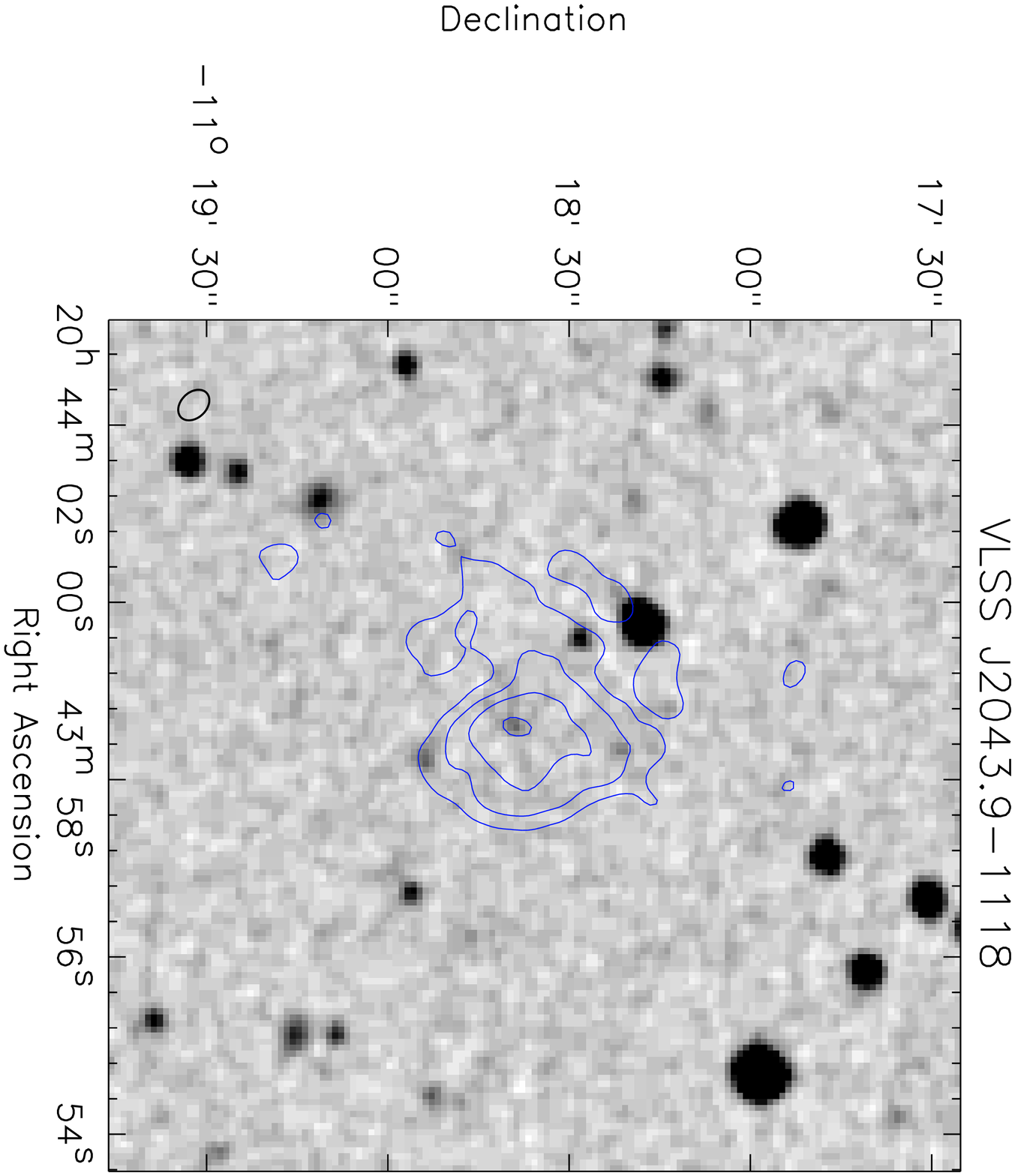}
      \includegraphics[angle = 90, trim =0cm 0cm 0cm 0cm,width=0.3\textwidth]{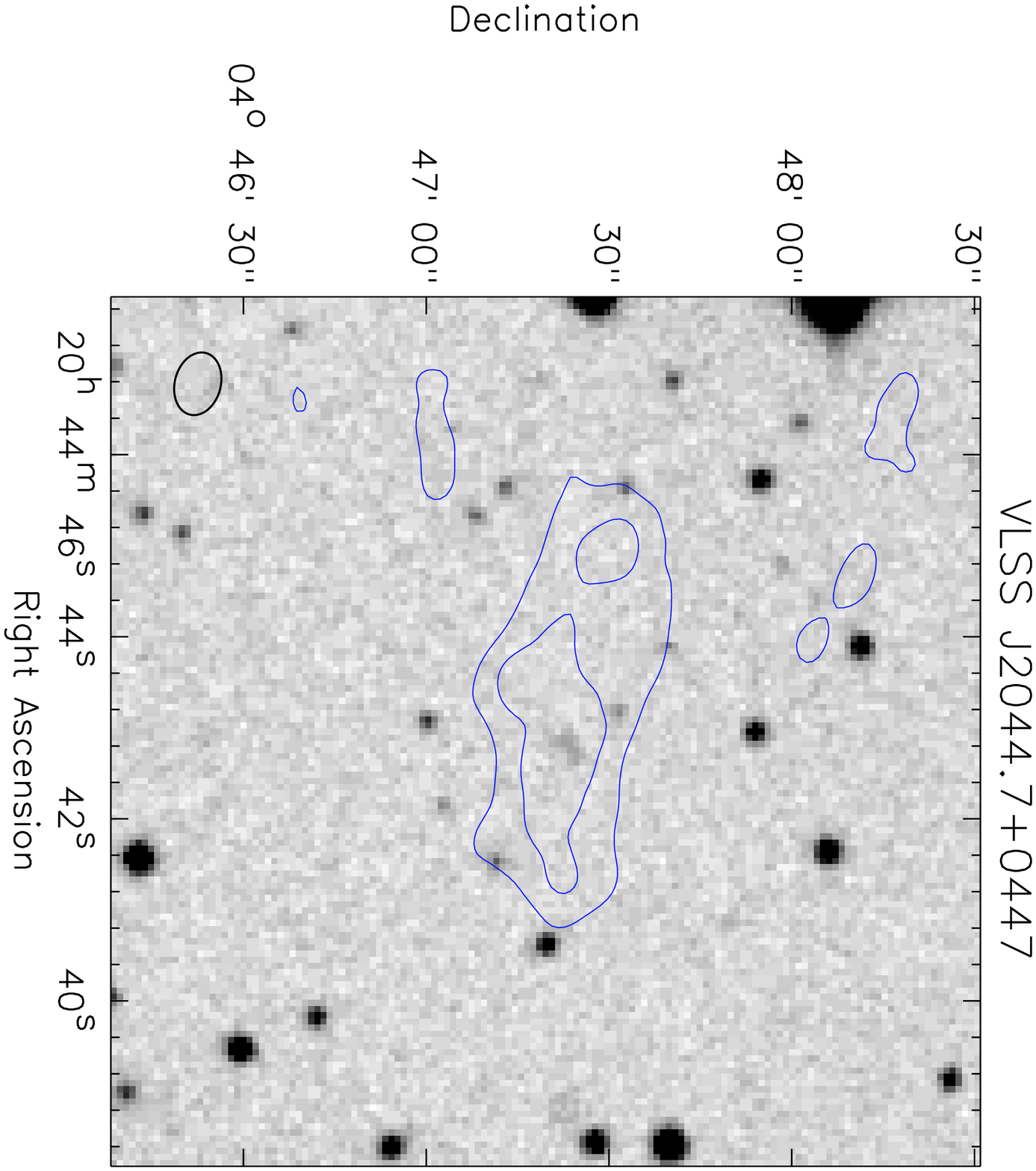}
      \includegraphics[angle = 90, trim =0cm 0cm 0cm 0cm,width=0.3\textwidth]{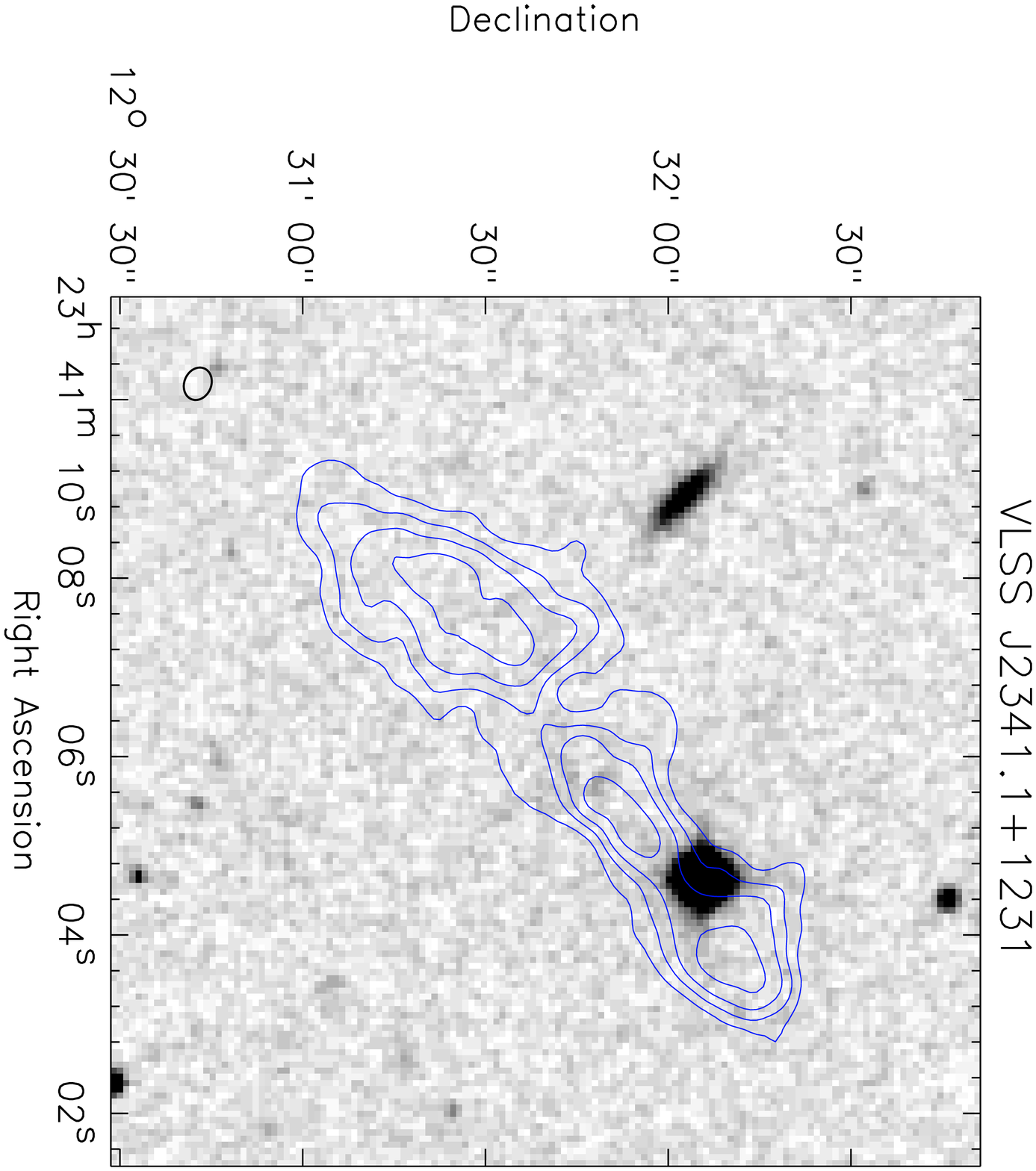}
      \includegraphics[angle = 90, trim =0cm 0cm 0cm 0cm,width=0.3\textwidth]{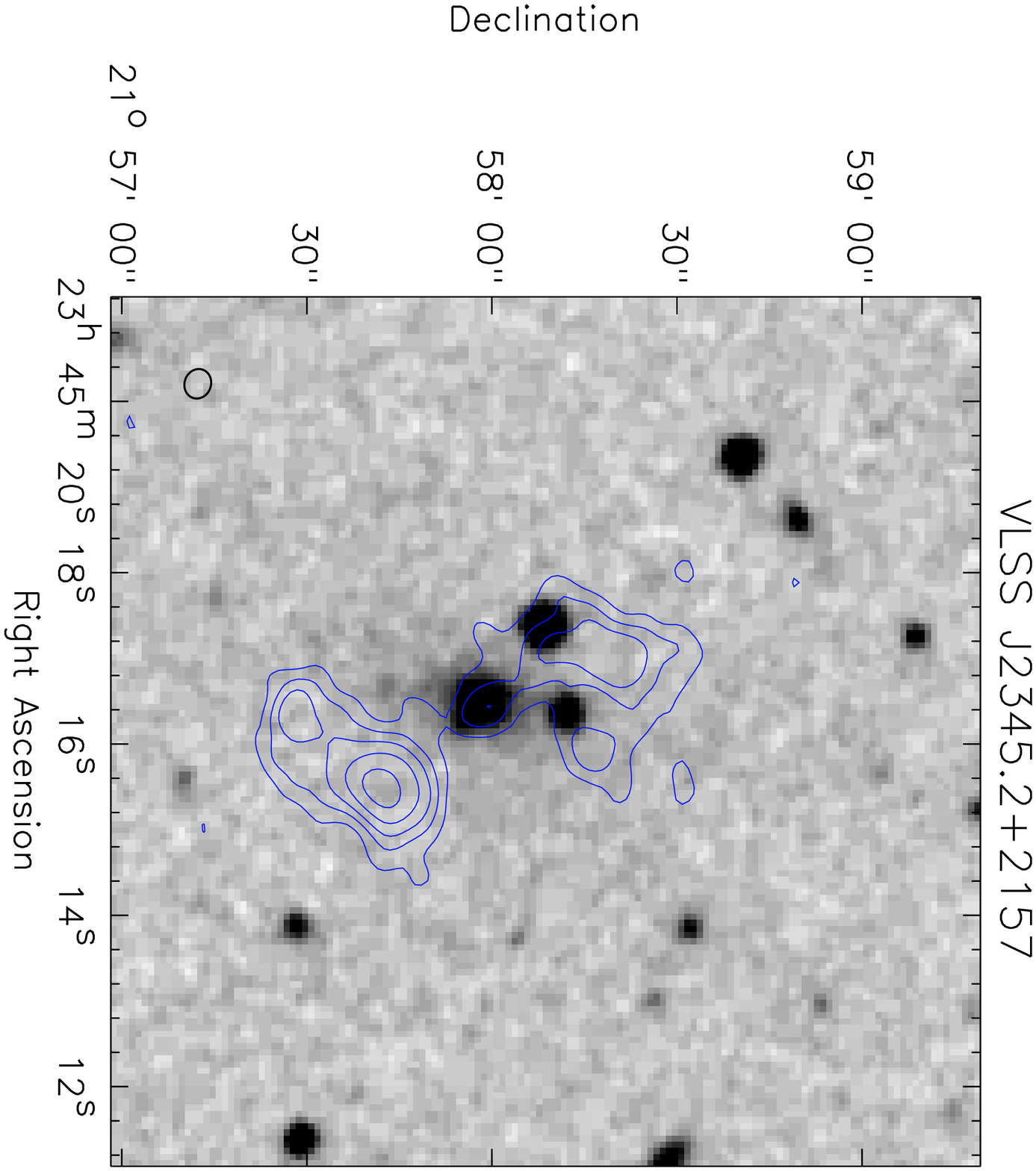}
      \includegraphics[angle = 90, trim =0cm 0cm 0cm 0cm,width=0.3\textwidth]{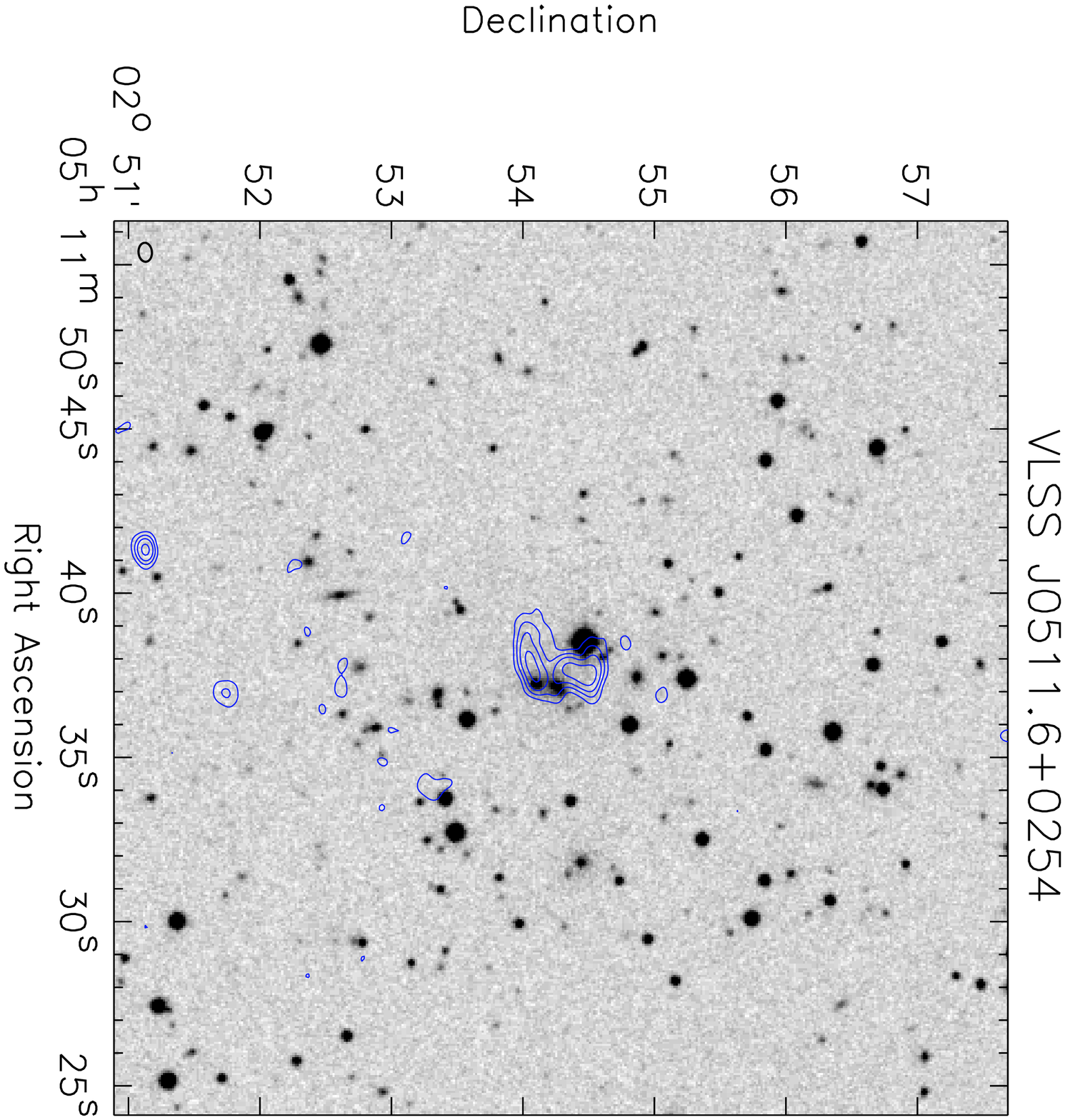}
       \end{center}
      \caption{Optical images for radio sources with counterparts. Images are from SDSS or POSS-II (red),  except for VLSS~J1710.5+6844 for which we took seven $i$-band images from the INT Wide Field Camera with a total exposure time of 2520~sec. GMRT 610~MHz contour levels are drawn as in Fig. \ref{fig:radiomap5_o}. }
              \label{fig:radiomap1_o}
 \end{figure*}

 \end{appendix}
\end{document}